\documentclass[a4paper,fleqn,usenatbib]{mnras}

\usepackage{newtxtext,newtxmath}

\usepackage[T1]{fontenc}
\usepackage{ae,aecompl}
\usepackage{tabularx}
\usepackage{tabularx,ragged2e,booktabs}

\usepackage{graphicx}	
\usepackage{amsmath}	
\usepackage{natbib}
\usepackage{hyperref}
\usepackage[font=scriptsize]{caption}
\usepackage{times}

\title[MW \& M83: opposite extremes of SF cycle]{The centres of M83 and the Milky Way: opposite extremes of a common star formation cycle}

\author[D. Callanan et al.]{Daniel Callanan,$^{1,2}$\thanks{E-mail: D.Callanan@2016.ljmu.ac.uk}
Steven~N. Longmore,$^{1}$
J.~M.~Diederik Kruijssen,$^{3}$
\newauthor
Andreas Schruba,$^{4}$
Adam Ginsburg,$^{5}$
Mark~R. Krumholz,$^{6,7}$
Nate Bastian,$^{1}$
\newauthor
Jo\~{a}o Alves,$^{8}$
Jonathan~D. Henshaw,$^{9}$
Johan~H. Knapen$^{1,10,11}$
and M{\'e}lanie Chevance$^{3}$
\\
$^{1}$Astrophysical Research Institute, Liverpool John Moores University, 146 Brownlow Hill, Liverpool L3 5RF, UK\\
$^{2}$Centre for Astronomy, Harvard University, 60 Garden Street, Cambridge MA 02138, USA\\
$^{3}$Astronomisches Rechen-Institut, Zentrum f{\"u}r Astronomie der Universit{\"a}t Heidelberg, M{\"o}nchhofstra{\ss}e 12-14, 69120 Heidelberg,\\ Germany\\
$^{4}$Max-Planck Institut f{\"u}r Extraterrestriche Physik, Giessenbachstra{\ss}e 1, 85748 Garching, Germany\\
$^{5}$National Radio Astronomy Observatory, Socorro, NM 87801, USA\\
$^{6}$Research School of Astronomy and Astrophysics, Australian National University, Canberra 2611, A.C.T., Australia\\
$^{7}$Centre of Excellence for Astronomy in Three Dimensions (ASTRO-3D), Australia\\
$^{8}$University of Vienna, T{\"u}rkenschanzstra{\ss}e 17, A-1180 Vienna, Austria\\
$^{9}$Max-Planck Institut f{\"u}r Astronomie, K{\"o}nigstuhl 17, 69117 Heidelberg, Germany\\
$^{10}$Instituto de Astrof\'{i}sica de Canarias, 38205 La Laguna, Tenerife, Spain\\
$^{11}$Departamento de Astrof\'\i sica, Universidad de La Laguna, E-38205 La Laguna, Tenerife, Spain
}

\date{Accepted XXX. Received YYY; in original form ZZZ}

\pubyear{2021}

\begin{document}
\label{firstpage}
\pagerange{\pageref{firstpage}--\pageref{lastpage}}
\maketitle

\begin{abstract}
In the centres of the Milky Way and M83, the global environmental properties thought to control star formation are very similar. However, M83's nuclear star formation rate (SFR), as estimated by synchrotron and H$\alpha$ emission, is an order of magnitude higher than the Milky Way's. To understand the origin of this difference we use ALMA observations of HCN ($1-0$) and HCO$^{+}$ ($1-0$) to trace the dense gas at the size scale of individual molecular clouds (0.54$\arcsec$, 12pc) in the inner $\sim$500\,pc of M83, and compare this to gas clouds at similar resolution and galactocentric radius in the Milky Way. We find that both the overall gas distribution and the properties of individual clouds are very similar in the two galaxies, and that a common mechanism may be responsible for instigating star formation in both circumnuclear rings. Given the considerable similarity in gas properties, the most likely explanation for the order of magnitude difference in SFR is time variability, with the Central Molecular Zone (CMZ) currently being at a more quiescent phase of its star formation cycle. We show M83's SFR must have been an order of magnitude higher $5-7$\,Myr ago. M83's `starburst' phase was highly localised, both spatially and temporally, greatly increasing the feedback efficiency and ability to drive galactic-scale outflows. This highly dynamic nature of star formation and feedback cycles in galaxy centres means (i) modeling and interpreting observations must avoid averaging over large spatial areas or timescales, and (ii) understanding the multi-scale processes controlling these cycles requires comparing snapshots of a statistical sample of galaxies in different evolutionary stages.

\end{abstract}




\section{Introduction}\label{sec:INT}

Determining how star formation varies with environment is a key step towards understanding how galaxies build their stellar mass over time. Most of what is known about the detailed processes of star formation on proto-stellar core scales comes from observations of star forming regions in the Solar neighbourhood \citep{2014prpl.conf..125M}. From studies of star formation regions on larger scales within our own Galaxy and external galaxies, we have learned that there exists a strong correlation between star formation rate surface density and gas surface density, although the exact form of this correlation is debated \citep{1998ARA&A..36..189K, 2008AJ....136.2846B, 2008AJ....136.2782L, 2011AJ....142...37S, 2011ApJ...739...84G, 2012ApJ...745..190L, 2012ApJ...753...16K}. Such relations are fundamental in the context of galaxy evolution because they dictate the location and rate at which galaxies grow their stellar mass. A major goal of star formation research is to build a bottom-up understanding of how these global  star formation relations are shaped by the physics of star formation on proto-stellar scales.

One particularly interesting region in this regard is the Central Molecular Zone (CMZ). This is the largest reservoir of dense molecular gas in the Galaxy, extending to a galactocentric radius of 250 pc. The CMZ contains roughly 5\% of our Galaxy's molecular gas \citep{1998A&A...331..959D}, putting the surface density at two orders of magnitude higher than the Milky Way average, and is subject to some of the most extreme conditions for star formation in our Galaxy. With pressures several orders of magnitude larger than those found in the Galactic disk \citep{1996ARA&A..34..645M}, temperatures reaching several hundreds of Kelvin \citep{2013ApJ...772..105M, 2016A&A...586A..50G, 2017ApJ...850...77K}, and densities of > 10$^{4}$ cm$^{-3}$ on spatial scales of 1 pc \citep{2013MNRAS.433L..15L}, the properties of the molecular gas found within this region are similar to those in galaxies at redshift $1 < z < 2$ \citep{2013MNRAS.435.2598K}. The proximity of this gas to the supermassive black hole (SMBH) Sagitarrius A* and the nuclear star cluster \citep{2010RvMP...82.3121G} means it has potentially been exposed to significant active galactic nuclei (AGN) and star formation feedback in the past, despite the SMBH currently being in a quiescent state \citep{Sofue1984, 2010ApJ...724.1044S}. This region of our Galaxy therefore provides a unique laboratory to study the star formation process in an extreme environment, similar to those commonly found in the early Universe.

Studies of the CMZ have advanced our understanding of how extreme environments can impact star formation \citep{2014prpl.conf..291L, 2019BAAS...51c.220G, Barnes2019, Barnes2020}. Recent and ongoing Galactic plane surveys across the electromagnetic spectrum \citep{2011ApJS..192....4A, 2018A&A...612A...1H}, large scale surveys of the Galactic Centre such as HOPS \citep{2011MNRAS.416.1764W}, SWAG \citep{2017ApJ...850...77K}, CMZoom \citep{Battersby2020, Hatchfield2020}, CHIMPS2 \citep{Eden2020} and SOFIA/FORCAST \citep{Hankins2020} as well as more targeted observations \citep{2012ApJ...746..117L, 2016MNRAS.463L.122H, 2018MNRAS.474.2373W, 2018ApJ...864L..17G, 2019MNRAS.485.2457H, Lu2019, Lu2020} continue to elucidate these processes in more detail.

However, future progress in this area is hampered by the difficulty in unambiguously constraining the three-dimensional geometry of the gas and young stars \citep{2015MNRAS.447.1059K,2016MNRAS.457.2675H,2018Galax...6...55L}. In addition, the CMZ only represents a single snapshot of the star formation/feedback and AGN feeding/feedback baryon cycle, which may vary in time \citep{2014MNRAS.439.3239K, 2015MNRAS.453..739K, 2017MNRAS.466.1213K, 2019MNRAS.490.4401A}. Both detailed simulations of gas flows in CMZ-like environments and observations of gas in the CMZ suggest the inflow is clumpy, supporting the notion of high variability with time \citep{2015MNRAS.449.2421S, 2015MNRAS.451.3437S, 2015MNRAS.454.1818S, 2018MNRAS.475.2383S, 2019MNRAS.484.1213S}.

Many of these problems can be overcome by studying the centres of other galaxies, with favourable (close to face-on) orientations and varying levels of star formation and AGN activity. In this paper we look to extend our understanding of star formation in extreme environments and test recent models of baryon cycles in the centres of barred spiral galaxies. To do this, we use high resolution ALMA data to study the distribution of dense gas and young stars in the central few hundred parsecs of the nearby Milky Way-like galaxy, M83 (NGC 5236).

In this paper we will use ALMA observations of dense gas tracers HCN and HCO$^{+}$ to measure the kinematics and dense gas properties of the central region of M83, and in turn compare these observations to the CMZ. In \autoref{sec:comp} we summarize previous observations of the centre of M83. \autoref{sec:obs} presents the ALMA observations and data reduction. In \autoref{sec:results} we derive the physical and kinematic properties of dense gas in the centre of M83 down to the size scale of individual molecular clouds. In \autoref{sec:mw_m83_comp} we compare the properties of dense gas clouds and young stellar clusters in the centre of M83 and the Milky Way. In \autoref{sec:broken_or_extreme} we seek to explain the order of magnitude offset in star formation rate between these two galaxy centers and discuss the implications of our findings for understanding star formation and feedback in the centres of galaxies. Finally, we summarise our conclusions in \autoref{sec:conclusion}.

\section{The centres of M83 and the CMZ: twins at heart?}\label{sec:comp}

The centre of M83 was selected for comparison with the CMZ for four key reasons: (1) M83 is nearby (4.6 Mpc, Table~\ref{table:M83}) allowing us to make comparisons with the CMZ at the scales of individual molecular clouds; (2) it has a similar physical structure, morphology, metallicity and gas/star content as the Milky Way within the central few hundred parsecs (Table~\ref{table:comparison}); (3) its moderate inclination of 24$^{\circ}$ \citep{1979ApJ...229...91T} allows for an almost unobscured view of the galactic centre; (4) while the CMZ is currently under-producing stars compared to dense gas relations \citep{2013MNRAS.429..987L, 2017MNRAS.469.2263B}, the centre of M83 is over-producing stars when compared to \citet{2012ApJ...745..190L}. This order of magnitude difference in star formation rate, despite having similar stellar and gas properties when averaged on hundred-pc scales, simultaneously provides a key test of star formation theories in extreme environments and of models of baryon cycles in galaxy centres.

\begin{figure}
	\includegraphics[width=0.48\textwidth]{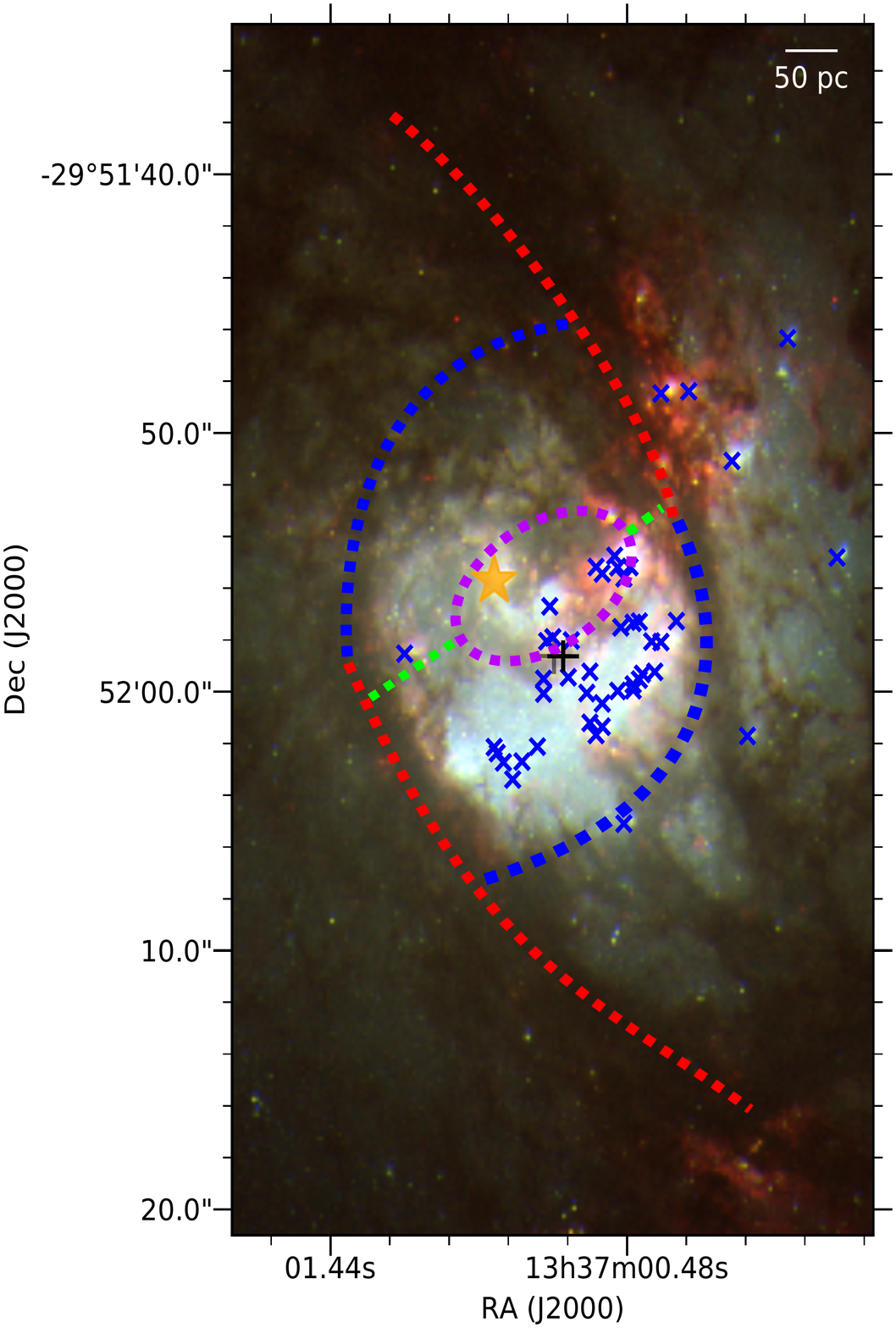}
    \caption{\normalsize Hubble Space Telescope three-colour image of the inner 1 kpc $\times$ 0.6 kpc of M83 (red = H$\alpha$, green = WFPC2 V, blue = WFPC2 B). Also labelled are the massive stellar cluster positions (blue $\times$) as observed by \citep{2001AJ....122.3046H} as well as M83's visible nucleus (orange $\star$). The weighted average location of the kinematic centre from \citet{2010MNRAS.408..797K} is shown as a black plus. A schematic of the main structural components observed by \citet{1998AJ....116.2834E} is overlaid and separated by colour. Red represents the dust lanes, blue is the outer circumnuclear ring, purple the inner circumnuclear ring and green represents the narrow bar or `bridge' connecting these latter two components.}
    \label{fig:schem_circ}
\end{figure}

As one of the nearest, face on ($i = 24^{\circ}$), massive \citep[M$_{*} = 6.4 \times 10^{10}$~M$_{\odot}$,][]{2004A&A...422..865L} spiral galaxies, M83 has been studied in detail across the electromagnetic spectrum: X-ray \citep{2017AAS...22914418C}, visible \citep{2014ApJ...788...55B}, near-infrared \citep{2015IAUGA..2255001W}, mid-infrared \citep{2005A&A...441..491V} and radio \citep{2006AJ....132..310M}. Observations of the Br$\alpha$ (4.05 $\mu$m) and Br$\gamma$ (2.17 $\mu$m) recombination lines of ionised hydrogen within the circumnuclear region of M83 by \citet{1987ApJ...313..644T} have shown that there is significant dust extinction within the region ($A_{\nu} \gtrsim 14$ mag), though the dust distribution is observed to be patchy. Sub-arcsecond angular resolution J and K band observations of the circumnuclear region of M83 show two prominent dust lanes (red dotted line in Figure~\ref{fig:schem_circ}) spiralling into a circumnuclear dust ring at a galactocentric radius of a few hundred pc. As shown in Figure~\ref{fig:schem_circ}, the outer circumnuclear ring (blue dotted line) is connected to an inner circumnuclear ring (purple dotted line) via a narrow inner bar or `bridge' (green dotted line) perpendicular to the primary stellar bar \citep{1998AJ....116.2834E}. The area between the two rings was identified as being a region of intense star formation. It is hypothesised that these two circumnuclear rings coincide roughly with the locations of the two inner Lindblad resonances \citep{1996FCPh...17...95B}.

\citet{2001AJ....122.3046H} identified 45 massive star clusters within the central region of the galaxy using equivalent widths of H$\alpha$ emission, with 90\% lying within the outer circumnuclear ring. 75\% of these clusters above the mass of $2 \times 10^4$M$_\odot$ are younger than 10 Myr old, and of the clusters younger than 10 Myr and more massive than $5 \times 10^3$M$_\odot$, 70\% are between 5-7 Myr. The remaining 25\% of clusters above the mass of $2 \times 10^4$M$_\odot$ range from 13-47 Myr. Of the 45 clusters, 9 are younger than 5 Myr, though 6 of these have anomalous photometry, potentially caused by dust attenuation. \cite{2001AJ....122.3046H} estimate the catalogue is complete to clusters of mass  $\geq 2 \times 10^4$M$_\odot$ for ages between $0-40$\,Myr.

There are two possible explanations for this age distribution of the clusters: (1) there was a burst of cluster formation between 5 and 7 Myr ago, with little formation occurring between 7 and 50 Myr ago; or (2) clusters did form prior to 7 Myr ago but have since dissolved into the field population. The sharp cutoff in the age distribution \citep{2001AJ....122.3046H} would suggest the former is far more likely \citep{2005A&A...429..173L, 2011MNRAS.414.1339K}. A majority of these star clusters are located within the star formation arc identified by \citet{1998AJ....116.2834E}, they are shown as blue crosses in Figure~\ref{fig:schem_circ}. The population of clusters is highly asymmetric with respect to the optical nucleus (orange star), in the south-western space between the inner and outer circumnuclear rings.

As the brightest, most compact young stellar systems, young massive clusters (YMCs) are the best tracer of star formation activity in the centre of M83 over the last $<$10\,Myr. Work by \citet{2001AJ....122.3046H} has shown a clear azimuthal age gradient in the population of YMC's in the inner $\sim$200\ pc of M83. The mass of the clouds in the circumnuclear gas stream derived by \citet{2017MNRAS.468.1769F} of $\sim10^4 - 10^6$\,M$_\odot$ provides the mass reservoir expected from a progenitor to $\sim10^3 - 10^4$\,M$_\odot$ stellar clusters, assuming a typical GMC star formation efficiency of $\sim$10\% \citep{2014prpl.conf..291L, 2019Natur.569..519K, 2020MNRAS.493.2872C}.

\citet{2004ApJ...616L..59S} confirmed the structure of the gas within the centre of M83, as first observed by \citet{1998AJ....116.2834E}, with SMA observations in CO (J=$2-1$) and CO (J=$3-2$) lines, suggesting that the dust lanes and the nuclear rings (red, blue and purple dotted lines in Figure~\ref{fig:schem_circ}) are following $x_1$ and $x_2$ orbits respectively due to their distance from the centre and orientation to the galactic bar \citep{1992MNRAS.259..328A}. They also found that while the K band isophotal centre lies on the systemic velocity contour, and as a result is likely the dynamical centre, the visible nucleus is offset from this contour. While \citet{2004ApJ...616L..59S} suggest this may be evidence of a second, hidden nucleus, \citet{2010MNRAS.408..797K} rule out a hidden nucleus due to a lack of enhancement in optical or near-IR emission among other reasons. They instead conclude that it is more probable that the visible nucleus is the only nucleus of M83 and that the offset of the kinematic centre is due to some extreme past event such as a merger or a galaxy-galaxy interaction. Following this argument, in this paper we focus only on the optical nucleus, illustrated by the star in Figure~\ref{fig:schem_circ}.

The structural components of the gas in the centre of M83 and the CMZ are similar, particularly the circumnuclear rings and the dust lanes feeding gas into the region, but they differ in some aspects that may play a role in how the gas evolves in these regions. Models seeking to interpret the 3D geometry and kinematics of the dense gas in the CMZ find the data is well fit by a gas stream orbiting the centre at a radius of $\sim100$ pc \citep{2011ApJ...735L..33M, 2015MNRAS.447.1059K}. However, due to our intrinsic edge-on view of the CMZ, these fits are model dependent, complicating direct comparisons. While the supermassive black hole of our own Galaxy, Sagittarius A*, is also known to be displaced from the geometric centre of symmetry of this orbit in much the same way we observe the visual nucleus within M83 to be, M83's visual nucleus is significantly more offset which will effect the gravitational potential within the region. \citet{2019MNRAS.484.1213S} recently proposed that much of the gas in the inner kpc of the Milky Way outside of the $\sim 100$ pc stream belongs to dust lanes feeding gas into the CMZ, which are analogous to the dust lanes seen in the centre of M83 (see red dashed lines in Figure~\ref{fig:schem_circ}).

\begin{table}
	\centering
	\caption{\textbf{M83 Characteristics}}
	\begin{tabularx}{0.48\textwidth}{p{3cm} p{3cm} p{3cm}}
		\hline
		Parameter & Value & Reference \\
		\hline
		$\alpha_{\text{J2000}}$ & 13:37:00.91 & (1) \\
		$\delta_{\text{J2000}}$ & -29:51:55.7 & (1) \\
		v$_{\text{lsr}}$ & 519 km s$^{-1}$ & (2) \\
		RC3 Type & SAB(s)c & (3)\\
		Inclination & 24$^{\circ}$ & (4) \\
		Position Angle & 225$^{\circ}$ & (5) \\
		Distance & 4.6 Mpc (1$" = 22.3$ pc) & (6)\\
		\hline
	\end{tabularx}
	\\[0.2cm]
	(1) \citet{Houghton2008}; (2) \citet{2004MNRAS.350.1195M}; (3) \citet{2013MNRAS.428.1927C}; (4) \citet{1981A&AS...44..441C}; (5) \citet{2012MNRAS.421.2917F}; (6) \citet{2013AJ....146...86T};
    \label{table:M83}
\end{table}

\begin{table}
	\caption{\textbf{CMZ-Inner M83 Comparison}}
	\begin{tabularx}{0.48\textwidth}{p{4.0cm} p{1.8cm} p{1.8cm}}
		\hline
		Parameter & CMZ & Inner M83 \\
		\hline
		Gas Content, M$_{\odot}$ & $5 \times 10^{7(1)}$ & $5 \times 10^{7(2)}$ \\
		Stellar Content M$_{\odot}$ & $10^{9(1)}$ & $5 \times 10^{8(3)}$ \\
		Circular Velocity, km s$^{-1}$& $\sim 100^{(4)}$ & $\sim 100^{(5)}$ \\
        Velocity Dispersion, km s$^{-1}$ & $\sim 20^{(6)}$ & $\sim 20^{(5)}$ \\
		Gas Surface Density, M$_{\odot}$pc$^{-2}$ & $10^{2-3(1)}$ & $10^{2-3(7)}$\\
		Metallicity & twice solar$^{(8)}$ & twice solar$^{(9)}$ \\
		Star Formation Rate, M$_{\odot}$yr$^{-1}$ & $0.08^{(10)}$ & $0.8^{(11)}$ \\
		Gas depletion time, Gyr & $0.6^{(1,10,12)}$ & $0.06^{(2,11)}$ \\
		\hline
	\end{tabularx}
	\\[0.2cm]
	A comparison of key physical, chemical and kinematic characteristics of the inner few hundred parsecs of the Milky Way and M83. Every characteristic except the star formation rate and gas depletion time is the same to within a factor of a few.\\ References: (1) \citet{2002A&A...384..112L}; (2)  \citet{2001A&A...371..433I}; (3) \citet{2008ApJ...675L..17F}; (4)
	\citet{2019ApJ...870L..10M}; (5)
	\citet{2004A&A...422..865L}; (6)
	\citet{2012MNRAS.425..720S}; (7)
	\citet{2004A&A...413..505L}; (8) \citet{2016A&A...585A.105L};
	(9) \citet{2014ApJ...787..142G}; (10) \citet{2017MNRAS.469.2263B}; (11) \citet{2007PASJ...59...43M}; (12) \citet{kruijssen14b}
	\label{table:comparison}
\end{table}

\section{Observations}\label{sec:obs}

The data presented in this work are Atacama Large Millimetre Array (ALMA) Cycle 3 observations targeting M83 over three nights between April 18th to September 22nd, 2016 (project ID 2015.1.01177.S, PI: S. Longmore). Observations covered an area of $100^{\prime\prime} \times 120^{\prime\prime}$ centred on the nucleus of M83. The typical cloud scale within the CMZ is $\sim$10 pc \citep{2013MNRAS.433L..15L, 2016MNRAS.463L.122H} so an angular resolution of 0.54$^{\prime\prime}$ was selected, corresponding to a physical scale of 12 pc at the distance of M83. Observations were taken in configurations C36-2 and C36-7 to reliably recover spatial scales from 0.54$^{\prime\prime}$ to 25$^{\prime\prime}$ (12.4 pc to 600pc). Callisto and Titan were used as flux calibrators on the first and second nights respectively, and J1427-4206 was observed as a bandpass calibrator on all three nights. The observations consist of 4 spectral windows, centred on 86.7 GHz, 88.5 GHz, 98.6 GHz and 100.5 GHz, each with 1.875 GHz of total bandwidth. These spectral windows were chosen to include ground state rotational transitions of bright, dense gas tracers: HCN ($1-0$), HCO$^{+}$ ($1-0$), and CS ($2-1$). 

\begin{table}
	\caption{\textbf{Observations}}
	\begin{tabularx}{0.48\textwidth}{p{1cm}|p{2cm} p{2cm} p{2cm}}
	\hline
    Spectral & Channel Width & Beam Size & Sensitivity \\
    Line & [km s$^{-1}$] & [$^{\prime\prime}$] & [mJy bm$^{-1}$] \\
    \hline
    HCN & 3.8 & (0.49x0.45) & 0.47 \\
    HCO$^{+}$ & 3.9 & (0.51x0.48) & 0.48 \\
    CS & 3.5 & (0.51x0.48) & 0.28 \\
    CCH & 5.8 & (0.51x0.49) & 0.12 \\
    \hline
	\end{tabularx}
	\\[0.2cm]
	\label{table:obs_params}
\end{table}

The data were calibrated using the standard ALMA pipeline. Visual inspection of the calibrated visibilities showed that no further steps beyond the pipeline reduction were needed before imaging. The observations were then concatenated to generate a final calibrated dataset using the \textbf{C}ommon \textbf{A}stronomy \textbf{S}oftware \textbf{A}pplications (CASA) package, version 4.3 \citep{2007ASPC..376..127M}, which was then used to image the data. 

\begin{figure*}
	\centering
	\begin{tabular}{cc}
		\hspace{-2cm} \includegraphics[width=0.68\textwidth]{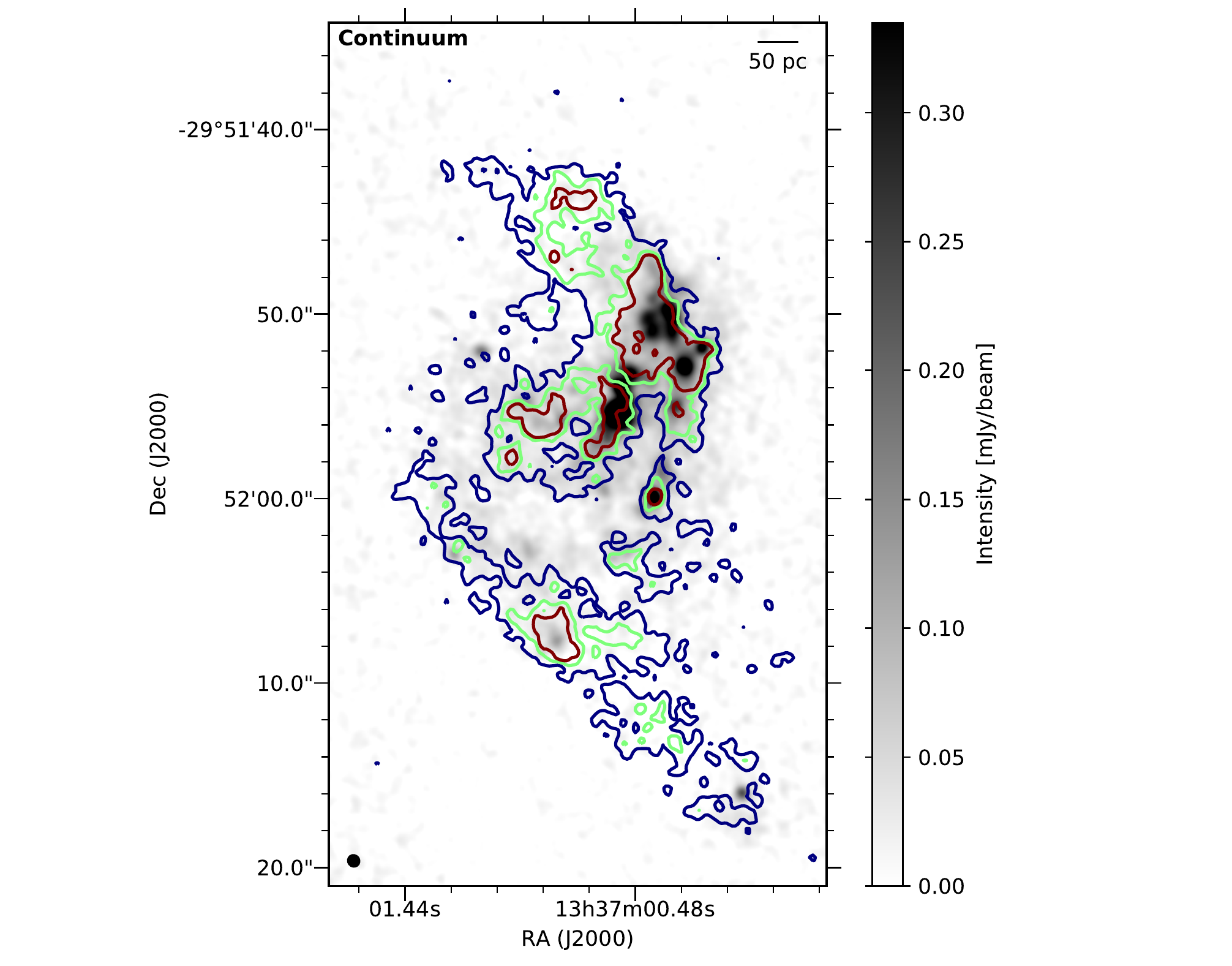} & \hspace{-2.7cm}
		\includegraphics[width=0.68\textwidth]{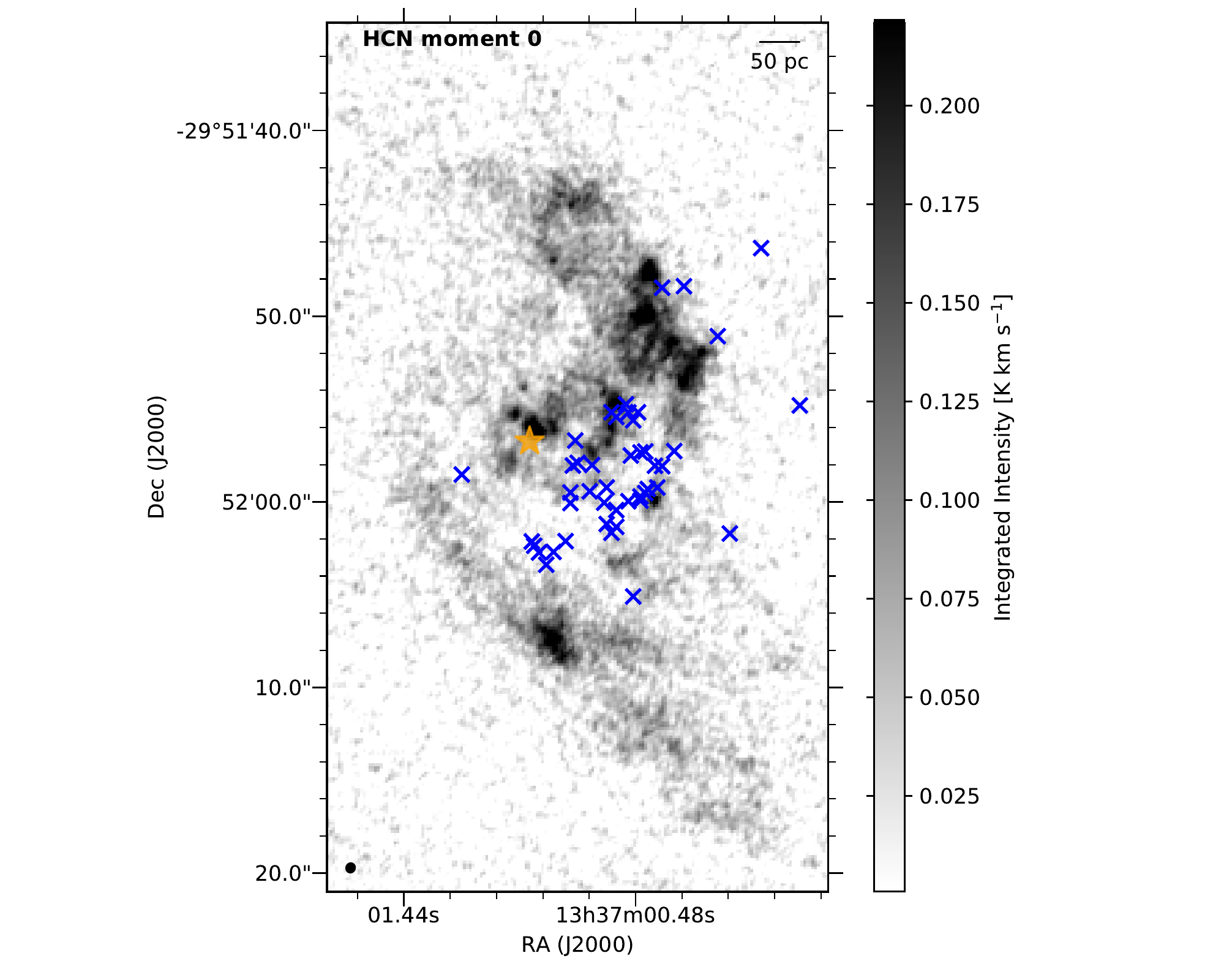} \\
		\hspace{-2cm} \includegraphics[width=0.68\textwidth]{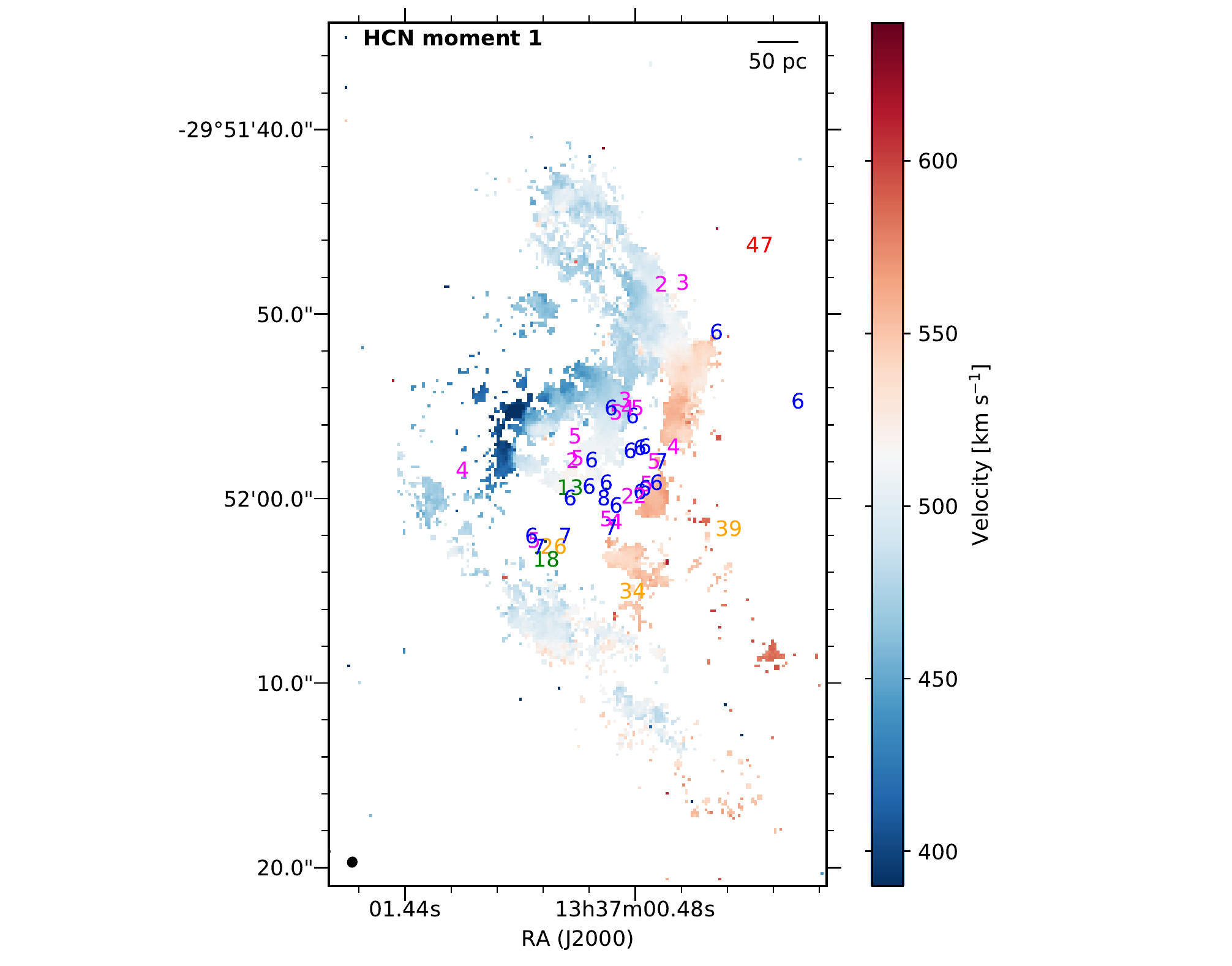}  & \hspace{-2.7cm}
		\includegraphics[trim={9cm 5cm 9cm 5cm},clip,width=0.58\textwidth]{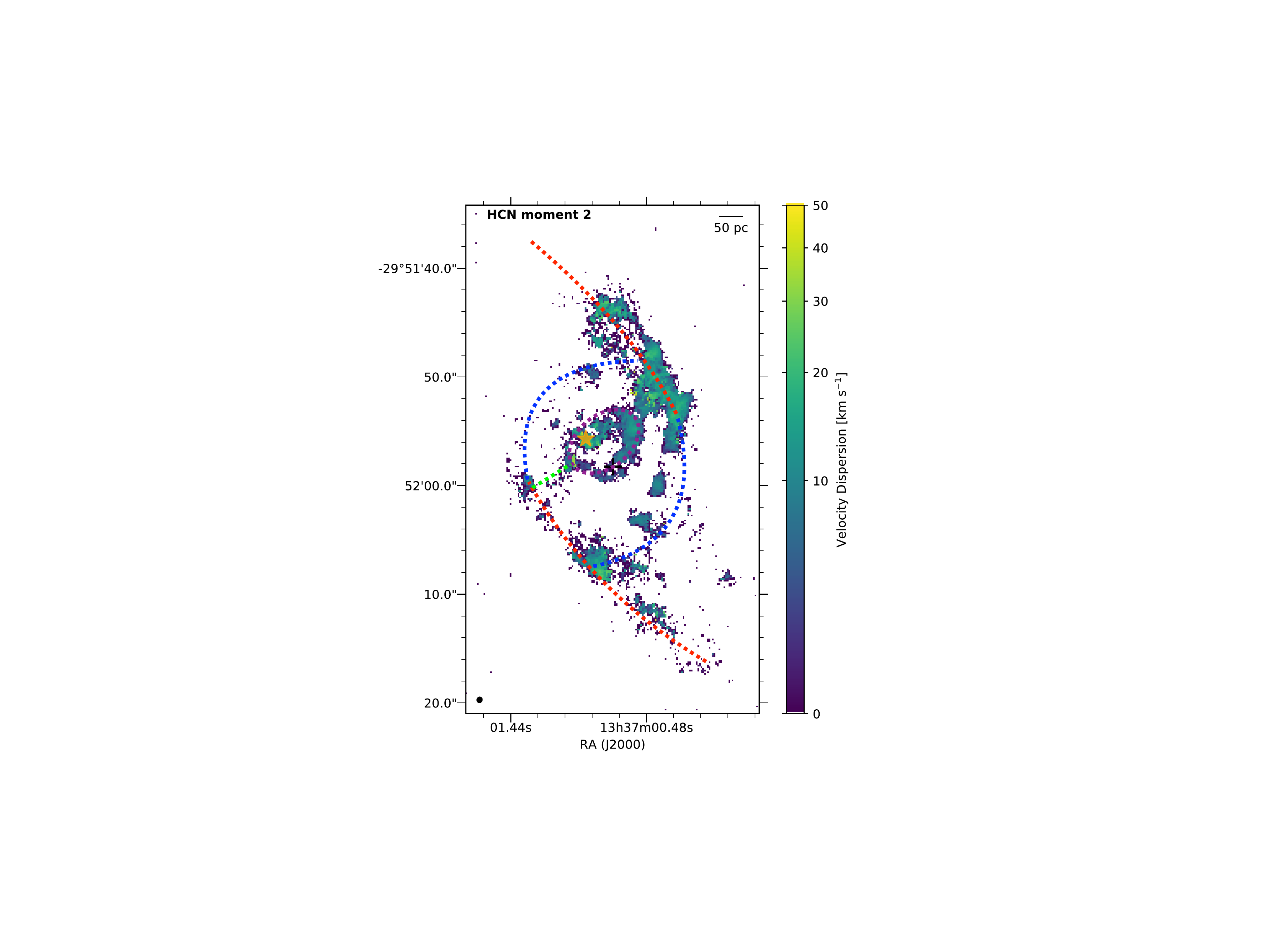} \\
	\end{tabular}
\caption{\normalsize [Top Left] ALMA 95\,GHz continuum emission map produced by averaging all line-free channels over the four spectral windows. HCN ($1-0$) integrated intensity contours are overlaid at (30, 60, 90) K km s$^{-1}$. There is a reasonable correlation between the continuum emission and the brightest HCN ($1-0$) emission. [Top Right] Integrated intensity map of HCN ($1-0$). Blue crosses show positions of massive stellar clusters as found by \citet{2001AJ....122.3046H}. The orange star indicates the visual centre of M83. [Bottom Left] First order moment (intensity weighted velocity) map of HCN ($1-0$). Here we replot the clusters with their corresponding cluster age. [Bottom Right] Second order moment (intensity weighted velocity dispersion) map of HCN ($1-0$). In all panels the synthesised beam is shown as the filled ellipse in the bottom left corner.}
\label{fig:moms}
\end{figure*}

A preliminary clean of each spectral window was performed using CASA's \textit{clean} task to allow for easier identification of key lines within each spectral window. These cleans were performed in an uninteractive mode, with an averaging over every 10 channels. Continuum subtraction was then performed on these datacubes by highlighting line-free channels within all four spectral windows within the \textit{uvcontsub} task in CASA. Cleaning of the continuum was done in an interactive mode using a Briggs weighting with a robust parameter of 0.5 and primary beam correction. An image of the 95\,GHz continuum is shown in the top left panel of Figure~\ref{fig:moms}.

Four lines were detected in the first phase of data reduction: HCN ($1-0$), HCO$^{+}$ ($1-0$), CS ($2-1$) and CCH (N = 1 - 0). Interactive cleaning was performed on each of those lines, cleaning down to the level of $2.5 \sigma$ (intensity of $5$ mJy/beam) with a cell size of 0.15$^{\prime\prime}$.

Figure~\ref{fig:moms} shows the integrated intensity, intensity weighted velocity and intensity weighted velocity dispersion maps of HCN ($1-0$), and Figure~\ref{fig:HCN_chan} shows the channel maps of the emission.  The morphology and velocity structure of the HCO$^{+}$ ($1-0$) emission (Figures~\ref{fig:moms_HCO} \& \ref{fig:HCO_chan}) is very similar to that of the HCN ($1-0$). This similarity provides confidence in the robustness of these lines as tracers of the dense gas morphology and kinematics. The CS and CCH transitions are much weaker than the HCN ($1-0$) and HCO$^{+}$ ($1-0$) lines (Figure~\ref{fig:add_mom0}). Since we are primarily interested in the kinematics of the gas, and both HCN ($1-0$) and HCO$^{+}$ ($1-0$) trace the same structure and kinematics, the following analysis therefore focuses on HCN ($1-0$). 

\section{Deriving gas properties at individual molecular cloud scales}\label{sec:results}

\subsection{Dense gas morphology}
The general structure of the dense gas traced by HCN ($1-0$) and HCO$^+$ ($1-0$) emission is similar to that reported in previous observations at J- and K-band by \citet{1998AJ....116.2834E} and CO by \citet{2004ApJ...616L..59S} (see Figures~\ref{fig:moms} \& \ref{fig:HCN_chan}). Two streams of gas from the north and south of the maps tracing the dust lanes in the HST map are connected to M83's outer circumnuclear ring which in turn is connected to the inner circumnuclear ring by a narrow inner bar.

\subsection{Continuum Emission and Spectral Index Maps}

\begin{figure}
    \centering
	\includegraphics[width=0.48\textwidth]{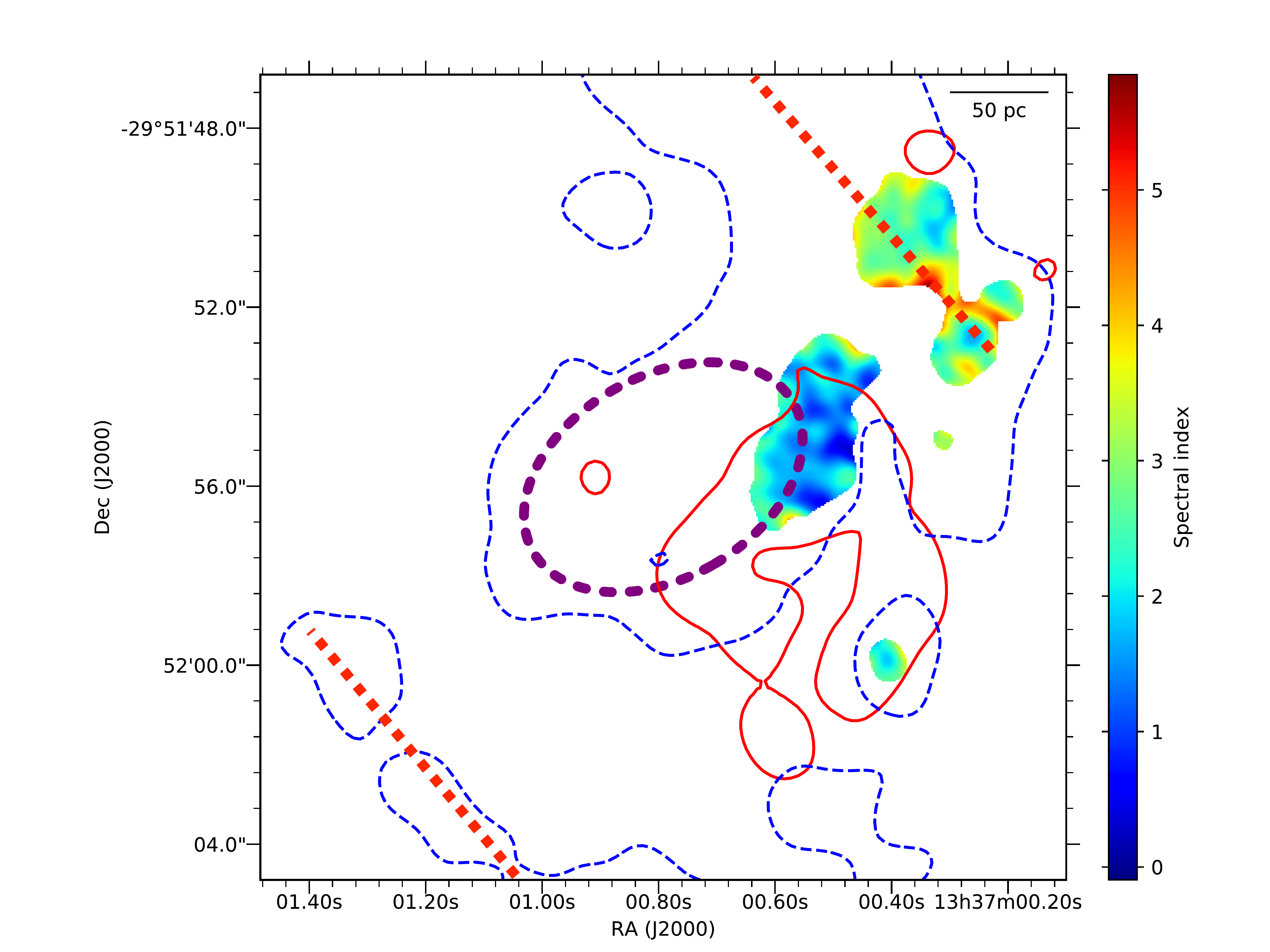}
	\caption{\normalsize Spectral index map calculated from the flux density ratio at 86.5 GHz and 100.5 GHz after masking each continuum image to a threshold of 5$\sigma$. For spatial context H$\alpha$ contours (red) and the HCN ($1-0$) emission (blue) at a level of 10 K are overlaid. The circumnuclear ring and the dust lanes are denoted by the purple dotted ellipse and red dotted lines respectively. Each contour is smoothed with a gaussian kernel of 3x3 pixels. The purple ellipse and red lines indicate the inner circumnuclear ring and dust lanes respectively. The continuum emission associated with the 100 pc circumnuclear ring has a spectral index of $0-2$, as expected from free-free emission of gas photoionised by high mass stars. This region of the ring is associated with star formation, as well as a large fraction of the clusters associated with the high intensity H$\alpha$ emission, as is expected for a region with young, high-mass stars.}
	\label{fig:spec}
\end{figure}

In order to derive the physical properties of the dense gas clouds we first need to assess the contribution to the flux which may come from free-free emission. The continuum emission is mostly confined to the northern dust lane and the western side of the circumnuclear ring (Upper Left panel of Figure~\ref{fig:moms}). To determine the source of this continuum emission we derive the continuum spectral index, i.e. the dependence of radiative flux density on frequency within each spectral window, for each pixel across the map. We do this by generating maps of the continuum emission using only the lowest and highest frequency spectral windows (spw1 and spw3) centred at 86.7\,GHz and 100.5\,GHz, respectively, masking all pixels with emission less than five times the RMS noise level in each image, and then determining the flux density ratio between these maps. Figure~\ref{fig:spec} shows the spectral index of this continuum emission, with contours of HCN and H$\alpha$ emission overlaid.

As shown in Figure~\ref{fig:spec}, the spectral index of the continuum emission in the circumnuclear ring and at the southern part of the northern dust lane vary between $0-2$, and $3-4$, respectively. Although there is scatter due to a combination of the uncertainty in flux measurements and the small frequency range over which the spectral index is calculated, the spectral index of continuum emission in the circumnuclear ring is consistent with free-free emission from gas photoionised by young, high-mass stars \cite[which lies between $-0.1$ and 2 for optically thin and thick emission, respectively;][]{1997pism.book.....D, 2005IAUS..227..111K}. We therefore conclude that there are embedded (i.e., recently formed) high-mass stars in this region.

The spectral index of the continuum emission at the bottom of the dust lane is consistent with that expected from thermal dust emission. We postulate that this comes from warm dust that has been heated by embedded star formation activity at this location, which is at an early evolutionary stage before free-free emission from young high-mass stars begins to dominate.

\begin{figure}
\centering	\includegraphics[width=0.46\textwidth]{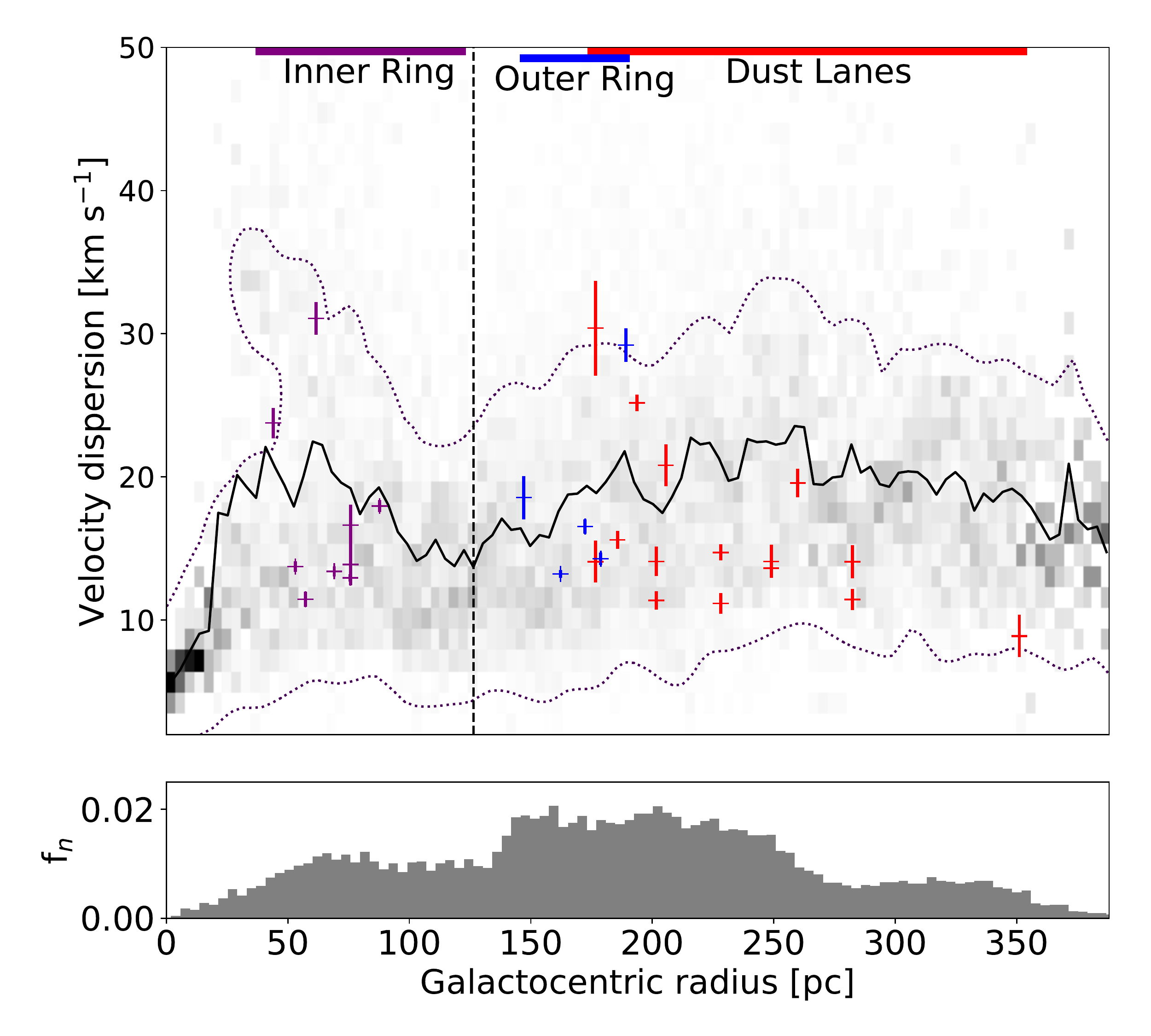}
	\caption{\normalsize Top: Velocity dispersion as a function of galactocentric radius. The background shows a 2D histogram of the maximum velocity dispersion per line of sight calculated via SCOUSE against galactocentric radius. The solid line shows the average velocity dispersion per radial bin. The dotted contours above and below this show a value of 0.02 of the normalized histogram for each 12 pc bin. These therefore contain approximately 98\% of the data. The individual data points show the velocity dispersion measurements from individual peaks shown in Figures~\ref{fig:in} (red),~\ref{fig:outer} (blue) and~\ref{fig:inner} (purple). The vertical dashed black line pinpoints the minimum of the velocity dispersion at $\sim130$ pc. Bottom: Histogram of fraction of total pixels within each galactogentric bin.}
	\label{fig:radius}
\end{figure}

\subsection{Dense gas kinematics}

Figure~\ref{fig:radius} shows a 2D histogram of the velocity dispersion per pixel determined by \textbf{S}emi-automated multi-\textbf{CO}mponent \textbf{U}niversel \textbf{S}pectral-line fitting \textbf{E}ngine \citep[SCOUSE; ][]{2016ascl.soft01003H} with respect to galactocentric radius as measured from the visible nucleus as indicated by the orange star in Figure~\ref{fig:schem_circ}, normalized per galactocentric radius bin of width 12 pc. The spread in the measured velocity dispersions at all galactocentric radii is substantially larger than the uncertainty in individual measurements (typically $\sim1$\,kms$^{-1}$). There are some clear trends in the range of the measured velocity dispersions with galactocentric radius. The velocity dispersion decreases within increasing galactocentric radius from the galactic centre to $\sim130$\,pc, where it reaches a minimum. It then increases to around $\sim200-250$\,pc before decreasing again towards a radius of 400\,pc. We discuss the possible origin of this in Section 5.

\section{Comparison of dense gas and young stars in the centre of M83 and the Milky Way}
\label{sec:mw_m83_comp}

We now compare the properties of dense gas and young stars in M83 and the Milky Way at individual cloud scales in order to determine why there is an order of magnitude difference in star formation rate in the inner few hundred pc of both galaxies, when the volume-averaged gas and stellar properties are similar to within a factor two (Table~\ref{table:comparison}). 


\subsection{Centre of M83 \& MW: similar morphology of gas and young stars}
\label{sub:morph_comp}

We start with a comparison of morphological structures as a function of radius. Unfortunately, as we sit in the plane of the Galaxy, we do not have a top-down view of the gas and stellar structure in the Milky Way. We therefore rely on observational distance constraints and numerical modelling to convert the position-position-velocity data into a 3D structure \citep{2015MNRAS.447.1059K, 2016MNRAS.457.2675H, 2018Galax...6...55L}. 

The properties of gas structures derived in this way are qualitatively very similar to those in M83, with gas falling towards the centre along `dust lanes' in the bar \citep{1991MNRAS.252..210B, 2019MNRAS.484.1213S} and a circumnuclear gas stream orbiting the centre at a similar galactocentric radius \citep{2011ApJ...735L..33M, 2015MNRAS.447.1059K, 2019MNRAS.484.5734K, 2019MNRAS.486.3307D}.  It has been known for a long time that the dense gas mass distribution in the inner few hundred pc of the Milky Way is highly asymmetric, with three-quarters of $^{13}$CO and CS emission at positive longitudes \citep{1988ApJ...324..223B}.  We see a similar degree of asymmetry in the distribution of dense gas structure in M83. A significant fraction of the gas is in-falling from the northern dust lane, with roughly two-thirds of the gas within the inner circumnuclear ring found on the western side. This asymmetry was predicted in simulations of the CMZ by \citet{2018MNRAS.475.2383S}, who highlighted M83 as an example of an external galaxy showing similar structure, though this was largely time dependent within the simulations.

Outside of the nuclear cluster in the inner few pc of the Milky Way, the 3D structure of young, high-mass stars and stellar clusters is even more difficult to ascertain than in the gas \citep{2018Galax...6...55L}. However, it is clear that the recent star formation activity in the CMZ is constrained to the inner $\sim$150\,pc -- the same galactocentric radius range of recent star formation in M83.

As the 3D structures of both dense gas and young stars in the inner few hundred pc of the Milky Way and M83 are similar to within the constraints provided by current Milky Way models, we conclude that differences in morphology cannot explain the order of magnitude difference in star formation rate within the inner kpc of both galaxies.


\subsection{Centre of M83 \& MW: galactocentric trends in velocity dispersion}
\label{sub:vel_disp}
Returning now to Figure~\ref{fig:radius}, we compare the dependence of velocity dispersion with galactocentic radius. Due to our relative position to the centre of our Galaxy, we do not have a face-on view of the velocity dispersion with galactocentric radius. Therefore, we compare to recent 1D models of gas inflows in the inner few hundred parsecs of barred spiral galaxies.

These models predict a relationship between the gas velocity dispersion and galactocentric radius that depends on the rotation curve of the galaxy \citep{2015MNRAS.453..739K, 2017MNRAS.466.1213K}. For galaxies with a rotation curve like the Milky Way, these models predict that the gas velocity dispersion should increase monotonically with decreasing galactocentric radius while the rotation curve is flat, and then decrease sharply as the rotation curve transitions to more solid body like rotation. In the Milky Way, this transition occurs at a galactocentric radius of $\sim100-200$\,pc \citep{2015MNRAS.453..739K}. No direct predictions have been made for the relationship between the gas velocity dispersion and galactocentric radius of M83, due to the unavailability in the literature of a rotation curve at sufficiently high spatial resolution. However, given the similarity in the properties of the gas and stellar distribution in the inner few hundred pc of both galaxies, it seems reasonable to expect a similar qualitative trend in M83 as that predicted for the Milky Way \citep{2018MNRAS.475.2383S}.

Due to the 1D nature of the \citet{2015MNRAS.453..739K} and \citet{2017MNRAS.466.1213K} gas inflow models, each galactocentric radius bin only has a single velocity dispersion assigned to it by definition. Therefore, a direct comparison with Figure~\ref{fig:radius} is non-trivial. Nevertheless, we note that the sharp drop in the mean and range of the measured velocity dispersion occurs at the same galactocentric radius at which recent star formation has occured ($\sim$100$-$200\,pc). Comparing this location with M83's velocity curve \citep[][]{2008ApJ...675L..17F}, we see a correspondence between the minimum velocity dispersion and the turnover in the velocity curve at roughly $\sim 130$ pc.

The coincidence of the minimum in gas velocity dispersion and maximum in star formation activity at the radius where the velocity curve turns over is consistent with the predictions of the 1D dynamical models \citep{2015MNRAS.453..739K,2017MNRAS.466.1213K}. \citet{2020MNRAS.494.6030S} recently tested whether the formation of nuclear rings \textit{requires} a shear minimum, by performing numerical simulations of barred potentials with a flat rotation curve (i.e.\ without a shear minimum). They find that a nuclear ring forms in their simulations regardless, demonstrating that a shear minimum is not a necessary condition. However, by adopting a flat rotation curve they do not address the main point of the prediction by 1D dynamical models, which is that in the presence of a shear minimum, the location of the nuclear ring would correlate with the position of the shear minimum. Further modelling of the M83 gravitational potential is needed for a quantitative comparison to the model predictions, but is beyond the scope of the current paper.

\subsection{Centre of M83 \& MW: similar average dense gas properties}
\label{sub:gas_comp}

Table~\ref{table:params} shows the dense gas properties in the centre of the Milky Way and M83, averaged over the main morphological components for the mass, and on the size-scales of individual molecular clouds ($\sim$12\,pc) within those morphological components for the velocity dispersion. The total mass of gas, the velocity dispersion, and the orbital period of the circumnuclear gas streams in M83 and the Milky Way are similar to within a factor of 2. The mass of gas within the circumnuclear ring area of M83 is calculated using $M_{\text{dense}} = \alpha_{\text{HCN}} L_{\text{HCN}}$, where $\alpha_{\text{HCN}}$ is a conversion factor, which we took to be $\alpha_{\text{HCN}} = 14$ M$_{\odot}$ / K km s$^{-1}$ pc$^{2}$ \citep{2018MNRAS.479.1702O}. While $\alpha_{\text{HCN}}$ can vary significantly, we assume it to be the same in both galactic centre environments. $L_{\text{HCN}}$ is given by multiplying the integrated brightness temperature by the area of the pixel. 

X-ray studies of M83 suggest that the AGN is either highly obscured, or emitting at a very low luminosity \citep{2016ApJ...824..107Y}. If obscuration is not the cause, this puts the AGN at a similar level of emission as Sgr A*, which is the faintest SMBH known \citep{2010A&A...512A...2S}. \citet{2013JMPh....4...55F} estimates the mass of the optical nucleus to be $\left(1 - 4\right) \times 10^{6}$ M$_{\odot}$, putting it well within the range of the highly accurately known mass of Sgr A* at $4 \times 10^{6}$ M$_{\odot}$ \citet{2016ApJ...830...17B}.

We conclude that neither the average properties of dense gas (n$_{H}$ = 10$^{4}$ cm$^{-3}$), nor AGN activity, can explain the order of magnitude difference in star formation rate in the inner few hundred pc of the Milky Way and M83. 

\begin{table}
	\caption{\textbf{Gas properties of CMZ and M83's circumnuclear ring.}}
\centering
\begin{tabularx}{0.47\textwidth}{p{1.5cm} | p{1cm} p{1.5cm} p{1cm} p{1.5cm}}
\hline
Galaxy & Mass & Velocity dispersion & Orbital Period & Rotational Velocity \\
 & (M$_{\odot}$) & (km s$^{-1}$) & (Myr) & (km s$^{-1}$) \\
\hline
Milky Way & 3$\times 10^{7(1)}$ & 17$^{(2)}$ & 3.1$^{(3)}$ & 150$^{(4)}$ \\
M83 & $6.5 \times 10^{7}$ & 15 & 3.7 & 120 \\
\hline
\end{tabularx}
\\[0.2cm]
\begin{flushleft}
Comparison of several key dense gas properties within the central circumnuclear rings of Milky Way and M83. Properties for M83 are derived from our HCN data. The velocity dispersion in both cases was calculated on the same scale of 12 pc, using the linewidth-size relation from \citet{2012MNRAS.425..720S}. (1) \citet{2011ApJ...735L..33M}; (2) \citet{2012MNRAS.425..720S}; (3) \citet{2015MNRAS.447.1059K}; (4) \citet{2017A&A...599A.136L}
\end{flushleft}
\label{table:params}
\end{table}


\subsection{Comparison of SFR measurements}
\label{sub:comp_SFR}

We now compare SFR measurements in the same regions of both galaxies to make sure the magnitude, spatial area, and timescales probed by the SFR measurements are as consistent as possible.

The SFR in the centre of the Milky Way has been studied in detail by \citet{2017MNRAS.469.2263B} using all available diagnostics and data in the literature. They find that all measurements are consistent with the SFR in the inner 500\,pc of the Milky Way being $\sim$0.08\,M$_\odot$\,yr$^{-1}$ for the last $\sim$5\,Myr.

We could find no similar compilation of nuclear SFR measurements for M83, so performed a literature search of recently reported SFR estimates. The most directly comparable SFR measurement with \citet{2017MNRAS.469.2263B} in terms of area is that of \citet{2007PASJ...59...43M}. They used 6\,cm continuum emission to infer a SFR in the inner 500\,pc of M83 of 0.8\,M$_\odot$\,yr$^{-1}$. The assumption used to convert the measured 6\,cm continuum luminosity to a SFR is that all of the flux is non-thermal emission from supernova remnants. If true, the representative timescale probed by this SFR measurement will be related to the supernovae responsible for generating the emission, as discussed below.

However, cm continuum emission can also arise from free-free emission caused by the ionising luminosity of high mass stars. The representative timescale for free-free emission is only a few Myr, so much shorter than the timescale for non-thermal emission. Given that we are interested in  the potential variability of M83's SFR, it is important to associate the correct timescale to the 6\,cm continuum SFR measurement. In their review on this topic, \citet{2012ARA&A..50..531K} state that the non-thermal emission should overwhelmingly
dominate the integrated radio emission at frequencies $\le$ 5\,GHz (wavelengths $\ge$6\,cm). This suggests that the SNe timescale is the correct one to use.

To determine a more accurate representative SFR timescale, we consider two effects: the timescales over which synchrotron-producing cosmic ray (CR) electrons are injected, and the timescales over which they persist once created. On the former, SNe will start anywhere between 3 and 9\,Myr post-star formation \citep[e.g.][]{2014ApJS..212...14L}, depending on exactly which stars succeed in blowing up and which fail and collapse directly to a black hole. For super-solar metallicity, where winds are expected to be more efficient and thus envelope loss makes it easier for the stars to explode, the timescale is probably closer to the younger end of the possible range, though with significant uncertainty. The SN explosions will continue until $\sim$40\,Myr, with a fairly flat rate between the beginning and end. Thus to first order the rate of CR electron injection represents an average of the SFR over the past $\sim 5 - 40$\,Myr.

On the latter question of persistence times, the synchrotron cooling timescale for electrons with a critical frequency $\nu_c$ is
\begin{align}
    &\sim 1 \text{Gyr} \times \,\, (B / \mu \rm G)^{-3/2} (\nu_c / \rm GHz)^{-1/2}
\end{align}
where B is the magnetic field \citep{1992ARA&A..30..575C}. In the absence of a direct measurement of the magnetic field strength in the centre of M83, we take the Solar neighbourhood mean of $\sim 5$\,$\mu$G as a likely lower limit, which sets an analogous lower limit on the cooling time. For $\nu_c = 5$\,GHz and $B=5$\,$\mu$G, the corresponding cooling time is 40 Myr. Increasing the magnetic field strength by an order of magnitude to a more likely value of 50\,$\mu$G would reduce the cooling timescale to $\sim$1\,Myr. The cooling time is therefore comparable to or shorter than the SN delay time. We therefore take the SFR based on synchrotron emission as representing an average over a timescale of order tens of Myr, making it comparable to FUV \citep[0-10-100\,Myr;][]{2012ARA&A..50..531K}, for example, as a SFR indicator.

Inverse Compton (IC) losses also set a limit on the CR electron lifetime that is probably much shorter than the upper limit of 40\,Myr. The IC loss time is equal to the magnetic loss time multiplied by the ratio U$_{B}$ / U$_{R}$, where U$_{B}$ = magnetic energy density and U$_{R}$ = radiation energy density. In the Solar neighbourhood, IC and synchotron loss times are about the same, but the SFR per unit area, and thus the radiation intensity, must be much higher in the centre of M83. Therefore, even assuming that the B field is no stronger than in the Solar neighbourhood, the CR loss time must be well under 40\,Myr as a result of IC losses. A zeroth-order estimate would be that the IC loss time just scales as the inverse of the SFR per unit area.

Other measurements of M83's nuclear SFR are determined over larger areas, so less directly comparable to \citet{2017MNRAS.469.2263B}. The most recent measurements are from \citet{2011ApJ...731...45H}, and \citet{2012MNRAS.421.2917F}, who determined SFRs of 0.8 and 0.7\,M$_\odot$\,yr$^{-1}$, respectively, for the inner $\sim$800\,pc of M83 using H$\alpha$ emission. The slight difference in their SFR values is due to the use of different  corrections to account for dust obscuration. H$\alpha$ emission traces star formation over the last $3-10$\,Myr \citep{2012ARA&A..50..531K, 2020MNRAS.498..235H}.

Based on the above measurements, we conclude that the SFR in the centre of M83 is $\sim$0.8\,M$_{\odot}$ yr$^{-1}$ averaged on several to tens of Myr timescales. This is an order of magnitude larger than the SFR in the centre of the Milky Way over the last $\sim$5\,Myr. Given the similar gas mass reservoirs between both centres, this translates to gas depletion times, of 0.6~Gyr in the Milky Way \citep[also see table~1 of][]{kruijssen14b} and 0.06~Gyr in the centre of M83 (see Table~\ref{table:comparison}).

As a sanity check, we also compare the star formation in M83 and the Milky Way at larger galactocentric radii. Figure~9 of \citet{2012MNRAS.421.2917F}, shows that the depletion time from $1-4$\,kpc in M83 is roughly constant at $1-2$\,Gyr -- consistent with `normal' star formation in galaxies \citep[see, e.g.][]{leroy13}. It is only within the inner few hundred pc that the depletion time deviates substantially from this, dropping to an order of magnitude smaller ($\sim$0.25\,Gyr), consistent with a "starburst" episode. Figure~7 in \citet{2012ARA&A..50..531K} shows that the depletion time in the Milky Way at 4\,kpc ($\sim 1-2$\,Gyr) is similar to that at the same galactocentric radius in M83, consistent with `normal' star formation. As the depletion time shows that the star formation in M83 is `normal' outside of the inner few hundred pc, it is only the comparison of the starburst vs suppressed star formation at galactocentric radii within a few hundred pc that we are interested in and trying to explain in this paper.

\section{Conundrum: broken star formation theories or extreme time variability?}
\label{sec:broken_or_extreme}

The conclusions of the previous section bring us to an interesting, intermediate result. We have shown that the morphology, total gas mass reservoir, and average properties of gas in that reservoir are the same in both galaxies, and yet the star formation rate differs by an order of magnitude. This means that from the time probed by the star formation rate measurements (up to $5-7$\,Myr for the young stellar clusters), either (i) the star formation efficiency per unit mass of dense gas varies by an order of magnitude between the galaxies, or, (ii) the star formation rate has varied by the same amount. Scenario (i) causes severe problems for theories of star formation, as one implicit assumption of all theories is that parcels of gas with similar properties should produce similar stellar populations. Given the extreme variation in star formation rate over a short period of time, scenario (ii) provides strong constraints on the time variability of feeding and feedback, with important implications for the baryon cycles in galactic centres.

We now try to distinguish between these possibilities by focusing in detail on the properties of dense gas and young stars in the inner $\sim$150\,pc of both galaxies, where all the current star formation activity is located.

The inner circumnuclear ring is the main morphological component of both galaxies containing all the recent star formation activity in the central regions. The relationship between the dense gas and young stars in the Milky Way's circumnuclear gas stream has been investigated in detail on the size scales of individual molecular clouds and stellar clusters \citep{2011ApJ...735L..33M, 2013MNRAS.433L..15L, 2017MNRAS.469.2263B}. Using the above data and analysis, we can now compare the properties of gas and young stars on similar scales in M83. 

Below we first investigate the likelihood that the gas we are observing within M83's circumnuclear ring exists in a stable orbit  ($\S$\ref{sub:ring_stab}). We then study variations in kinematic properties of the gas in the ring ($\S$\ref{sub:ring_kin}), its gravitational stability ($\S$\ref{sec:analysis}), and how this might be affected by the galactic gravitational potential ($\S$\ref{sub:grav_pot}). Next we compare the properties of the gas with the surrounding young stellar cluster population to see if they may be causally related ($\S$\ref{sub:ring_cluster_cradle}), and understand what this means for the implied star formation rate as a function of time ($\S$\ref{sub:comp_mw_m83}). Finally, in $\S$\ref{sub:resolving_conundrum}, we try and bring all this information together to  understand whether the comparison of gas and young stars in the Milky Way and M83's nuclear regions implies broken star formation theories or extreme time variability.

\subsection{Orbital stability within M83's circumnuclear ring}
\label{sub:ring_stab}
Before approaching this conundrum, it is important to assess whether the gas in the circumnuclear ring is in a stable\footnote{Note that here we are not referring to the formal definition of a `stable orbit'. Here `stable' is simply intended to mean a coherent gas stream that remains in orbit around the centre for at least one revolution.} orbit around the centre. Given the offset nucleus, the suggestion that the galaxy may have undergone a recent interaction, and evidence of an $m =1$ perturbation, it is plausible that the gas in the circumnuclear ring is strongly dynamically disturbed and not in a stable orbit. However, several lines of reasoning suggest the gas has been in stable orbits for at least an orbital time.

\begin{figure*}
    \centering
  \includegraphics[width=0.99\textwidth,trim={0 0.8cm 0 2cm},clip]{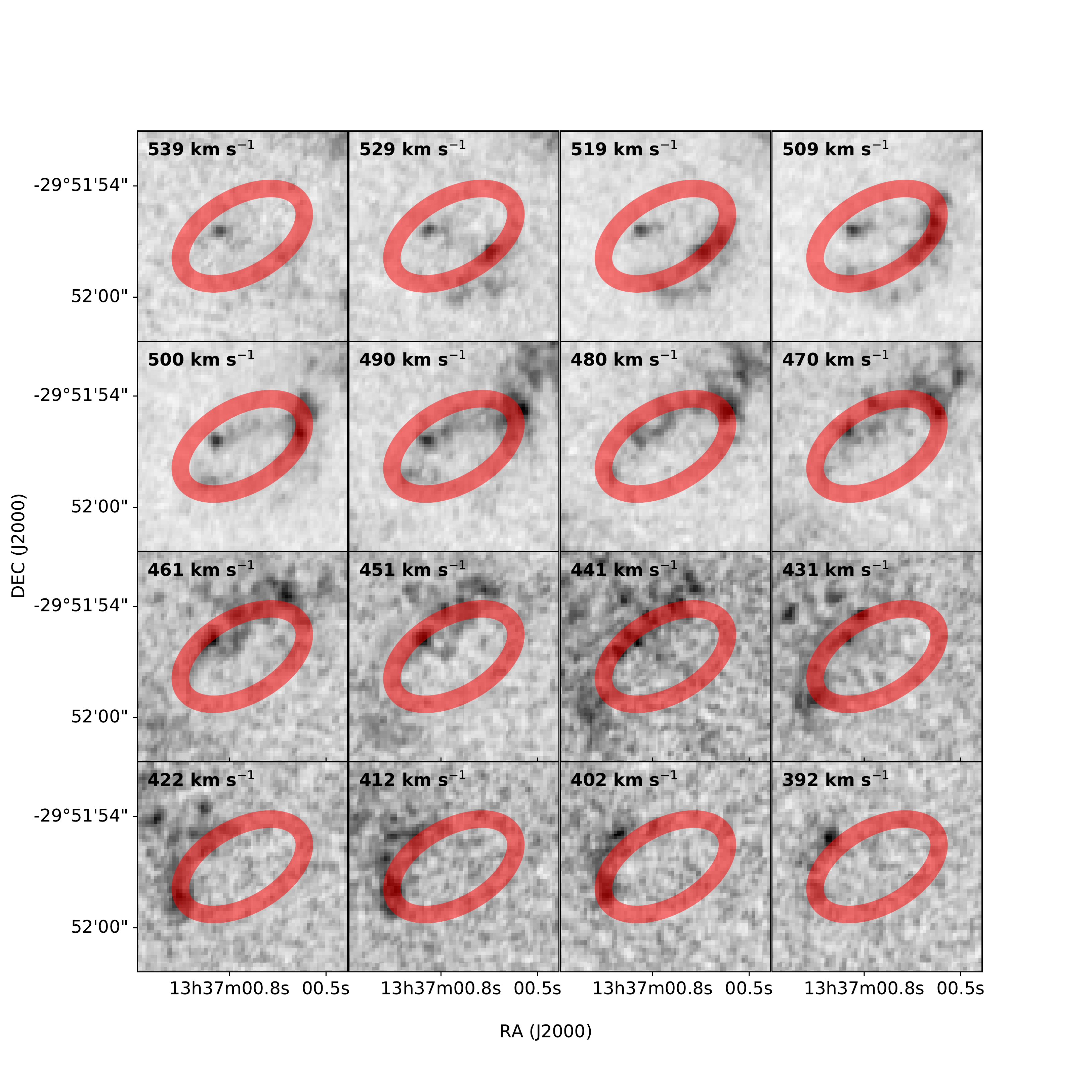}
\caption{\normalsize Channel maps of HCN ($1-0$) emission, with every $\sim$10\,km\,s$^{-1}$ averaged together between 392\,km\,s$^{-1}$ and 539\,km\,s$^{-1}$. The central velocity of each velocity bin is shown.}
	\label{fig:ring_chanmap}
\end{figure*}

Figure~\ref{fig:ring_chanmap} shows the HCN channel map of gas in the circumnuclear ring. The observations show that the gas morphology, density and kinematics vary smoothly and trace the gas in a ring around the nucleus in PPV space. We then seek to construct a simple toy model to investigate at a very basic level whether this PPV structure is in any way similar to expected motions of gas on orbits around centre. To do this, we constructed elliptical orbits in the x-y plane (inclination $=0^\circ$, position angle $=0^\circ$) of model galaxies, and then transformed the position and velocity vectors to M83's inclination (24$^\circ$) and position angle (45$^\circ$) using 3$\times$3 rotation matrices. 

\begin{figure*}
    \centering
    \begin{tabular}{ccc}
  \includegraphics[width=0.3\textwidth,clip]{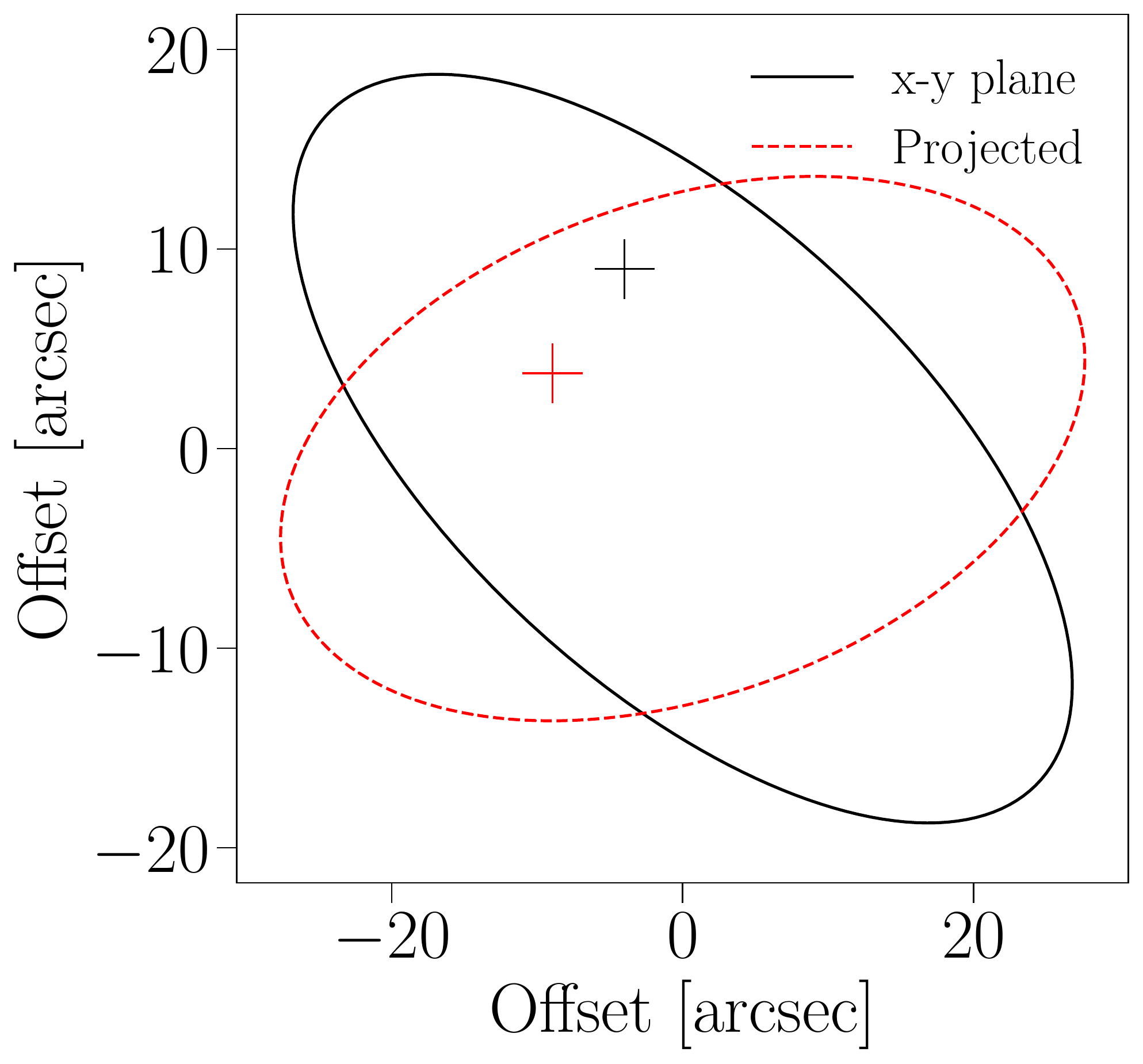} &
  \includegraphics[width=0.3\textwidth,clip]{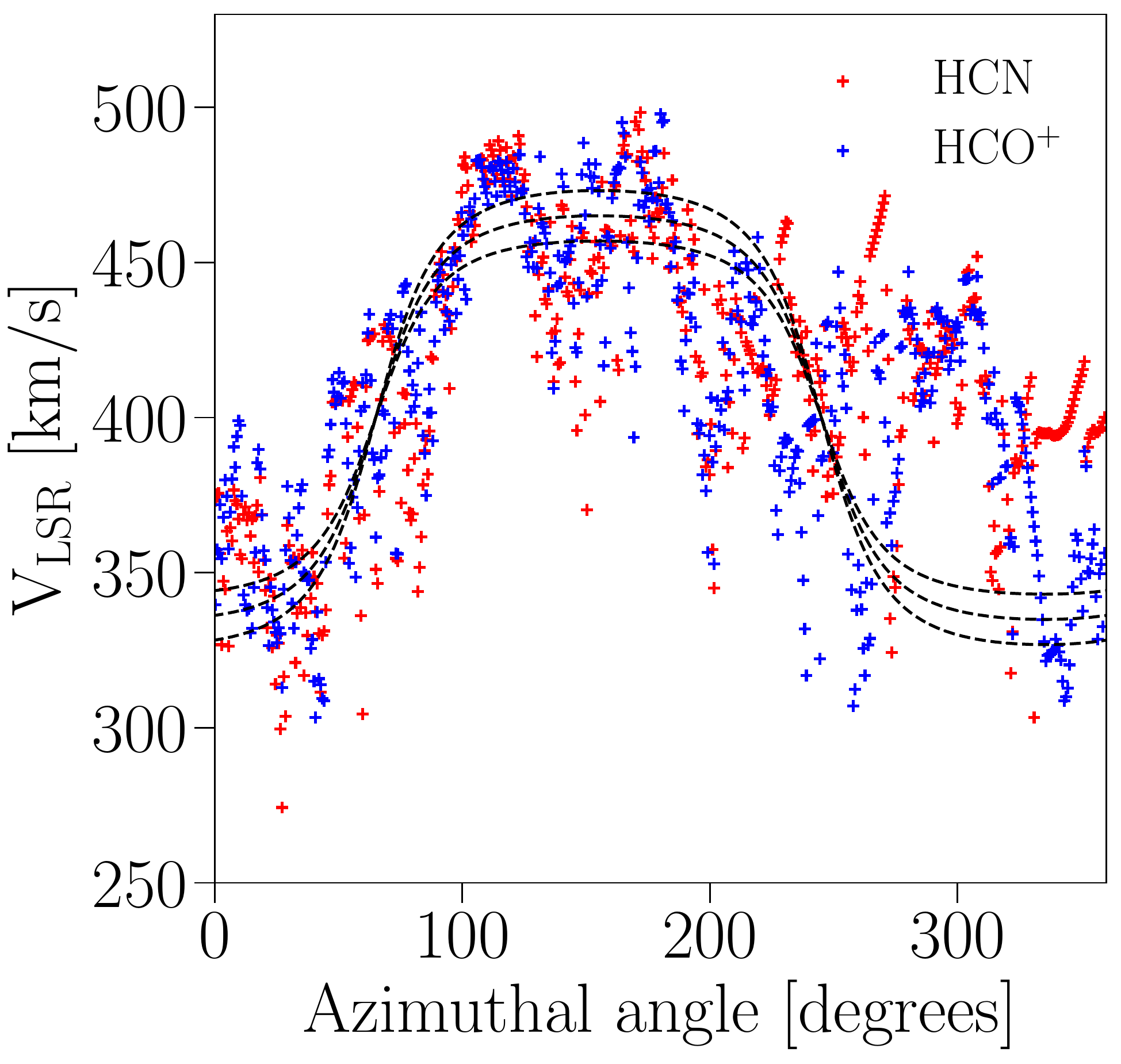} &
  \includegraphics[width=0.3\textwidth,clip]{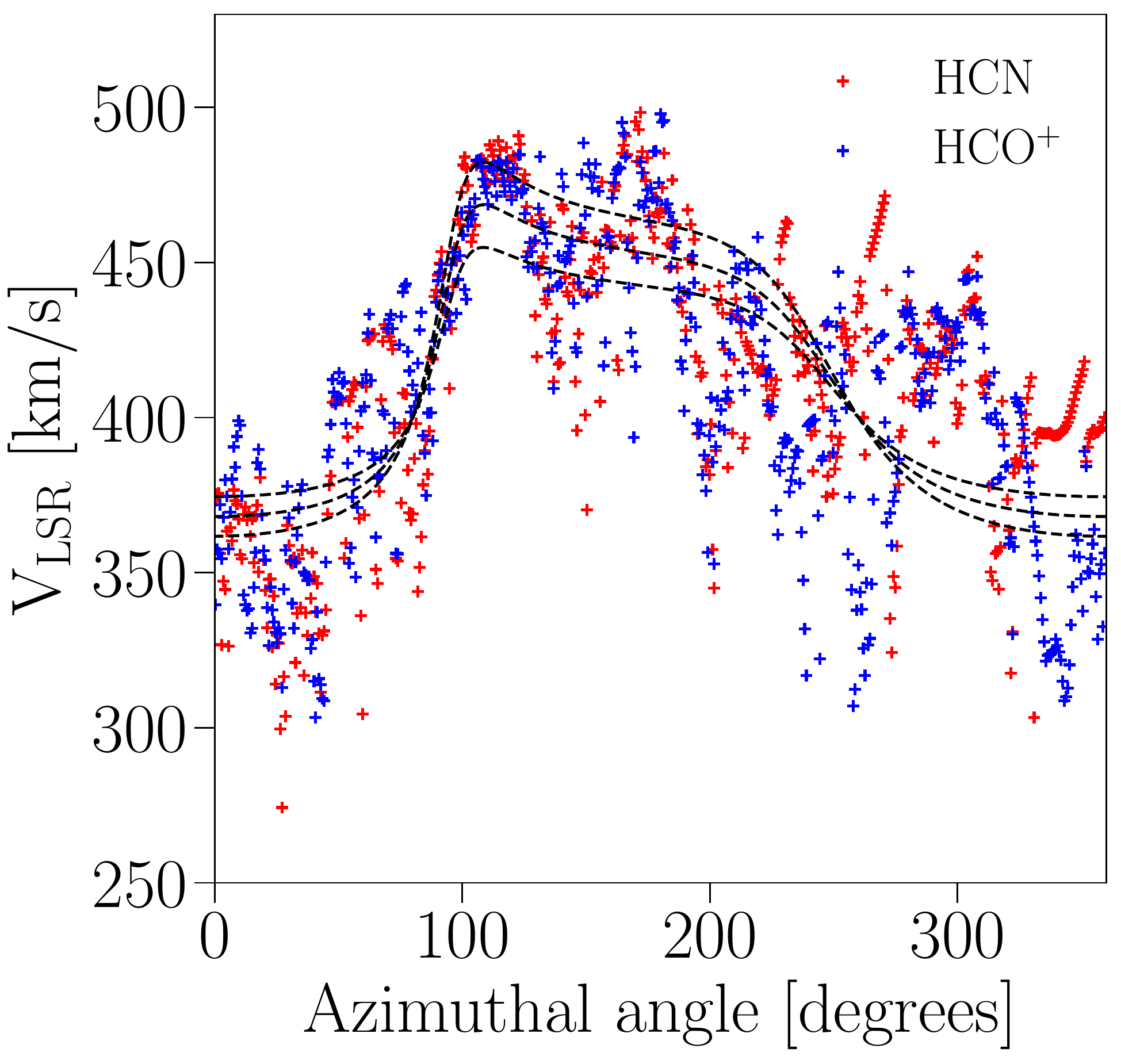} \\
  \end{tabular}
\caption{\normalsize [Left] The observed orientation of the circumnuclear ring and location of the visible nucleus (red), and the ellipse and location of nucleus when de-projected into the x-y plane of the galaxy (black) using the known inclination and position angles of M83.  [Centre] V$_{\rm LSR}$ of the HCO$^+$ (blue crosses) and HCN (red crosses) emission determined from the {\sc SCOUSE} fitting for all pixels in the ellipse used to define the circumnuclear ring. These observed velocities are plotted as a function of azimuthal angle around the ring. The dashed black lines show the expected line-of-sight velocity as a function of azimuth determined from models of gas on elliptical orbits in the galaxy's x-y plane when projected to M83's inclination and position angle. For the centre panel, the zero radius point is defined as the centre of the ellipse, and the circular velocity at semi-major axis is 140, 160, and 180\,kms$^{-1}$, respectively, for the dashed lines with increasing velocity amplitude. [Right] Same as the centre panel, but with the zero radius point defined as the location of the visible nucleus, and the circular velocity at semi-major axis of 80, 100, and 120\,kms$^{-1}$, respectively, for the dashed lines with increasing velocity amplitude.   }
	\label{fig:ring_orbit}
\end{figure*}

The left panel of Figure~\ref{fig:ring_orbit} shows the observed orientation of the circumnuclear ring and the location of the visible nucleus in red. The black ellipse and cross show the geometry of the ellipse and nucleus when de-projected into M83's x-y plane. Using this geometry, we constructed models of gas on the elliptical orbit assuming the conservation of angular momentum. Once the circular velocity at one point on the ellipse has been specified, the velocity at every other point on the ellipse is known because the velocity times the radius is constant. 

The centre and right panels of Figure~\ref{fig:ring_orbit} show the V$_{\rm LSR}$ of the HCO$^+$ and HCN emission determined from the {\sc SCOUSE} fitting for all pixels in the ellipse used to define the circumnuclear ring. These observed velocities are plotted as a function of azimuthal angle around the ring. The dashed black lines show the line-of-sight velocity as a function of azimuthal angle from the models of gas on the elliptical orbits described above, when projected to M83's inclination and position angle. The dashed lines in the centre and right panels of Figure~\ref{fig:ring_orbit} show the expected velocity structure when using the centre of the ellipse and the location of the visible nucleus in the x-y plane, respectively, to define the zero radius location.  

Given the simplicity of the orbital model with so few free parameters, it is interesting that the PPV structure in Figure~\ref{fig:ring_orbit} can be reasonably well reproduced by elliptical orbits moving under the conservation of angular momentum for models where the radius equals zero location is defined either at the centre of the ellipse or the location of the visible nucleus. The scatter on the observed V$_{\rm LSR}$ as a function of azimuthal angle is too large to immediately distinguish which model best fits the data.  

We caution that these simple orbital toy models have several shortcomings. Firstly, the orbital models are strictly unphysical and inconsistent with Newton's laws. The equations of motion of a particle moving in an external gravitational field are time-reversible, so any solution seen backwards in time is still a solution. This implies some basic symmetry properties that are violated by the orbital models. Secondly, the 80\,pc offset between the kinematic and photometric nucleus shows that the nucleus of M83 is currently out of equilibrium. The relative motions and resulting changes in the location of the gravitational sphere of influence of both nuclei is not included in the toy model.

The fact that the velocity structure in M83's circumnuclear ring is comparable to the motions expected from the simple toy orbits suggests that the potential must be stable enough that it is not wildly varying on the orbital timescale. If the gas kinematics were deeply disturbed one would not expect to see a closed ring of gas. 

Another possibility is that the bar potential is preventing stable orbits from existing over the galactocentric radii encompassing the circumnuclear ring. In general, gas in a bar potential should only be able to orbit without self-colliding if it is on an $x_1$ or an $x_2$ orbit, and there is a range of galactocentric radii where no such orbits exist. Given the above considerations, and the fact that the bar potential does not prevent the existence of closed, non-intersecting orbits, it seems plausible that the inner circumnuclear ring does in fact follow an $x_2$ orbit, and the exterior dust lanes are in the forbidden zone where no such orbits exist.

In summary, given all the potential complexities in the environment, the fact that the PPV structure of the data can be reasonably well fit by a simple orbital model is intriguing, and suggests the idea that this gas may be in a stable orbit is worthy of further investigation. Future dedicated simulations that can self-consistently model the complexities of the environment are needed to fully understand the long-term evolution of the gas stream.


\subsection{Variation in kinematic properties within M83's circumnuclear gas ring}
\label{sub:ring_kin}
While studying individual peaks around the circumnuclear gas ring gives us an insight into trends within the gas, it does not provide a complete picture.

\begin{figure*}
    \centering
    	\begin{tabular}{c}
 \includegraphics[trim={5.5cm 5.5cm 5.5cm 5cm},clip,width=0.7\textwidth]{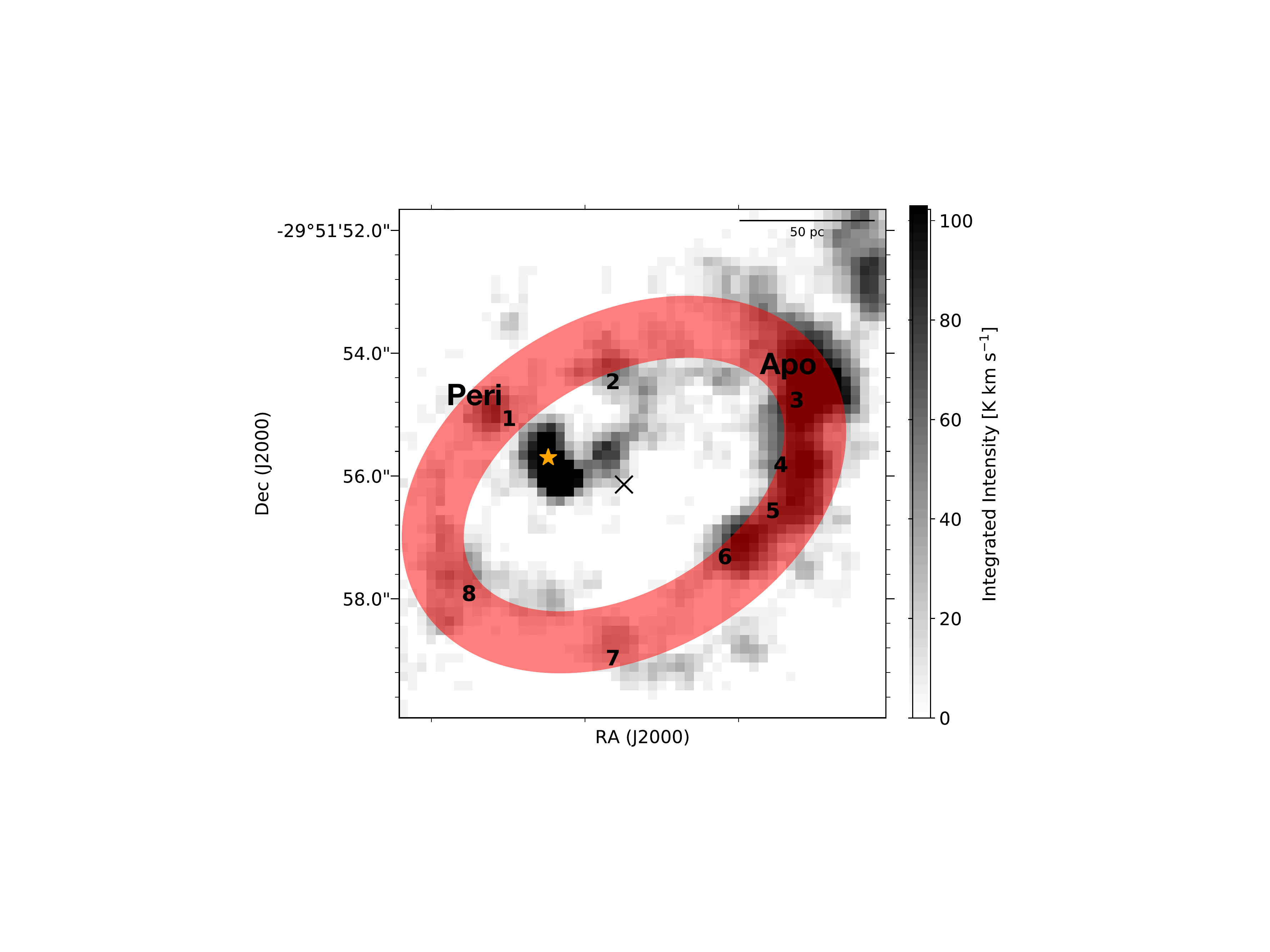} \\
 \vspace{-0.5cm} \\
 \includegraphics[width=0.99\textwidth,trim={0 0.8cm 0 2cm},clip]{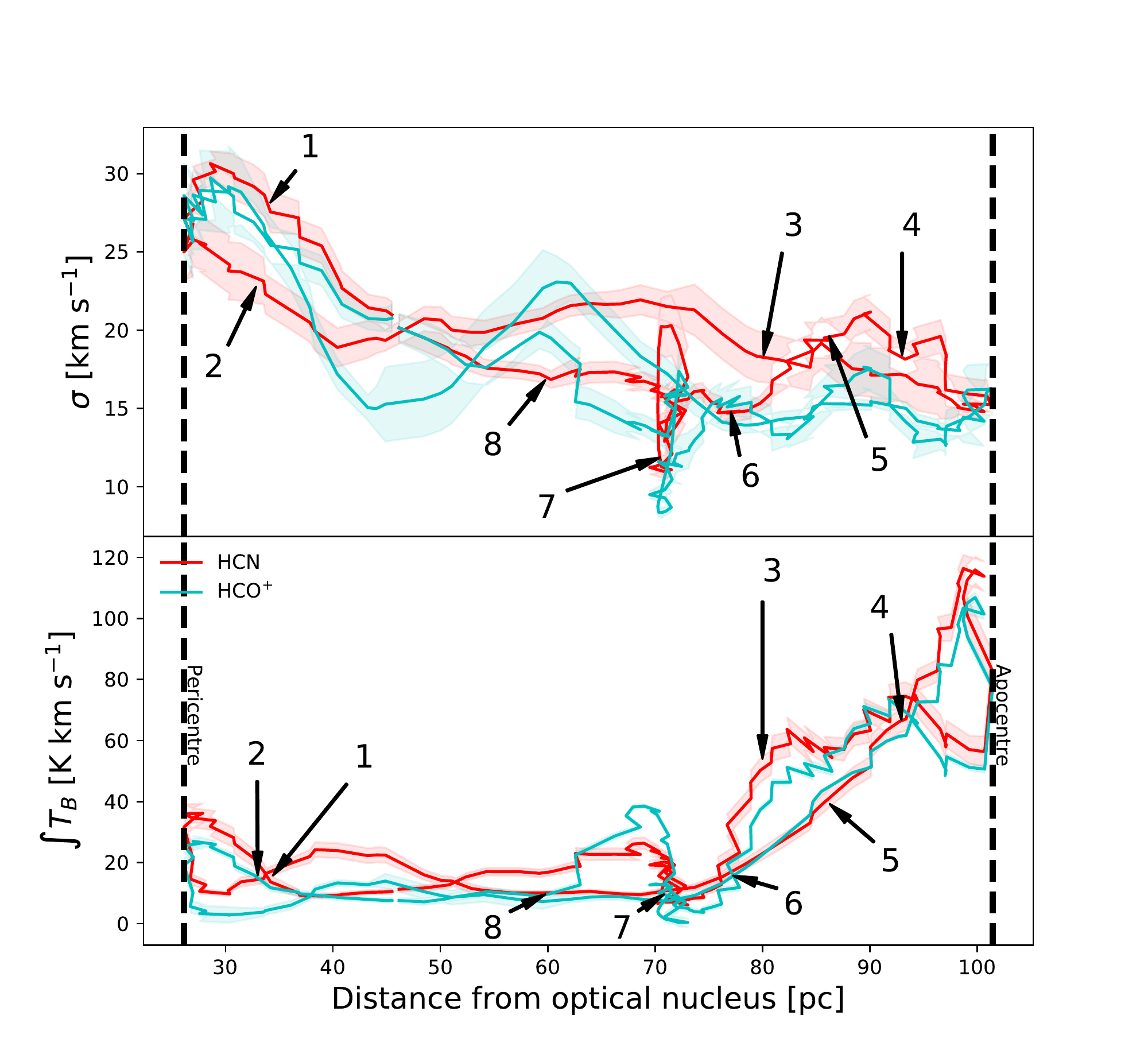}
 \end{tabular}
\caption{\normalsize [Top]: Integrated brightness temperature map. The manually generated ellipse is shown in red. The numbers indicate the positions for which individual spectra were taken (see Figure~\ref{fig:inner}). The orange star is the nucleus and the black cross is the centre of the fitted ellipse. [Middle]: Variation in velocity dispersion as a function of distance from the optical nucleus around the circumnuclear ring for HCN ($1-0$) (red) and HCO$^{+}$ ($1-0$) (blue). We do this around the circumnuclear ring to avoid averaging the two sides of the ellipse together. [Bottom]: Variation in brightness temperature as a function of radius for the same two lines. The shaded regions denote the 1$\sigma$ uncertainty in velocity dispersion and brightness temperature. The positions of the spectra shown in Figure~\ref{fig:inner} are labelled.}
	\label{fig:ellipse}
\end{figure*}

To study how the gas evolves along the ring in M83, as opposed to within individual intensity peaks, we deproject the inner circumnuclear ring into cylindrical polar coordinates and average the polar image of the ring per azimuthal angle bin. As we are considering both radial and azimuthal trends, we also consider the azimuthal profile around the ellipse which has been defined in Figure~\ref{fig:ellipse} and average the velocity dispersion and integrated intensity over a region of $1^{\prime \prime}$ surrounding each pixel along this ellipse. This is because the deprojection method will blend together the radial bins across the entire ring, removing the subtleties of any potential radial trends. Figure~\ref{fig:ellipse} shows the variation in velocity dispersion and integrated brightness temperature with galactocentric radius for HCN ($1-0$) and HCO$^{+}$ ($1-0$) respectively. The shaded regions show the uncertainty on the velocity dispersion and integrated brightness temperature calculated by SCOUSE. These two transitions follow qualitatively similar trends, which gives confidence that the observed trends are accurately tracing the underlying gas kinematics. The observed velocity dispersion reaches a maximum at pericentre and a minimum at roughly 70pc for both transitions. While the average velocity dispersion appears to decrease with increasing distance from the optical nucleus, there are significant variations within this trend that appear unrelated to distance.

\begin{figure*}
    \centering
	\includegraphics[width=0.99\textwidth]{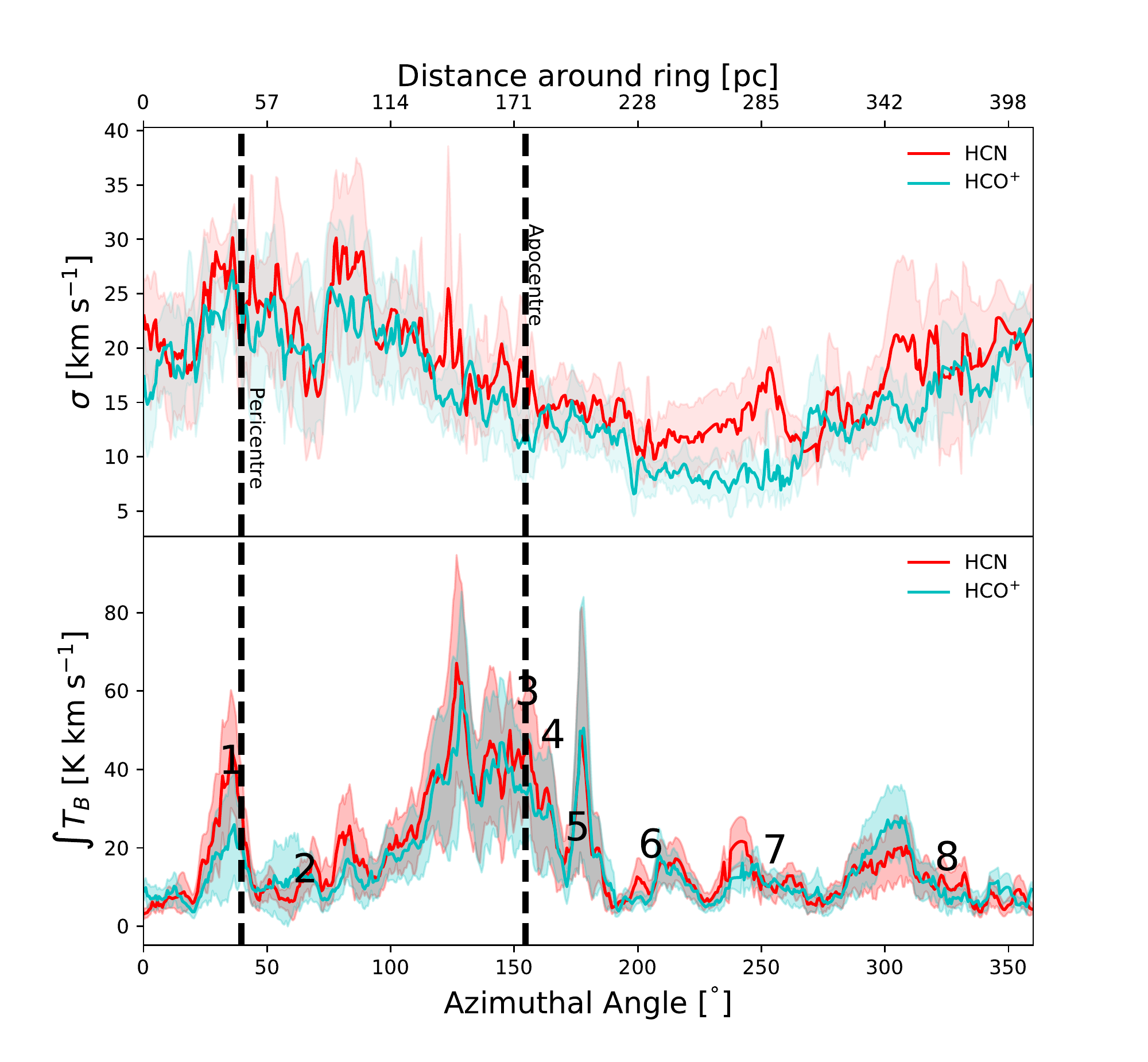}
	\caption{\normalsize Variation of velocity dispersion (top) and integrated brightness temperature (bottom) associated with the inner circumnuclear ring as a function of azimuthal angle, going clockwise from left to right. HCN emission is shown in red while HCO$^{+}$ is shown in blue. The shaded regions shows the standard deviation per azimuthal angle bin. The positions of the spectra shown in Figure~\ref{fig:inner} are labelled, as are the positions of pericentre and apocentre with respect to visible nucleus of M83, represented by the vertical shaded region and dashed line respectively. These are separated by less than $180^{\circ}$ due to the focus of the ellipse being slightly displaced from the visible nucleus of M83. The shaded region corresponds to the 2$\sigma$ uncertainty in pericenter due to the uncertainty in the position of the visible nucleus derived by \citet{2006ApJ...652.1122D} of 0.15$^{\prime\prime}$ or $\sim4$ pc. The upper x-axis shows the distance around the inner circumnuclear ring, assuming an ellipse with a semimajor axis of 50 pc.}
	\label{fig:azimuth}
\end{figure*}

We then investigate whether there are any trends with azimuthal angle of the gas as it orbits the centre. Figure~\ref{fig:azimuth} shows the evolution of brightness temperature and velocity dispersion around the circumnuclear gas ring as a function of azimuthal angle.

Focusing first on the integrated intensity, we see a significant peak at apocentre, and several local peaks around pericentre. By eye, the distribution of the peaks throughout the integrated intensity curve appears quasi-regular despite the peaks themselves showing considerable variation in brightness.

We calculate the structure function of the integrated intensity (Henshaw et al. 2019, sub.) to determine if there is a preferred separation between the observed peaks. The structure function of order $p$ is given by $SF \equiv \left<\left|I(x) - I(x+d)\right|^{p}\right>$ averaged in this case over azimuthal angle, where $I$ is the intensity (in this example), measured at a location $x+d$ relative to position $x$. In the following, we compute the first-order structure function and so $p=1$. Structure functions are traditionally used in studies of the interstellar medium to measure the scale-dependence of certain quantities  (e.g. velocity; \citealp{2002ApJ...580L..57P, 2004ApJ...615L..45H}). However, a property of the structure function, exploited mainly in time series analysis \citep{1985ApJS...59..343C, 2006MNRAS.368.1025L} but more recently in studies of the ISM (Henshaw et al. 2019, sub.), is its sensitivity to periodicity in data. The structure function of a periodic quantity will display a local minimum at the location of the corresponding wavelength.

\begin{figure}
    \centering
    \includegraphics[width=0.49\textwidth]{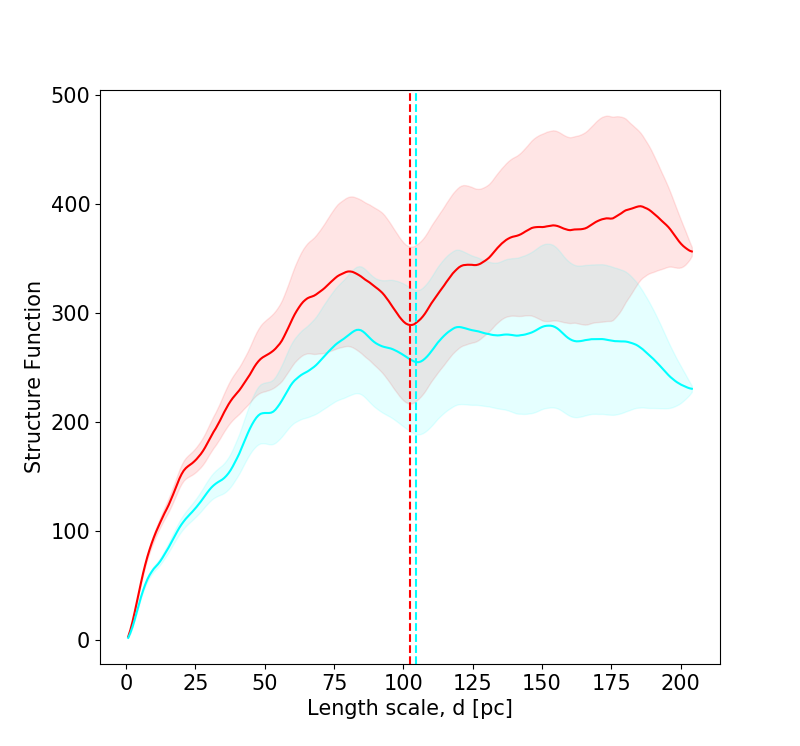}
    \caption{\normalsize The structure functions computed when varying the zero point in azimuth in steps of $0.5^{\circ}$. The blue line shows the average structure function and the shaded region shows the variation in structure function when changing the zero point. The dashed line shows the location of the minimum at 100 pc. The colours in both correspond to the same transitions as Figures~\ref{fig:ellipse} and~\ref{fig:azimuth}.}
    \label{fig:fft}
\end{figure}

We compute the structure function at $0.5^{\circ}$ increments in azimuth around the ellipse. This is to prevent any possible bias introduced by, for example, starting our measurement at a position which happens to intersect one of the intensity peaks. In Figure~\ref{fig:fft} we display the mean structure function measured at each location and the 1$\sigma$ dispersion about the mean. A clear dip in the profile of the structure function is observed at $d\approx100$\,pc. Visual inspection of the left hand panel of Figure~\ref{fig:fft} confirms that the most prominent peaks are indeed spaced by approximately $\sim100$\,pc.

\subsection{M83's circumnuclear gas ring: unstable to gravitational collapse?}\label{sec:analysis}

We now investigate what might cause this quasi-regular spacing in the gas properties.  \citet{2016ApJ...829...45K} model the gravitational instability of rotating isothermal rings at the centres of barred galaxies, like M83, to understand their star formation potential. Using the observed circumnuclear ring radius ($\sim$100\,pc), circular velocity ($\sim$75\,kms$^{-1}$), mass ($\sim5\times10^8$\,M$_\odot$) and velocity dispersion (17\,kms$^{-1}$, see below) we calculate the \citet{2016ApJ...829...45K} $\alpha$ (virial parameter) and $\hat{\Omega}_0$ (critical angular frequency) parameters for M83's circumnuclear gas ring through their Eq. 53 and 54 to be 0.023 and 0.7, respectively. Given their definitions of $\alpha$ and $\hat{\Omega}_0$, this places the ring in the regime of being marginally unstable against gravitational collapse \citep[][Fig. 12]{2016ApJ...829...45K}. In this model, the growth rate of the instabilities is always close to  $\sim0.81(G \rho_c)^{0.5}$, where $\rho_c$ is the central density of the ring. For reasonable values of $\rho_c$ ($>10^2$\,cm$^{-3}$), the instabilities are expected to develop within an orbital period and produce around $\sim$10 approximately evenly spaced clumps. Given the circumference of M83's circumnuclear gas ring, the clumps should be separated by $\sim$60\,pc. Considering the idealised nature of the \citet{2016ApJ...829...45K} model (e.g. uniform density, circular orbits) the similarity with the predicted clump spacing suggests gravitational instabilities are a plausible explanation for the observed quasi-regular gas spacing. 

To investigate this further, we also consider families of  physical models which have been constructed to understand what determines the spacing of gas fragments within a filament \citep{1953ApJ...118..116C, 1987PThPh..77..635N, 1992ApJ...388..392I, 1993PASJ...45..551N, 1995ApJ...438..226T, 1996PASJ...48..701T}. The most simplistic model is the ``sausage'' instability \citep{1987PThPh..77..635N}, in which the fragment spacing within filaments is roughly equal to the wavelength of the fastest growing unstable mode of the fluid instability. For isothermal cylinders of finite radius R, this wavelength depends on the ratio between the cylinder radius and the isothermal scale height $H = c_{s} \left(4 \pi G \rho_{c}\right)^{-1/2}$, where $c_{s}$ is the sound speed, G is the gravitational constant and $\rho_{c}$ is the gas mass density at $R=0$, where R is the radius of the filament or cylinder. In the case that the radius of the filament is much larger than the scale height, this wavelength is $\lambda_{\text{max}} = 22 H$. Taking $\lambda_{\text{max}}$ to be the scale length determined in the previous section, we calculate a scale height of $\sim 4.5$\,pc.

As the gas kinematics in the circumnuclear ring are dominated by non-thermal motions, we instead use the average velocity dispersion around the ring in place of the sound speed to determine the required density. However, we first must ensure that this velocity dispersion is not significantly impacted by velocity gradients within the circumnuclear ring. We approximate the  total measured velocity dispersion, $\sigma_{\rm tot}$, as the convolution of two Gaussians: the first, $\sigma_{\rm int}$, corresponding to the intrinsic velocity dispersion of the gas; the second, $\sigma_{\rm orb}$, is the contribution to the observed velocity dispersion caused by the local velocity gradients along the orbit. The magnitude and relative contribution of $\sigma_{\rm orb}$ to $\sigma_{\rm tot}$ will increase as the aperture over which the velocity dispersion is measured increases. To quantify the magnitude of $\sigma_{\rm orb}$ as a function of size scale, we first calculated the velocity gradient around the orbit at the highest, intrinsic angular resolution of the observations ($\sim$12\,pc). We found a linear fit with a gradient of $\pm$ 3 km\,s$^{-1}$\,pc$^{-1}$ provides a good approximation of the orbit. Using this gradient we can determine $\sigma_{\rm orb}$ as a function of size scale by averaging over the required size scale and determining the intrinsic velocity dispersion as $\sigma_{\rm int} = \sqrt{\sigma_{\rm TOT}^{2} - \sigma_{\rm orb}^{2}}$. We find that the contribution of the orbital velocity gradient to the total velocity dispersion is negligible -- even averaging over a spatial scale of $\sim 30$ pc the quadrature-subtracted velocity dispersion only contributes $\sim$25$\%$ to the total velocity dispersion. As we determine the velocity dispersion at a size scale of $\sim 12$\,pc we conclude that the observed velocity dispersion is not impacted significantly by the orbital velocity gradient. As such we use $\sigma = 17$\,kms$^{-1}$ to calculate a critical density of $n = 4 \times 10^{3}$\,cm$^{-3}$.

While this critical density is broadly consistent with those measured on large scales in the CMZ and M83 which suggests that the gas could be subject to this instability, this is an over simplistic scenario for several reasons. Firstly, this model deals with a cylinder of infinite length, instead of a rotating ring. Additionally, while the model is in isolation, the rotating stream has gas being fed in at both extremes by the dust lanes further out. Finally, the displacement of the nucleus from the centre of the ring may produce additional perturbations within the gas in the ring.

For these reasons, we consider other possible mechanisms of gas fragmentation. It is possible that the observed fragment separation is a result of the `wiggle' instability \citep{2004MNRAS.349..270W, 2012ApJ...747...60K, Henshaw2020}, seen in hydrostatic simulations of galactic centres \citep{2015MNRAS.449.2421S, 2017MNRAS.469.2251R}. It is also plausible that the turbulence produces this quasi-periodicity somewhat sporadically, and we are merely observing it here by chance.

Although the observed fragmentation length is intriguing, more realistic analytical models or dedicated simulations are required to understand its origin.


\subsection{M83's circumnuclear gas ring properties: shaped by the gravitational potential?}
\label{sub:grav_pot}

We now seek to understand what may be causing the variations in integrated brightness temperature and velocity dispersion of HCN ($1-0$) and HCO$^{+}$ ($1-0$) along the circumnuclear ring.

In the absence of any numerical simulations with time-dependent chemistry, we make the assumption that the abundance of HCN ($1-0$) and HCO$^{+}$ ($1-0$) are constant throughout the circumnuclear ring. We see that the HCN ($1-0$) and HCO$^{+}$ ($1-0$) emission is well resolved, with the smallest cloud being 80\% larger than the beam size, so beam dilution should not be a major issue. Therefore, variations in integrated intensity can be a result of a change in excitation conditions, opacity, column density of material, or a combination of all three. 

If the emission were optically thick we would expect the line brightness temperature to equal the excitation temperature and the line profiles to become self-absorbed and non Gaussian. Given the brightness temperature of the spectra is $<$5K and the line shape is roughly Gaussian, we conclude opacity is not a serious issue at the scales probed by these observations. The variation in integrated intensity is therefore due to an increase in excitation conditions, column density, or volume density.

One potential explanation for the trends in integrated intensity and velocity dispersion is that we are witnessing conservation of mass flux as the gas orbits the galactic centre. Azimuthally, due to the elliptical orbit, as the gas moves further away from the nucleus it slows down and will tend to `pile up' at apocentre as it spends more time at that location. Radially, however, pileup occurs where the density of orbital streamlines is the highest, which is at pericenter. Assuming the HCO$^{+}$ ($1-0$) and HCN ($1-0$) integrated brightness temperature traces the dense gas mass on scales of $\sim 10$ pc \citep[as discussed by][]{2017ApJ...835...76M}, we would expect to observe a correlation between brightness temperature and radius. The bottom panel of Figure~\ref{fig:ellipse} shows that indeed the highest brightness temperature emission is at largest radii. However, the sudden increase in brightness temperature between position 2 and 3 (apocentre) and much slower drop off in brightness temperature from apocentre to position 5 and 6 is not consistent with the picture of orbital pile-up, which would require azimuthal symmetry.

Another potential explanation could be that the trend in velocity dispersion and integrated intensity is simply due to the clouds having a similar virial state, and clouds with a larger column density (brighter HCO$^{+}$ ($1-0$) and HCN ($1-0$) emission) will have larger velocity dispersions. A comparison of the two panels of Figure~\ref{fig:ellipse} shows there is an anti-correlation between brightness temperature and velocity dispersion at both peri- and apocentre; while the velocity dispersion peaks at the smaller radii, the integrated intensity peaks closer to apocenter.

Returning to Figure~\ref{fig:azimuth}, it is interesting to note that the location of the sharp rise in integrated intensity at apocentre in the circumnuclear ring's orbit also corresponds to the location at which the circumnuclear ring and the `bridge' intersect (see Figure~\ref{fig:schem_circ}). One explanation for the increase in integrated intensity close to apocentre would therefore be that this is the location at which gas from the dust lanes is deposited onto the circumnuclear gas ring though the `bridge'. In this scenario, the increased integrated intensity would then be due to an increase in column or volume density from the new material being added on to the ring.

However, several lines of evidence argue against this scenario. Firstly, if substantial quantities of gas were being deposited at the bridge-ring intersection we would expect to see a sudden jump in the integrated intensity of the ring at the intersection point. Figure~\ref{fig:azimuth} shows the integrated intensity increases steadily in azimuthal angle from significantly before the intersection point (number 3). In addition, the total mass of dense gas in the whole bridge inferred from the HCN ($1-0$) and HCO$^{+}$ ($1-0$) integrated intensity emission is much smaller than the increase in integrated intensity seen in the circumnuclear ring at the bridge-ring intersection point. If the current mass in the bridge region is representative of the time-averaged mass flow, then if mass \emph{is} transferred to the ring through the bridge, it is at a much smaller rate than can explain the increase in integrated intensity. The lack of extinction at this bridge suggests that the gas is not simply being transferred to the ring at lower densities.

Secondly, if substantial quantities of gas were being deposited at the bridge-ring intersection we would expect to see signs of this in the gas kinematics in the form of multiple spectral components, or broad line emission.  Figure~\ref{fig:inner} shows that the one location with unambiguous multiple velocity components is indeed at the bridge-ring intersection point. However, this is the \emph{opposite} intersection point from where we see the increase in integrated intensity. At the other intersection point (number 3) with the maximum integrated intensity, Figure~\ref{fig:azimuth} shows that in fact the velocity dispersion is closest to its minimum value.

An alternative explanation for the observed variation in gas properties is that the clouds in the circumnuclear gas stream are being shaped by the external gravitational potential. 3D hydrodynamical simulations of gas clouds orbiting the centre of the Milky Way at a similar galactocentric radius show that a combination of the background potential and eccentric orbital motion shape the morphological and kinematic evolution of the clouds \citep{2019MNRAS.484.5734K}. Specifically, strong shear, tidal and geometric deformation, and the passage through the orbital pericentre affect the cloud sizes, column densities, velocity dispersions, line-of-sight velocity gradients, angular momenta, and kinematic complexity. Although such simulations have not been run for gas clouds in the circumnuclear gas stream of M83, the similarity of the inner few hundred parsec of M83 and the Milky Way make it plausible that M83's external potential will exhibit similar behavior.

Furthermore, we note a strong increase in velocity dispersion around pericentre passage, with a maximum of $\sim$30\,kms$^{-1}$ at the location of position 1. The simulations of \citet{2015MNRAS.447.1059K}, \citet{2019MNRAS.484.5734K} and \citet{2019MNRAS.486.3307D} show that additional turbulence driven by motion in the shearing potential, which reaches a maximum as clouds move through pericentre, may be responsible for increasing the velocity dispersion. As the clouds pass pericentre, the rate of turbulent energy injection slows down, and the energy is expected to dissipate on a crossing time. While the impact of pericentre passage in these simulations is quite small, they are based on the gravitational potential of the CMZ. To determine how significant this effect is in M83, simulations would have to be run using a model of M83's gravitational potential.

From the observed galactocentric radius and velocity dispersion (Table~\ref{table:disp}) the crossing time for the cloud nearest pericentre (position 1) is $\sim$0.4\,Myr. Given the previously calculated orbital period of the inner ring of 3.1\,Myr and assuming a fixed orbital velocity, the cloud will have moved $\sim$50\,pc along the orbit by the time it has dissipated the additional energy. The cloud at position 2 lies $\sim$50\,pc along the orbit and the velocity dispersion has dropped from $\sim$30\,kms$^{-1}$ to $\sim$20\,kms$^{-1}$. These qualitative trends are consistent with that expected from the injection and dissipation of turbulent energy. However, care does need to be taken when comparing these observations with the simulations. The velocity dispersions reported in the simulations are determined from a viewing angle looking through the disk mid-plane, whereas our observations view the galaxy from above.

Given that the predominant age of clusters sits well within the most likely timeframe within which SNe will likely start \citep[3 - 9 Myr;][]{2014ApJS..212...14L}, we consider the likelihood of the energy liberated in these events being enough to impact the gas flow, perhaps even disrupting the ring entirely. Using 3D hydrodynamical simulations to understand how SNe affect surrounding molecular gas clouds, \citet{2013MNRAS.431.1337R} found that the energy released from SNe primarily escape from the region along lower density channels. While the very edges of their dense molecular clouds were ablated, the majority of the gas in their dense clouds were resistant to this process. Applying this to the centre of M83, this would imply that the energy from the SNe, which explode in a low density environment offset from the ring, will likely escape with minimal impact on the ring itself. However, the combined affect of many SNe may play an important role in the longer term star formation cycle, e.g. by making it more difficult for gas to enter the circumnuclear stream.


\subsection{M83's circumnuclear gas ring: cradle for the observed stellar clusters?}
\label{sub:ring_cluster_cradle}

We now look in more detail into the stellar clusters, and how trends observed in the gas may have imprinted onto the cluster population.

The \citet{2019MNRAS.484.5734K} simulations show that the transformative dynamical changes to the clouds as they orbit can lead to cloud collapse and star formation. This can generate an evolutionary progression of cloud collapse with a common starting point, which either marks the time of accretion onto the tidally-compressive region or of the most recent pericentre passage. Such an evolutionary progression should leave an imprint on the age distribution of recently formed stars as a function of their position with respect to the gas clouds. Specifically, they should exhibit an age gradient that increases with distance travelled from the common starting point for star formation (e.g. pericentre passage or the circumnuclear ring-bridge intersection point).

If the gas in the circumnuclear ring is to form stars, the imprint of the $\sim$100\,pc spacing in the gas should also be observable in the distribution of young stars (at least until they are disrupted by galactic dynamics). Returning to the distribution of YMC's in Figure \ref{fig:schem_circ}, there are too few clusters in \citet{2001AJ....122.3046H} to do a rigorous spatial clustering analysis. However, it is interesting to note that by-eye there are a few groups of clusters which are clearly separated from other groups by around 100\,pc, though we cannot state firmly that this is anything but confirmation bias.

\begin{figure}
	\centering
	\includegraphics[scale=0.45]{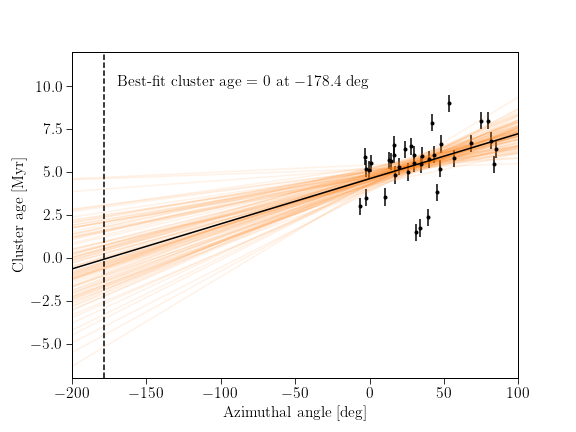} \\
	\includegraphics[scale=0.45]{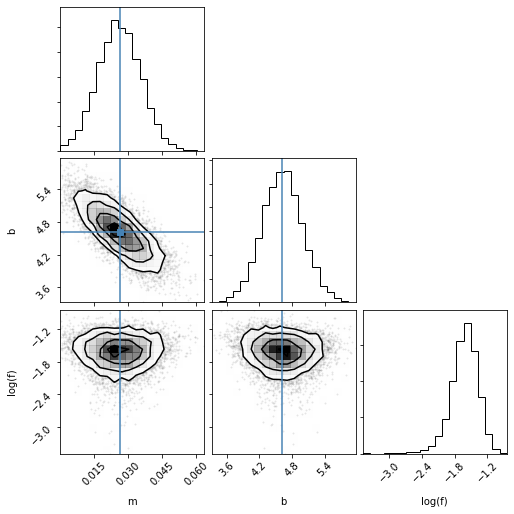}
	\caption{\normalsize [Top] Ages of the massive star clusters observed by \citet{2001AJ....122.3046H} [black dots] as a function of azimuthal angle around the circumnuclear ring. Error bars show a representative 1\,Myr uncertainty in cluster ages. The black line shows the result of Bayesian fitting of a straight line to the data points. The orange lines show opacity-weighted, randomly selected fits from the posterior probability distribution to provide a visual assessment of the line parameter uncertainties. [Bottom] Corner plot showing 1D and 2D projections of the posterior probability distribution parameters, where $m$ is the gradient, $b$ is the intercept, and $f$ is the fractional uncertainty in the cluster ages (see text for details). The blue horizontal and vertical lines show the best-fit $m$ and $b$ from least squares minimisation. 
  }
	\label{fig:ages}
\end{figure}

To see if there is any relationship between the YMCs and the gas in the circumnuclear ring, we first plot the distribution of clusters as a function of galactocentric radius, which shows a large gap in clusters between radii of roughly $220-350$\,pc. We assume the clusters with galactocentric radius $\ge350$\,pc are unassociated with the circumnuclear gas stream. We then take the ages\footnote{We note that there are discrepancies between the cluster ages in Table~2 and Fig. 11 of \citet{2001AJ....122.3046H}. We use the values in Table~2 but the results are robust when using either values.} of the stellar clusters with galactocentric radius less than 220\,pc and plot them in azimuth in the same way as for the gas\footnote{In doing this, we have checked for local outliers (i.e.\ with ages vastly different from the neighbouring clusters) and verified if these could be caused by age degeneracies in the colour-colour space used to determine the ages \citep[Fig.~6]{2001AJ....122.3046H}. For discrepant cluster ages that can be explained by this degeneracy, we have set the ages to that of their neighbours.}. Figure~\ref{fig:ages} shows that there is a linear relation between cluster age and azimuthal angle. We thus confirm the similar age gradient within these clusters reported by \citet{2001AJ....122.3046H} and \citet{2010MNRAS.408..797K}. 

We then used Bayesian analysis to determine the uncertainty on the line fit by constructing a model with three parameters: the gradient ($m$), intercept ($b$) and fractional error on the age uncertainty ($f$). Including parameter $f$ allows the modelling to take into account the fact that the age uncertainties are not reported in \citet{2001AJ....122.3046H}, so any uncertainties we choose may be over-/under-estimates. We assume flat priors in all three parameters in the ranges, $0.0 < m < 0.5$, $0.0 < b < 10.0$, and $-3.5 < {\rm log}(f) < 1.0$. We used the python package {\sc emcee} \citep{2013PASP..125..306F} to sample the posterior probability distribution. The initial positions of the posterior distribution sampling chains (or `walkers') were drawn from a narrow Gaussian centred on the maximum likelihood solution. After initialising the positions of 32 `walkers' across the posterior distribution in this way, we used {\sc emcee} to let the `walkers' independently sample the posterior distribution in 5000 Markov chain Monte Carlo (MCMC) steps. The autocorrelation time (how long it takes each `walker' to lose its memory of where it started, and hence begin fairly sampling the posterior distribution) was $\sim$40 steps. We therefore discarded the first 120 steps of each `walker' (commonly known as a `burn-in' time) to ensure the remaining steps sampling the posterior distribution were not affected by the choice of initial `walker' location.

Figure~\ref{fig:ages} shows the results of the Bayesian analysis where the cluster age uncertainty was assumed to be 1\,Myr. The 16$^{\rm th}$, 50$^{\rm th}$, and 84$^{\rm th}$ percentiles of the samples in the marginalized distributions are $m = 0.0259^{+0.0092}_{-0.0096}$\,yr\,degree$^{-1}$,
$b =  4.6196^{+0.366}_{-0.364}$\,yr,  and log$(f) = -1.60^{+0.245}_{-0.2186}$. The resulting angular frequency, $\Omega$, orbital period, and circular velocity, V$_{\rm circ}$, are 0.673, 9.3 Myr and 79 kms$^{-1}$, respectively. These values are consistent with the observed gas rotation curve at the galactocentric radius range of the ring and clusters \citep{2004A&A...422..865L}. We repeated the analysis varying the age uncertainty on the clusters from 0.1\,Myr to 4\,Myr. While the additional fractional error on the age uncertainty ($f$) increased as our assumed age uncertainty decreased, the values of $m$ and $b$ changed very little. We conclude that our results are robust against the unknown uncertainty on the cluster ages.

Extrapolating this relation back in azimuth to where the cluster age equals zero suggests that the progenitor clouds from which these clusters formed began collapsing at a common point. A natural explanation for this is that some event may be responsible for triggering star formation. The azimuthal angle at this `cluster age equals zero' point is $-178^\circ$, where $0^\circ$ is defined as directly West (to the right) of the center. The uncertainties on the line fit translate to large (tens of degrees) uncertainties on this `cluster age equals zero' angle. In addition, the simple orbital models discussed in $\S$\ref{sub:ring_stab} suggest the angular velocity of the gas is non-linear with azimuth, adding additional uncertainty on the `cluster age equals zero' point. Nevertheless, it appears likely that the `cluster age equals zero' point is constrained to lie in the quadrant of the orbit in which the circumnuclear gas stream passes closest to the bottom of the gravitational potential. If the YMCs formed in the circumnuclear gas stream, their age gradient is consistent with their formation having been triggered by pericentre passage. A similar scenario has been proposed for star formation in the circumnuclear stream of the Milky Way \citep[e.g.][]{2013MNRAS.433L..15L,2015MNRAS.447.1059K,2019MNRAS.484.5734K,2018MNRAS.478.3380J}. 

If pericentre passage triggers star formation in the circumnuclear ring, star formation is expected to occur over the next (few) free-fall time(s) along the orbit. Taking a density of $n \sim 10^{4-5}$ cm$^{-3}$, we find a free-fall time of 0.03 - 0.3 Myr, with the higher end of this range closely matching the free-fall times of clouds found in the CMZ \citep{2015MNRAS.447.1059K}. Given the orbital velocity, the orbital position corresponding to a few free fall times places the star formation at the following apocentre and beyond (locations 3, 4, 5 and 6). Figure~\ref{fig:moms} shows that these positions coincide with the continuum source in the circumnuclear ring with spectral indices consistent with free-free emission from young, high-mass stars. 
The locations in the circumnuclear stream with the brightest continuum emission coincide with the brightest HCN ($1-0$) and HCO$^{+}$ ($1-0$) integrated intensity emission. If the clouds at these positions have embedded star formation, as implied by the continuum emission, their densities and temperatures will be higher than in quiescent clouds. These conditions will result in brighter line emission, explaining the increased HCN ($1-0$) and HCO$^{+}$ ($1-0$) integrated intensity at these locations. The resulting feedback from young stars will eventually disperse the remaining gas, potentially explaining the lack of HCN ($1-0$) and HCO$^{+}$ ($1-0$) at position 7 and beyond and the bright H$\alpha$ emission at this location shown in Figure~\ref{fig:spec}.

The only other location outside the circumnuclear gas ring that is both near enough to the stellar clusters, and has a large enough gas reservoir to form stellar clusters, is the southern end of the western dust lane. In simulations of gas flows in barred spiral galaxies, individual gas streams can collide at these locations \citep{2015MNRAS.449.2421S}. The resulting strong shocks can lead to increased gas density -- a natural location for star and cluster formation. In this scenario, the continuum emission peaks at the end of the dust lane with spectral indices consistent with those of thermal dust emission would represent the youngest sites of star formation activity, as their continuum is not yet dominated by free-free emission. It is interesting to note that there is then a linear increase in star formation age from this location, through the free-free continuum sources at the western end of the circumnuclear gas ring to the well-known age gradient in the clusters. 


\subsection{Comparison of the dense gas and young stars with the Milky Way}
\label{sub:comp_mw_m83}

Regardless of the causal relationship between the circumnuclear gas ring, the southern end of the dust lane, and the stellar clusters, the observed properties of the dense gas along M83's circumnuclear ring are remarkably similar to those of the circumnuclear gas stream in the Milky Way. The total mass, mass distribution of clouds, orbital velocity, galactocentric radius and gas velocity dispersion are the same within the observational uncertainties (Table~\ref{table:params}). In addition, when comparing the gas velocity dispersion, column density and star formation activity as a function of azimuth around the ring, the magnitude of change in these properties in both galaxies is similar when the azimuth angle is measured from pericentre passage with the bottom of the galactic gravitational potential. Indeed, an observer located at the same distance from the centre of M83, and at the same angle with respect to M83's stellar bar as the Sun is in the Milky Way, would have a strikingly similar view of the gas and stars at their galactic centre as we do of ours. Even the observed locations and mean masses of M83's stellar clusters (few $10^4$\,M$_\odot$) are similar to the distribution of the 24 micron sources in the centre of the Milky Way.

Having conducted a detailed comparison of gas and young stars at similar spatial scales in both galaxies, the only significant difference we can find in these properties between the two galactic centres is the number, location and age distribution of the young stellar clusters. The inner 200\,pc of the Milky Way contains two clusters \citep[Arches and Quintuplet,][]{2010ARA&A..48..431P} and a distributed population of either very young or evolved high-mass stars \citep[e.g. `24$\mu$m point sources',][]{2008AAS...212.1807Y}. On the other hand, in the same galactocentric radius range, M83 has 45 clusters of similar mass. 

However, when we separate the clusters by age, a very different picture emerges. As mentioned in the introduction, the age distribution of clusters in M83 has a very strong peak at ages of 5-7\,Myr \citep{2001AJ....122.3046H}. If we only select clusters with a similar age range as the Arches and Quintuplet in the Milky Way ($\lesssim$4\,Myr), the number of clusters is roughly similar. Unfortunately it is particularly difficult to age such young clusters accurately, so a direct comparison is difficult, but we estimate that the centre of M83 only has a factor $\sim$2 more clusters in the age range $\lesssim$4\,Myr than the centre of the Milky Way.

Regarding the relative location of the clusters in the centre of the two galaxies, while most of the current star formation within the CMZ is occurring within the circumnuclear stream, the clusters in the centre of M83 are primarily distributed outside of the circumnuclear ring. While there is little to no current star formation at similar galactocentric radii in the CMZ, we note that there is a well known population of 24$\mu$m point sources \citep{2008AAS...212.1807Y}, which are thought to be related to a previous generation of star formation, at the same galactocentric radii range as the clusters in the centre of M83. Therefore, the galactocentric radii range of star formation over the last $\sim$10\,Myr appears similar in the centre of both galaxies and M83 contains a large population of clusters aged between 5-7\,Myr in the inner few hundred pc that are missing in the Milky Way.


\subsection{Resolution of the conundrum: time variability in the SFR, not broken star formation theories}
\label{sub:resolving_conundrum}

We now return to the conundrum posed at the beginning of this section and the original motivation for comparing the dense gas and young stellar populations in the centres of the two galaxies: what is causing the order of magnitude difference in star formation rate when the dense gas properties are almost indistinguishable?

The resolution of this conundrum, as also indicated by \citet{2001AJ....122.3046H}, is that the conundrum disappears almost completely when only the most recent SF (i.e within the last 4 Myr) and the current properties of the gas are considered. This implies that the SFR is strongly variable with time, and causes one to overestimate the SFR in M83 when using more standard estimates. While this result may seem obvious in hindsight, there are several important implications.

Firstly, it gives confidence that gas clouds with similar properties produce similar stellar populations, a key assumption of all star formation theories. The $\sim$Myr timescale for star formation to occur corresponds to several free-fall times at the average cloud density. This is often invoked as a natural time for star formation in gas clouds. It follows from the above points that what we are learning about the detailed physical processes shaping star formation in the centre of Milky Way can be directly applied to similar environments in nearby galaxies. 

The second implication is that M83 had a burst of star formation 5-7\,Myr ago. This possibility was previously pointed out in the original young massive cluster survey by \citet{2001AJ....122.3046H}. However, they were careful to make clear that they couldn't rule out an alternative possibility, that the elevated star formation episode had continued for much longer than 7\,Myr, and that the reason older clusters were not detectable in their data was due to disruption.

Given the remarkable similarity between the present-day properties of the gas, the youngest stellar clusters in the centres of the two galaxies and the SF estimates from free-free emission, it is far more likely that the elevated star formation episode had a very short duration, and that the present-day conditions are much more representative of the time-averaged conditions for both galaxies. 

If true, this suggests that galaxies like the Milky Way and M83 have a duty cycle for star formation. For much of the time they have a relatively low star formation rate, consistent with observations that show most nearby galaxy centres have much lower than average dense gas star formation efficiency \citep{2015AJ....150..115U}. The comparison of M83 and the Milky Way suggests that these periods of quiescence are punctuated by short episodes lasting for a few Myr where the star formation rate can increase by between one and two orders of magnitude. The young massive cluster population in M83 suggests that the star formation rate was an order of magnitude higher than average for a period of a few Myr. The relatively short starburst duration means finding a galaxy in this phase is statistically unlikely, so observational examples will be rare and large galaxy samples are needed to overcome this problem. Previous studies of star-forming nuclear rings by \citet{2006MNRAS.371.1087A} and \citet{2007MNRAS.380..949S} found strong evidence of this episodic star formation cycle.

The galaxy NGC~253 is particular interesting in this regard. Much like M83 and the Milky Way, NGC-253 also contains a circumnuclear gas ring with a similar radius \citep{2018ApJ...869..126L}. Recent observations have shown that NGC~253 has 14 extremely young ($<$1\,Myr old) super star clusters, which contain the bulk of the nuclear star formation activity. We postulate that 5-7\,Myr ago M83 went through a starburst phase qualitatively similar to that currently observed in NGC~253, which produced the majority of the clusters we see today in M83's centre.

If the centre of the Milky Way, NGC~253 and M83 represent the quiescent, starburst, and post-starburst phase, respectively, of a commomly shared duty cycle, future detailed comparison of their gas properties and young stellar populations will help understand key aspects of the duty cycle. For example, what controls the duration of quiescence between starbursts? What eventually triggers and then ends the starburst? What controls the increased magnitude in star formation? Is there any link between star formation and feedback to feeding of the central supermassive black holes?

Finally, we point out that the interpretation of a duty cycle with a long period of quiescence punctuated by short, extremely intense star formation episodes has important implications for the mass flows and energy cycles in galaxy centres, and thus galaxy evolution. As star formation is highly localised in both space and time, the resulting feedback will be much more efficient at driving galactic-scale nuclear outflows \citep[e.g. the outflow currently being driven in NGC~253,][]{2019ApJ...881...43K, 2018ApJ...867..111Z} than the same star formation integrated over the whole duty cycle.

\section{Conclusions}\label{sec:conclusion}
Using ALMA Band 3 HCN ($1-0$) and HCO$^{+}$ ($1-0$) observations we have studied the distribution and kinematics of the dense gas on $\sim$10\,pc scales in the inner few hundred parsec of the nearby spiral galaxy M83. The HCN and HCO$^{+}$ emission closely traces the previously known molecular gas features and dust absorption features. Visual inspection of the HCN and HCO$^{+}$ data cubes show that multiple velocity components in the spectra are prevalent, especially at the end of the dust lanes. We used SCOUSE to perform multi-component spectral line fitting of the HCN and HCO$^{+}$ line profiles. The resulting fits from both lines are remarkably similar, giving confidence in the robustness of these transitions to trace the dense gas distribution and kinematics accurately.

We find that the range in the measured velocity dispersion varies considerably with galactocentric radius. The drop in velocity dispersion at the same radius range containing all the recent star formation activity and the turn over in the rotation curve qualitatively matches the predictions of recent 1D models of gas transport and star formation in the centres of galaxies 
\citep{2015MNRAS.453..739K, 2017MNRAS.466.1213K}.

The gas in the inner circumnuclear gas ring (galactocentric radii $\lesssim$120\,pc) shows strong variations in HCN and HCO$^{+}$ velocity dispersion and integrated intensity. When averaged in azimuth around the ring, the integrated intensity emission shows quasi-periodic behaviour with a spacing between the oscillations of $\sim$100\,pc. Given the absence of an analytical model for the stability of orbiting gas in a circumnuclear ring, we use this to estimate the density required to produce an instability of this length in a self-gravitating cylinder, which was calculated to be $4 \times 10^{3}$ cm$^{-3}$. This is in reasonable agreement with the density of the CMZ and M83 measured at this scale, given the sources of error in this calculation.

The variation in the HCN and HCO$^{+}$ velocity dispersion and integrated intensity around the circumnuclear gas stream is consistent with a scenario in which a combination of an eccentric orbit through an axisymmetric potential is shaping the gas properties. Specifically, there is a strong increase in the velocity dispersion of gas at pericentre passage, consistent with the expectation of additional turbulence being added to the gas driven by motion in the shearing potential, which reaches a maximum at pericentre. The velocity dispersion peaks and then quickly drops off between pericentre and apocentre returning to the average value after approximately a crossing time, consistent with expectations of turbulence dissipation. The apocentre also corresponds to the peak in the HCN and HCO$^{+}$ integrated intensity emission, corresponding to an increase in the column density and/or excitation conditions. The detection of free-free continuum emission towards this location is consistent with these clouds containing embedded, recently formed, high mass stars.

Comparing the properties of the gas in the circumnuclear ring with the age and location of the nearby young massive clusters, we find a linear age gradient of the clusters with azimuthal angle around the galactic centre, suggesting there is a common location for their formation. If the clusters formed in the circumnuclear gas ring, their ages are consistent with the common location for the onset of star formation being close to pericentre passage at the bottom of the galactic gravitational potential. Though we note that the uncertainty in this location as a result of the rotation curve and its derivation is considerable, especially over more than one orbital cycle.

We put forward a scenario to explain the observed properties of the gas in the circumnuclear gas stream and the surrounding young massive clusters. In this scenario, gas in the circumnuclear stream is undergoing gravitational instabilities which determines the spacing and mass of individual clouds around the ring. A combination of the external gravitational potential and eccentric orbit then shape the gas properties, compressing the gas and adding turbulent energy into the gas as it approaches pericentre. The gas then dissipates its turbulent energy on a crossing time and begins to form stars. Over the next $\sim$Myr, feedback from the newly formed stars disperses the remaining molecular gas, leaving the observed young massive clusters.

Finally we show that the only way to reconcile the order of magnitude difference in SFRs between the two galaxies given their remarkably similar dense gas properties is with time variability. Isolating the youngest ($<$4\,Myr old) stellar populations, the inferred SFRs of both galaxies agree within a factor $\sim$2. This has important implications for interpreting observations of galaxy centres and understanding their mass flows and energy cycles.

M83's young massive cluster population suggests the SFR must have been an order of magnitude higher $5-7$\,Myr ago. The comparison of observed SFR with present day gas properties is therefore highly misleading,  and highlights the danger of interpreting dense gas vs SFR relations to understand the physics of star formation in galaxy centres. In addition, M83's `starburst' phase was highly localised, both spatially and temporally, greatly increasing the feedback efficiency and ability to drive galactic-scale outflows. 

This highly dynamic nature of star formation and feedback cycles in galaxy centres means (i) modeling and interpreting observations must avoid averaging over large spatial areas or timescales, and (ii) understanding the multi-scale processes controlling these cycles requires comparing snapshots of galaxies in different evolutionary stages.

\section*{Acknowledgements}

JMDK and MC gratefully acknowledge funding from the Deutsche Forschungsgemeinschaft (DFG) through an Emmy Noether Research Group (grant number KR4801/1-1) and the DFG Sachbeihilfe (grant number KR4801/2-1). JMDK gratefully acknowledges funding from the European Research Council (ERC) under the European Union's Horizon 2020 research and innovation programme via the ERC Starting Grant MUSTANG (grant agreement number 714907). NB gratefully acknowledges funding from the European Research Council (ERC-CoG-646928, Multi-Pop) and the Royal Society in the form of a University Research Fellowship. MRK acknowledges funding from the Australian Research Council through the Future Fellowship (FT180100375) and Discovery Projects (DP190101258) funding schemes. J.H.K. acknowledges financial support from the European Union's Horizon 2020 research and innovation programme under Marie Sk\l odowska-Curie grant agreement No 721463 to the SUNDIAL ITN network, from the State Research Agency (AEI-MCINN) of the Spanish Ministry of Science and Innovation under the grant "The structure and evolution of galaxies and their central regions" with reference PID2019-105602GB-I00/10.13039/501100011033, and from IAC project P/300724, financed by the Ministry of Science and Innovation, through the State Budget and by the Canary Islands Department of Economy, Knowledge and Employment, through the Regional Budget of the Autonomous Community.
A.G. acknowledges support from the National Science Foundation under grant No. 2008101




\bibliographystyle{mnras}
\bibliography{references} 
\nocite{*}




\newpage

\appendix

\section{Additional data}
\label{sec:app_additional_maps}

\begin{figure*}
	\centering
	\begin{tabular}{ccc}
		\includegraphics[trim={1cm 0 0 0},clip,width=0.4\textwidth]{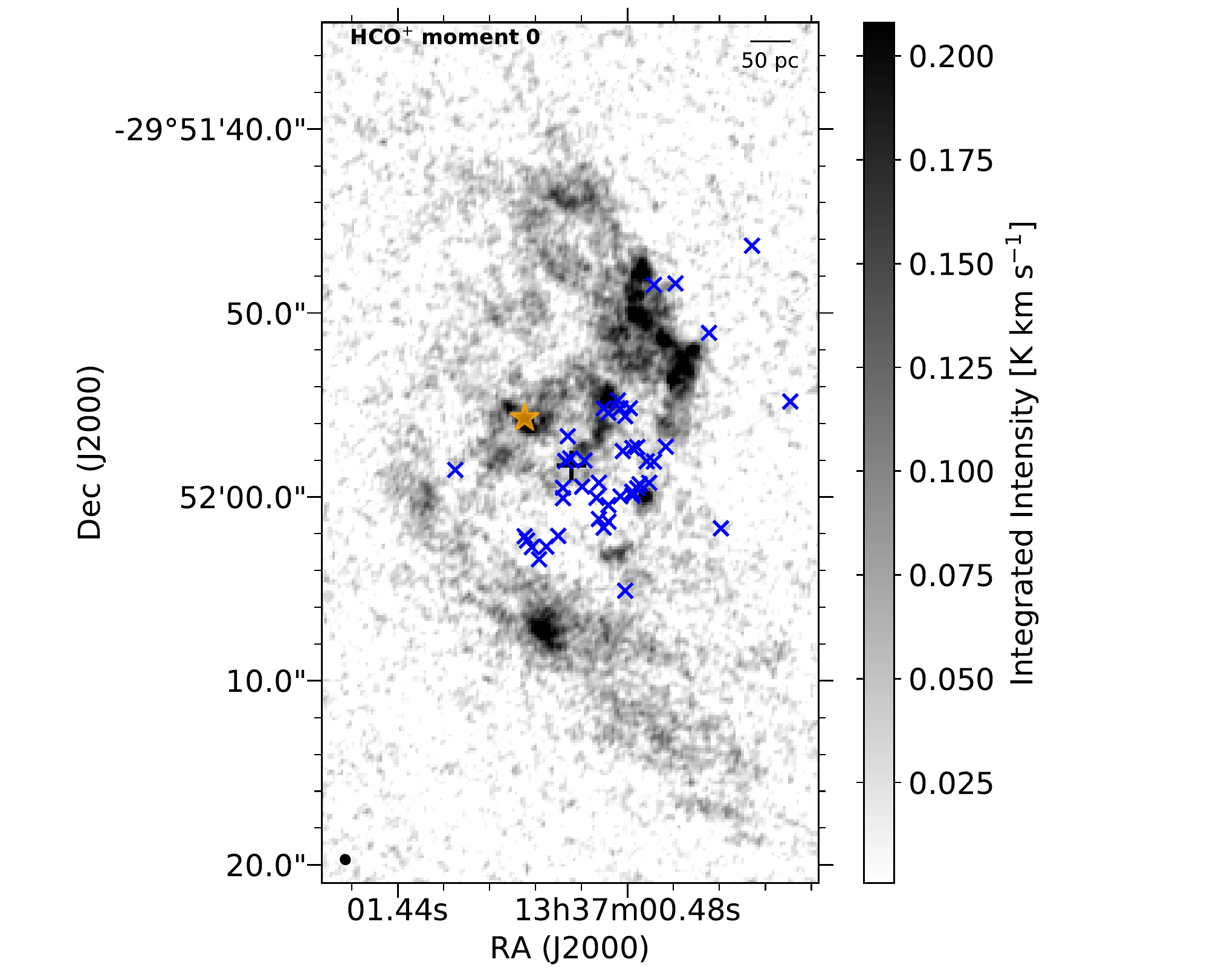} & \hspace{-1.6cm} \includegraphics[trim={1.5cm 0 0 0},clip,width=0.4\textwidth]{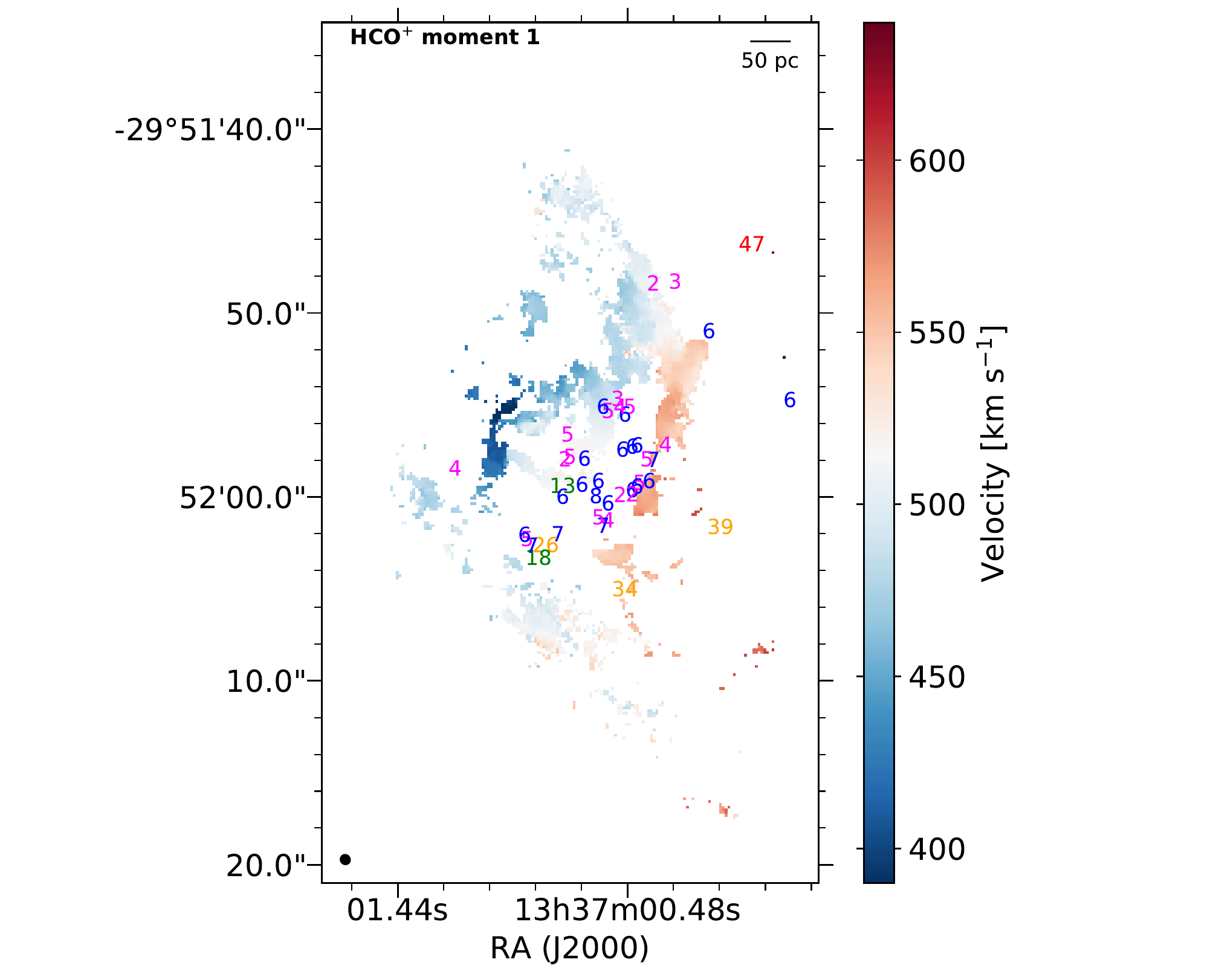} & \hspace{-1.6cm} \includegraphics[trim={1.5cm 0  0 0},clip,width=0.4\textwidth]{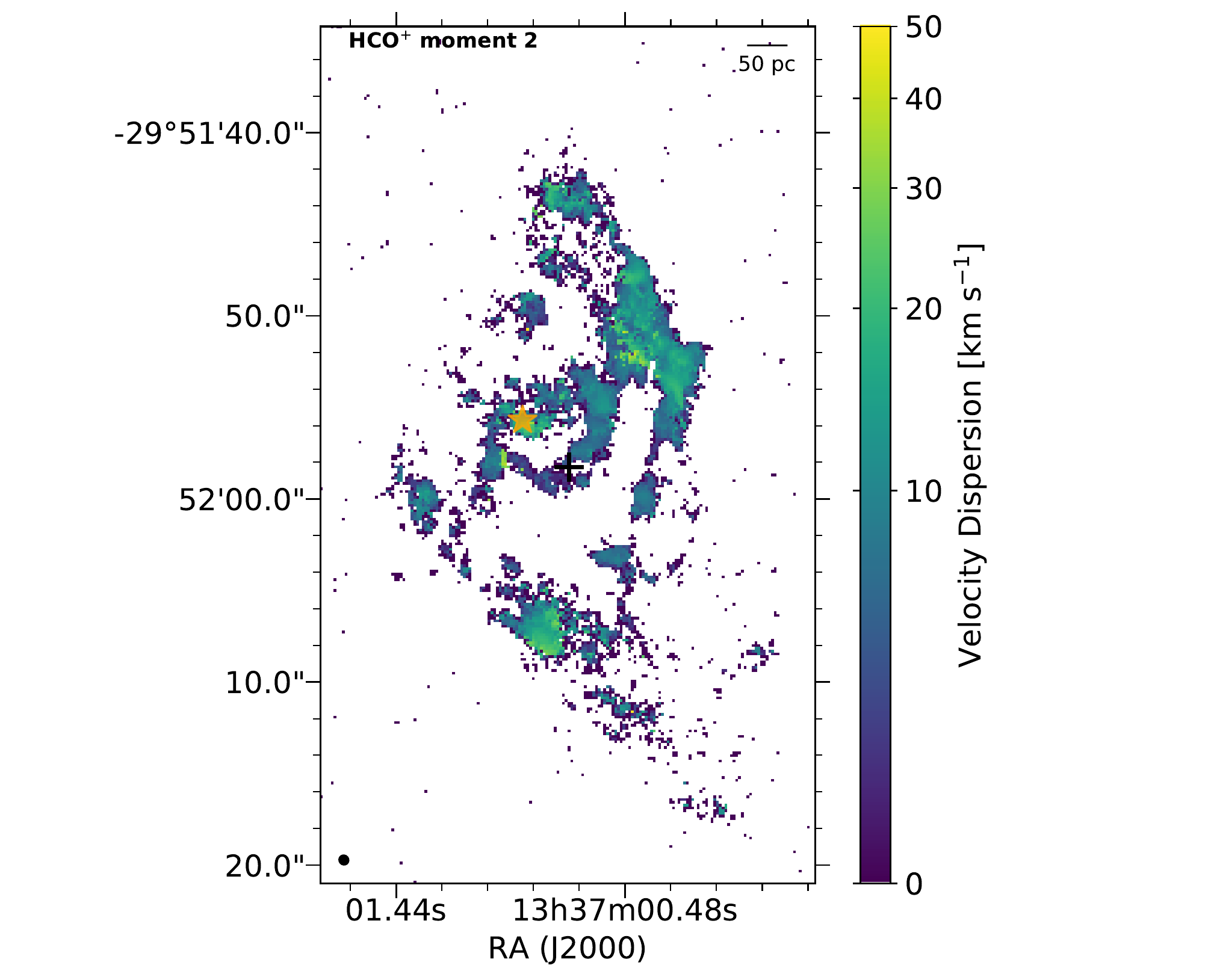}\\
	\end{tabular}
\caption{\normalsize [Left] Integrated intensity; [Middle] intensity weighted centroid velocity; [Right] intensity weighted velocity dispersion for HCO$^{+}$ ($1-0$). The structures and trends present in these maps are very similar to those in HCN ($1-0$), as shown in Figure~\ref{fig:moms}.}
\label{fig:moms_HCO}
\end{figure*}

\begin{figure*}
	\centering
	\includegraphics[trim={1.5cm 0 2cm 0},clip,width=1.1\textwidth]{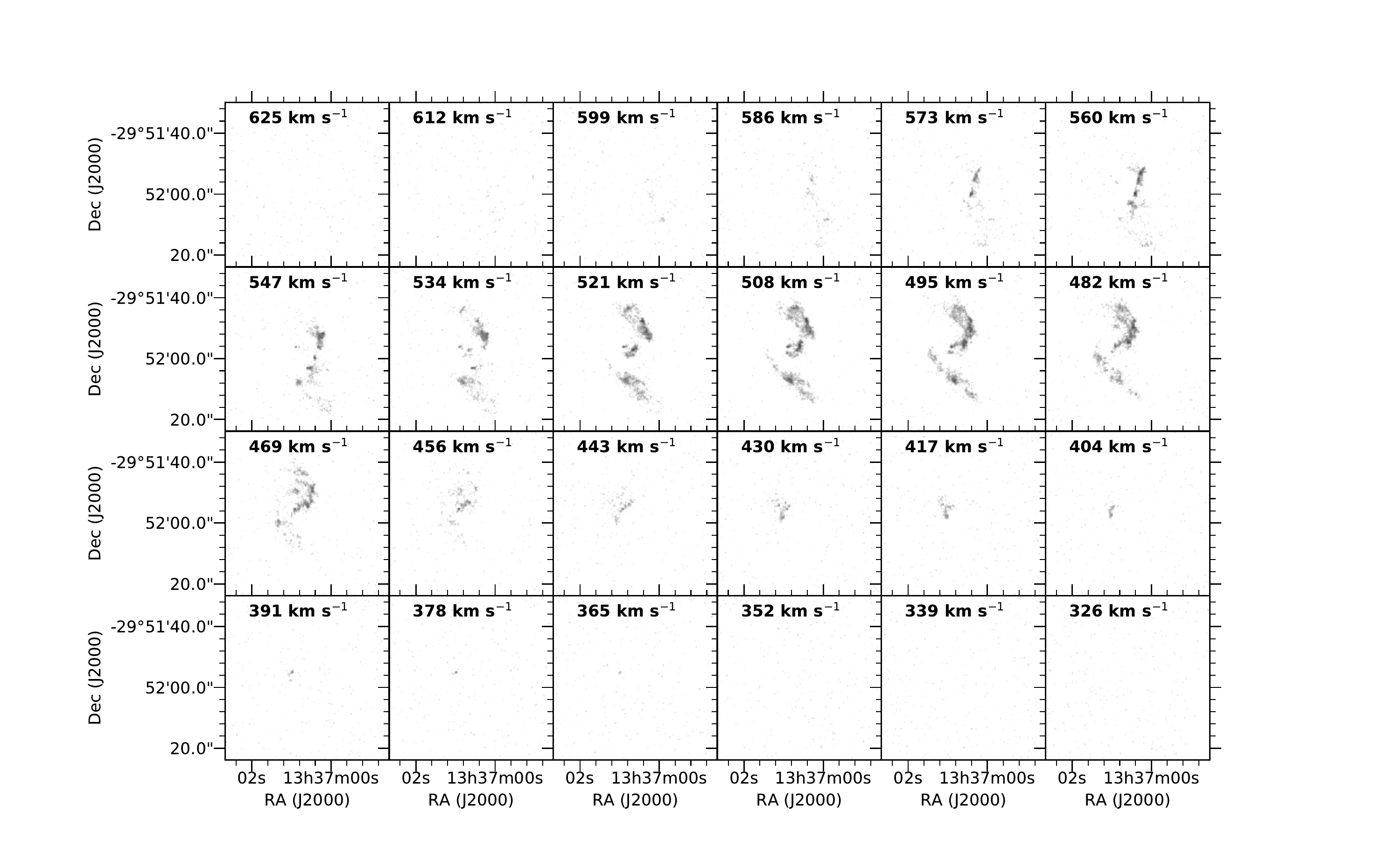}
	\caption{\normalsize Channel map of HCN ($1-0$) emission, with every 13\,km\,s$^{-1}$ averaged together between 326\,km\,s$^{-1}$ and 625\,km\,s$^{-1}$. The central velocity of each velocity bin is shown.}
	\label{fig:HCN_chan}
\end{figure*}

\begin{figure*}
	\centering
	\includegraphics[trim={1.5cm 0 2cm 0},clip,width=1.1\textwidth]{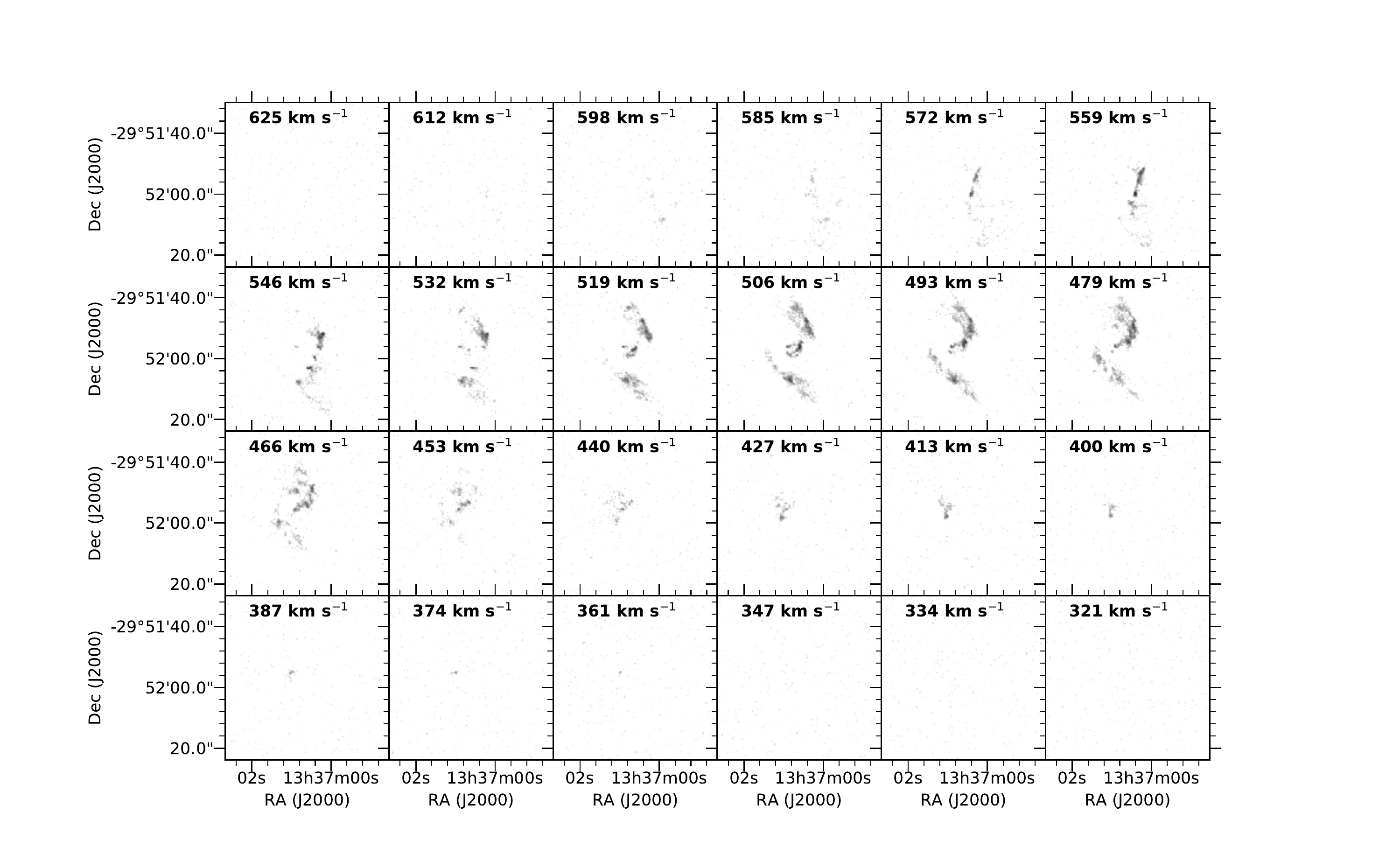}
	\caption{\normalsize Channel map of HCO$^{+}$ ($1-0$) emission, with every 13 km$^{-1}$ averaged together between 326 km s$^{-1}$ and 625 km s$^{-1}$. The central velocity of each velocity bin is shown.}
	\label{fig:HCO_chan}
\end{figure*}

\begin{figure*}
\centering
	\begin{tabular}{cc}
		\hspace{-1cm}\includegraphics[width=0.65\textwidth]{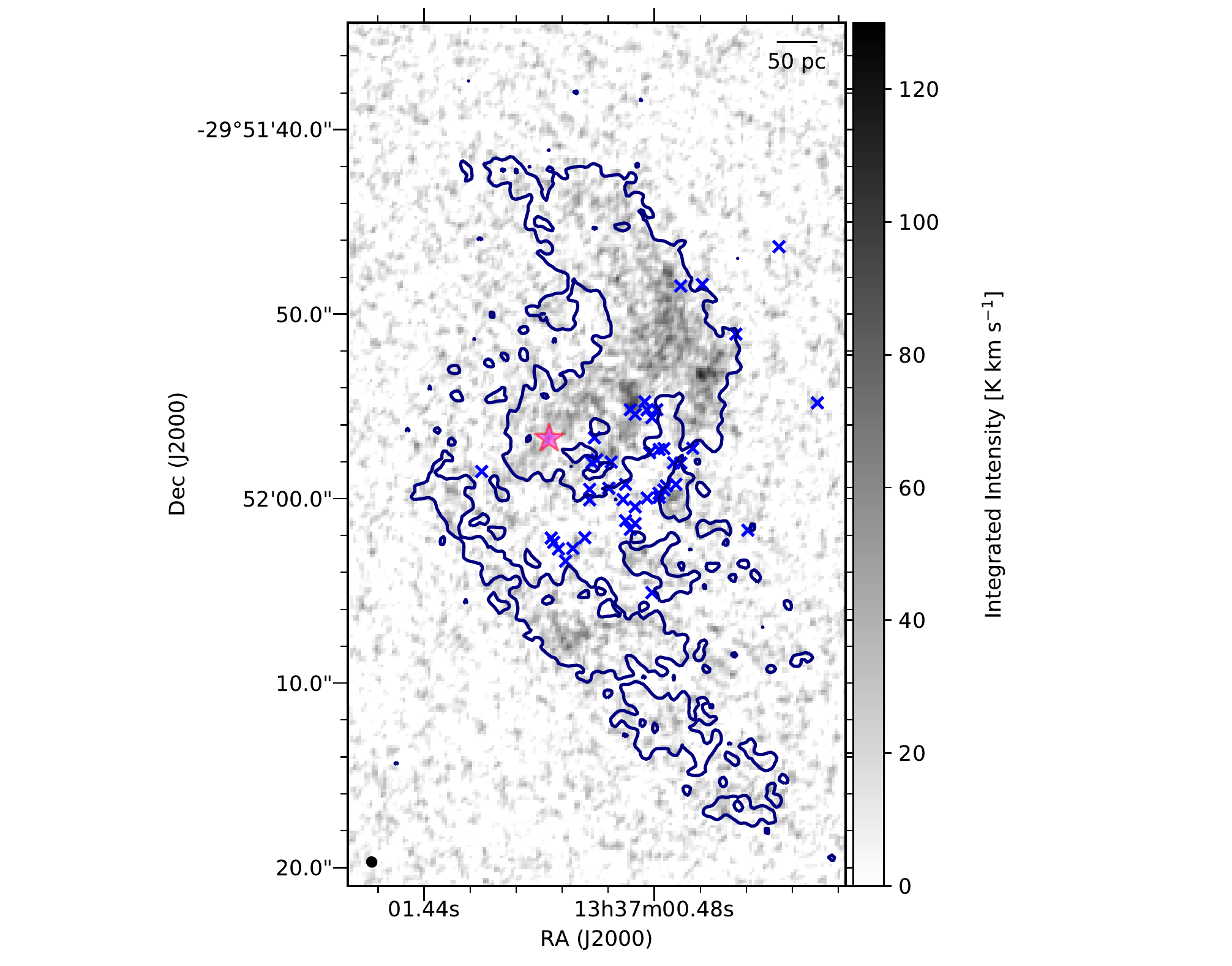} & \hspace{-2.2cm}\includegraphics[trim={2cm 0 2cm 0},clip,width=0.55\textwidth]{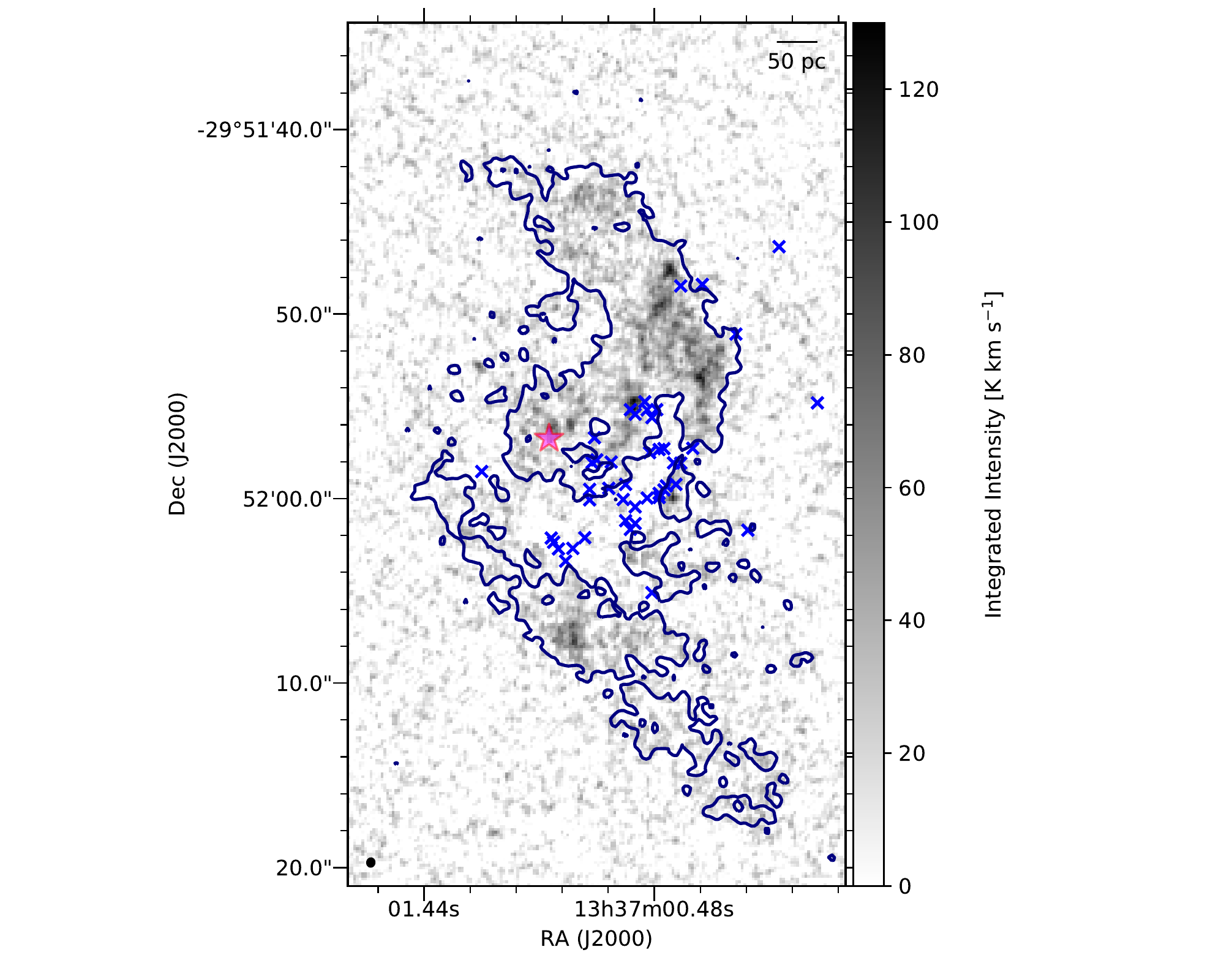}
	\end{tabular}
\caption{\normalsize Integrated intensity maps of the two other detected lines. [Left]: CCH (N = 1 - 0), [Right]: CS (J = 2 - 1). Blue crosses show positions of massive stellar clusters as found by \citet{2001AJ....122.3046H}. The orange star indicates the visual centre of M83, the green star indicates the location of the secondary nucleus observed by \citet{2000A&A...364L..47T}. CCH and CS maps also show HCN ($1-0$) contours overlaid at 30 K km s$^{-1}$ integrated intensity levels. Due to the significantly lower signal to noise ratio of these data, they were not used for analysis.}
\label{fig:add_mom0}
\end{figure*}

Figure~\ref{fig:moms_HCO} shows the integrated intensity, intensity weighted centroid velocity and intensity weighted velocity dispersion for HCO$^{+}$ ($1-0$). Figures~\ref{fig:HCN_chan} and~\ref{fig:HCO_chan} shows the HCN and HCO$^{+}$ ($1-0$) channel map respectively. Figure~\ref{fig:add_mom0} shows the integrated intensity maps of CCH (N = 1 - 0) and CS (J = 2 - 1).

Figures~\ref{fig:in},~\ref{fig:outer} and~\ref{fig:inner} show the spectra taken from intensity peaks along the dust lanes, the outer circumnuclear ring and the inner circumnuclear ring, respectively. To increase the signal-to-noise ratio of these spectra they were averaged over an area of 1$^{\prime \prime}$ ($\sim 24$ pc), the largest size scale at which it is still possible to reliably isolate individual clouds. Table~\ref{table:disp} shows the peak brightness temperature and velocity dispersions of these spectra.
\newpage
\begin{table}
	\caption{\textbf{SCOUSE Fit Data}}
\centering
\begin{tabularx}{0.5\textwidth}{p{2cm} p{1.5cm} p{2cm} p{2cm}}
\hline
\hline
Component & Spectrum & T$_{B}$ (K km s$^{-1}$) & $\sigma$ (km s$^{-1}$)  \\
\hline
Dust Lanes                              & 1 & 1.20 $\pm$ 0.05 & 14.1 $\pm$ 1.2 \\
(Figure~\ref{fig:in})               	&   & 1.56 $\pm$ 0.06 & 11.4 $\pm$ 0.7 \\
					  					& 2 & 0.84 $\pm$ 0.04 & 19.6 $\pm$ 1.0 \\
					  					& 3 & 1.62 $\pm$ 0.07 & 11.2 $\pm$ 0.7 \\
					  					&   & 2.88 $\pm$ 0.05 & 14.7 $\pm$ 0.6\\
					  					& 4 & 1.31 $\pm$ 0.19 & 14.1 $\pm$ 1.5 \\
					  					&   & 0.83 $\pm$ 0.09 & 30.4 $\pm$ 3.3 \\
					 					& 5 & 1.98 $\pm$ 0.04 & 25.2 $\pm$ 0.6 \\
					  					& 6 & 1.80 $\pm$ 0.06 & 14.1 $\pm$ 1.0 \\
					  					&   & 2.10 $\pm$ 0.10 & 11.4 $\pm$ 0.6 \\
					  					& 7 & 0.89 $\pm$ 0.08 & 8.9 $\pm$ 1.5 \\
                                        &   & 1.20 $\pm$ 0.08 & 8.9 $\pm$ 1.1 \\
					  					& 8 & 2.64 $\pm$ 0.06 & 13.6 $\pm$ 0.5 \\
					  					&   & 1.30 $\pm$ 0.06 & 14.1 $\pm$ 1.2 \\
					  					& 9 & 0.74 $\pm$ 0.05 & 20.8 $\pm$ 1.5 \\
					  					& 10 & 1.32 $\pm$ 0.05 & 15.6 $\pm$ 0.6 \vspace{1.0cm} \\
Outer Ring                              & 1 & 1.54 $\pm$ 0.04 & 16.5 $\pm$ 0.5 \\
(Figure~\ref{fig:outer})  				& 2 & 2.74 $\pm$ 0.05 & 13.2 $\pm$ 0.3 \\
				   						& 3 & 1.84 $\pm$ 0.05 & 14.3 $\pm$ 0.4 \\
				   						& 4 & - & - \\
				   						& 5 & 0.50 $\pm$ 0.04 & 18.6 $\pm$ 1.5 \\
				   						& 6 & 0.79 $\pm$ 0.03 & 29.2 $\pm$ 1.2 \vspace{1.0cm} \\
Inner Ring                               & 1 & 0.98 $\pm$ 0.03 & 31.1 $\pm$ 1.1 \\
(Figure~\ref{fig:inner}) 		    	 & 2 & 1.00 $\pm$ 0.04 & 23.8 $\pm$ 1.1 \\
					 				     & 3 & 2.60 $\pm$ 0.05 & 18.0 $\pm$ 0.4 \\
									     & 4 & 2.81 $\pm$ 0.05 & 13.0 $\pm$ 0.3 \\
									     & 5 & 2.51 $\pm$ 0.06 & 13.4 $\pm$ 0.4 \\
									     & 6 & 2.10 $\pm$ 0.05 & 13.7 $\pm$ 0.4 \\
								 	     & 7 & 1.21 $\pm$ 0.05 & 11.5 $\pm$ 0.5 \\
					 				     & 8 & 1.27 $\pm$ 0.04 & 16.6 $\pm$ 0.6 \\
					 				     &   & 0.51 $\pm$ 0.04 & 14.1 $\pm$ 1.4 \\
\hline
\end{tabularx}
\\[0.2cm]
\begin{flushleft}
Velocity dispersions  and brightness temperatures of Gaussian components fit to spectra taken at integrated intensity peaks throughout key regions in the observed gas structure. These values have been extracted in apertures of 1$^{\prime\prime}$. Two values are given for a single spectrum in cases where a two-component Gaussian fit was used. No value is reported if the fit was unreliable.
\end{flushleft}
\label{table:disp}
\end{table}

\newpage

\clearpage
\newpage
\begin{figure*}
	\centering
	\begin{tabular}{cc}
     \includegraphics[scale=0.27]{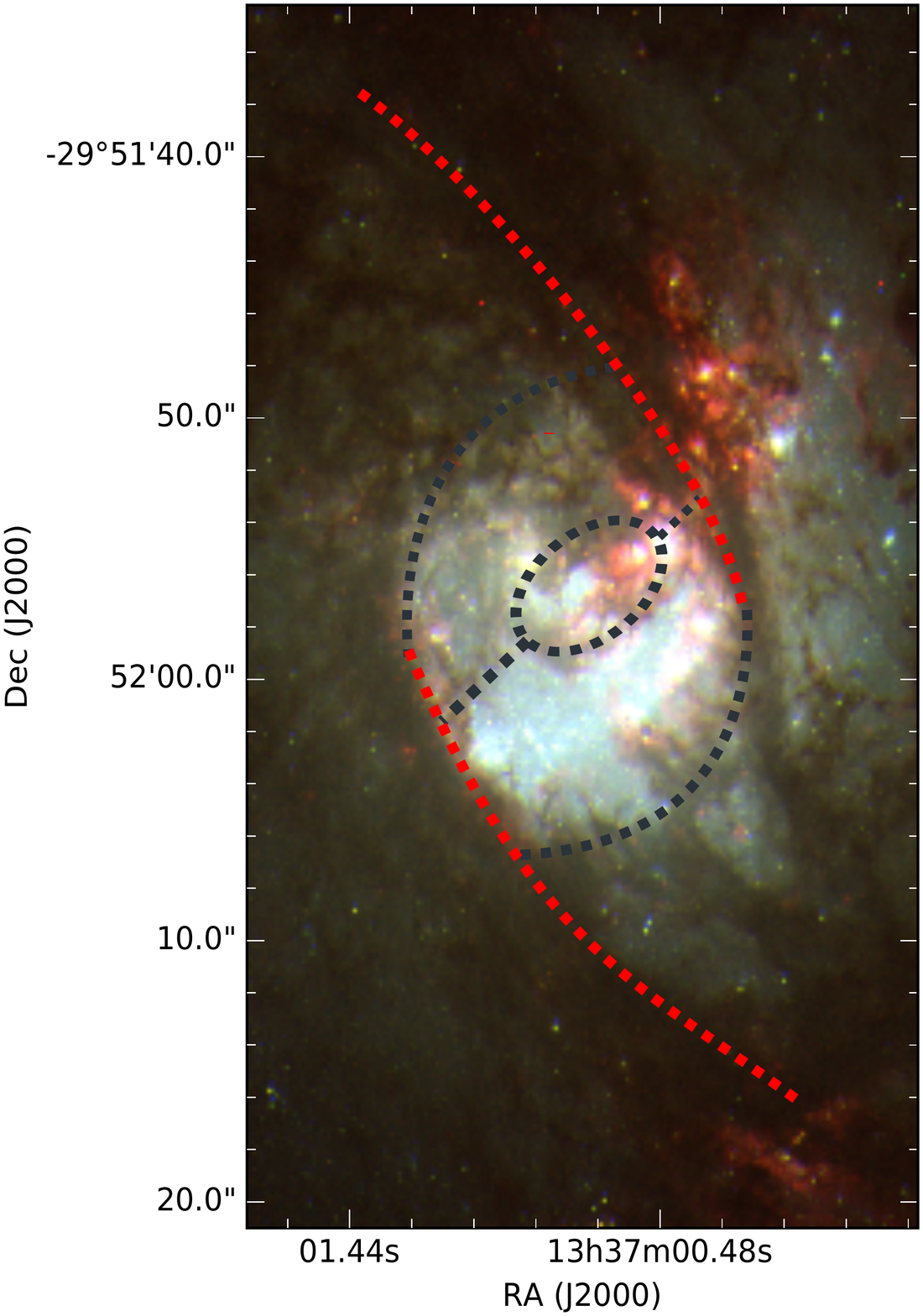} \hspace{3cm} & \hspace{-3cm} \includegraphics[scale=0.45]{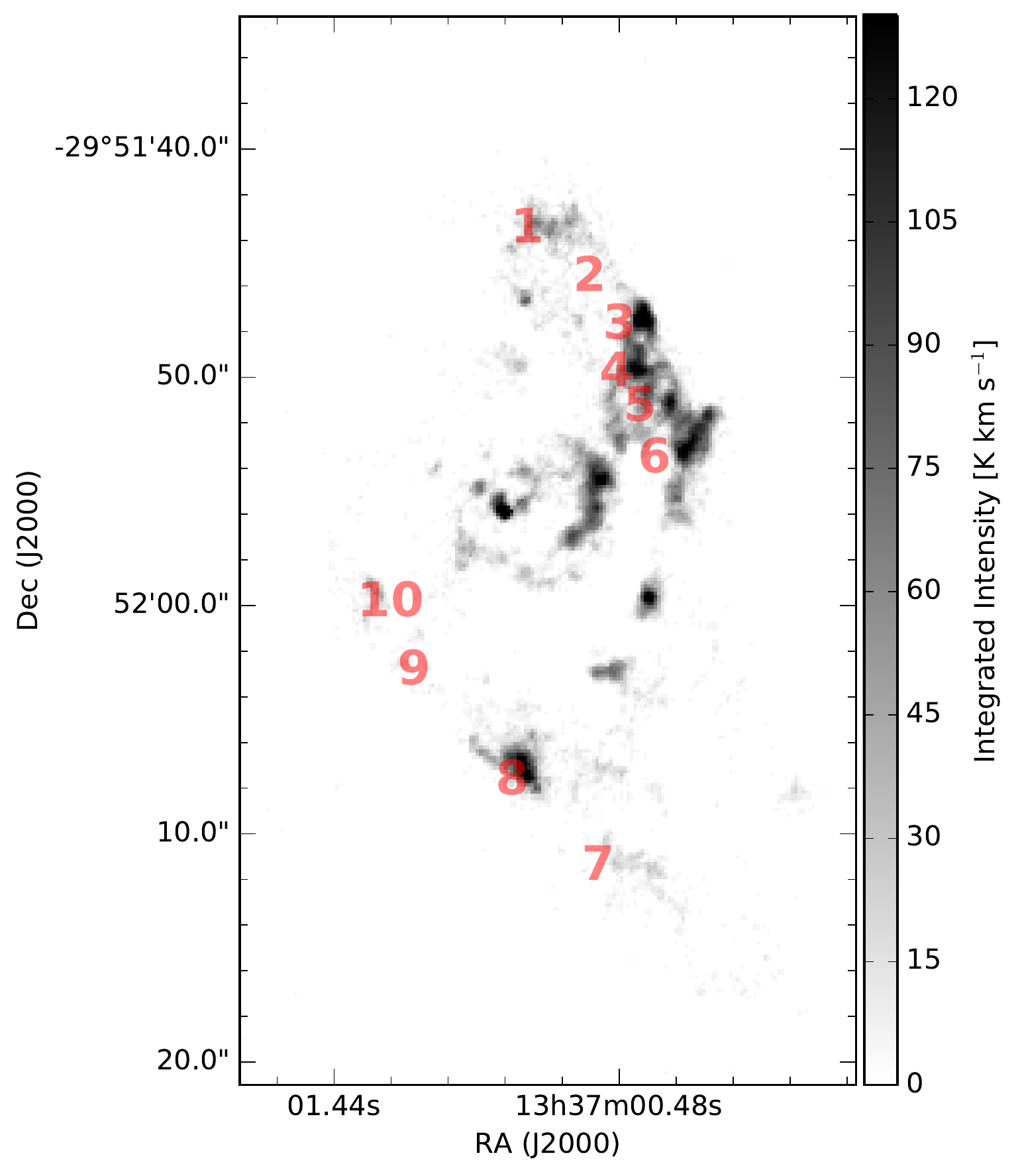} \\	
	\end{tabular}
\end{figure*}

\begin{figure*}
	\centering
	\includegraphics[width=0.90\textwidth]{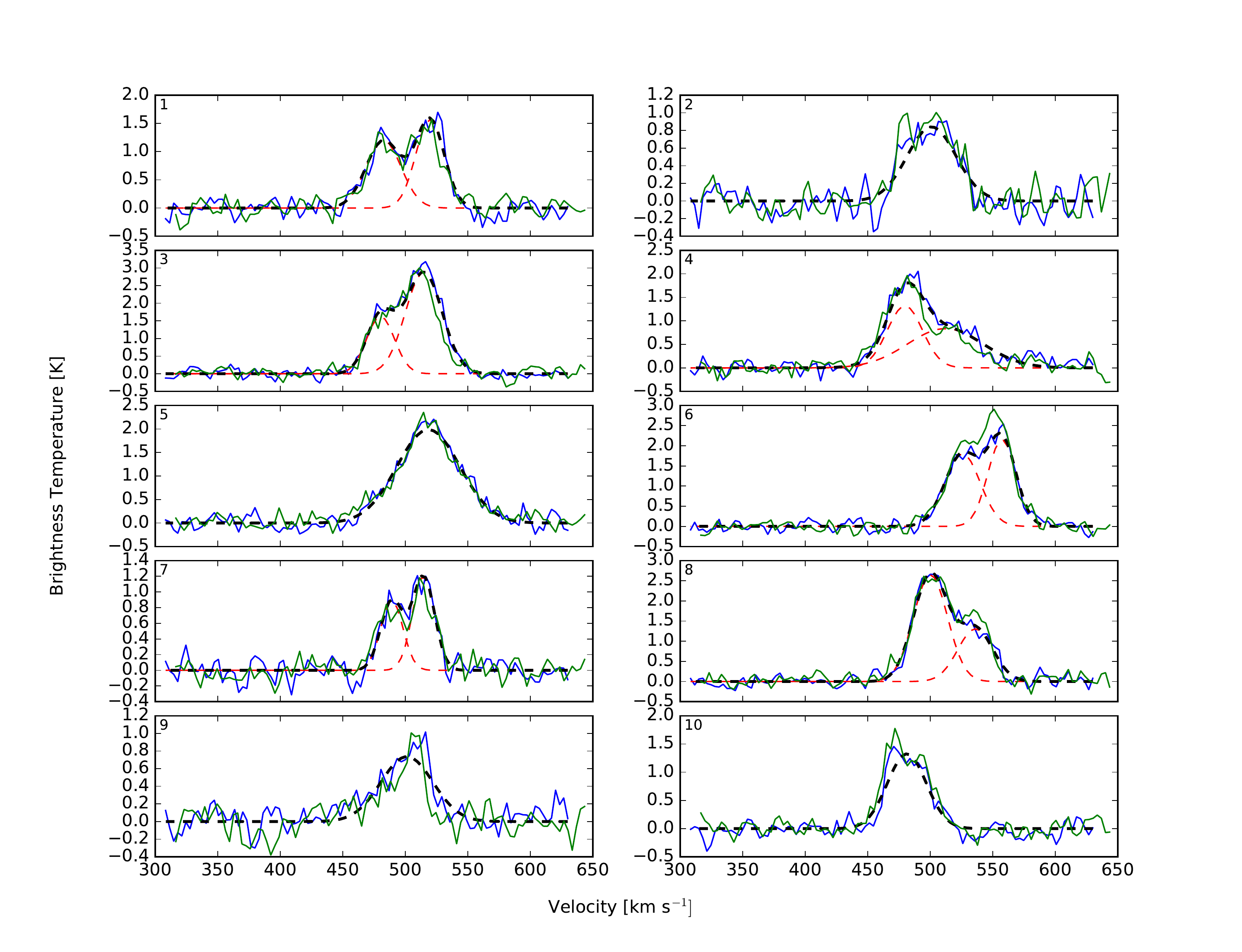}
		\caption{\normalsize Example spectra of HCN ($1-0$) [blue] and HCO$^{+}$ ($1-0$) [green] taken at HCN ($1-0$) integrated intensity peaks throughout key regions of the observed gas structure. [Top Left]: the same colour scale image as Figure~\ref{fig:schem_circ} overlaid with the overall schematic and the relevant region highlighted -- in this case, the dust lanes highlighted in red. [Top Right]: the HCN ($1-0$) integrated intensity image in grey scale overlaid with the locations at which each spectra was taken. The spectra shown start at 1 (upper left) and end at 10 (lower right), and were averaged over a region 1$^{\prime\prime}$ in size. The black dashed line shows the Gaussian component fit to the HCN ($1-0$) spectra. In cases where a multi-component Gaussian fit was used the red dashed line shows the properties of each Gaussian component individually. As shown in Table~\ref{table:disp}, there is no monotonic trend between gas velocity dispersion and galactocentric radius in the dust lanes.}
\label{fig:in}
\end{figure*}

\begin{figure*}
	\centering
	\begin{tabular}{cc}
     \includegraphics[scale=0.3]{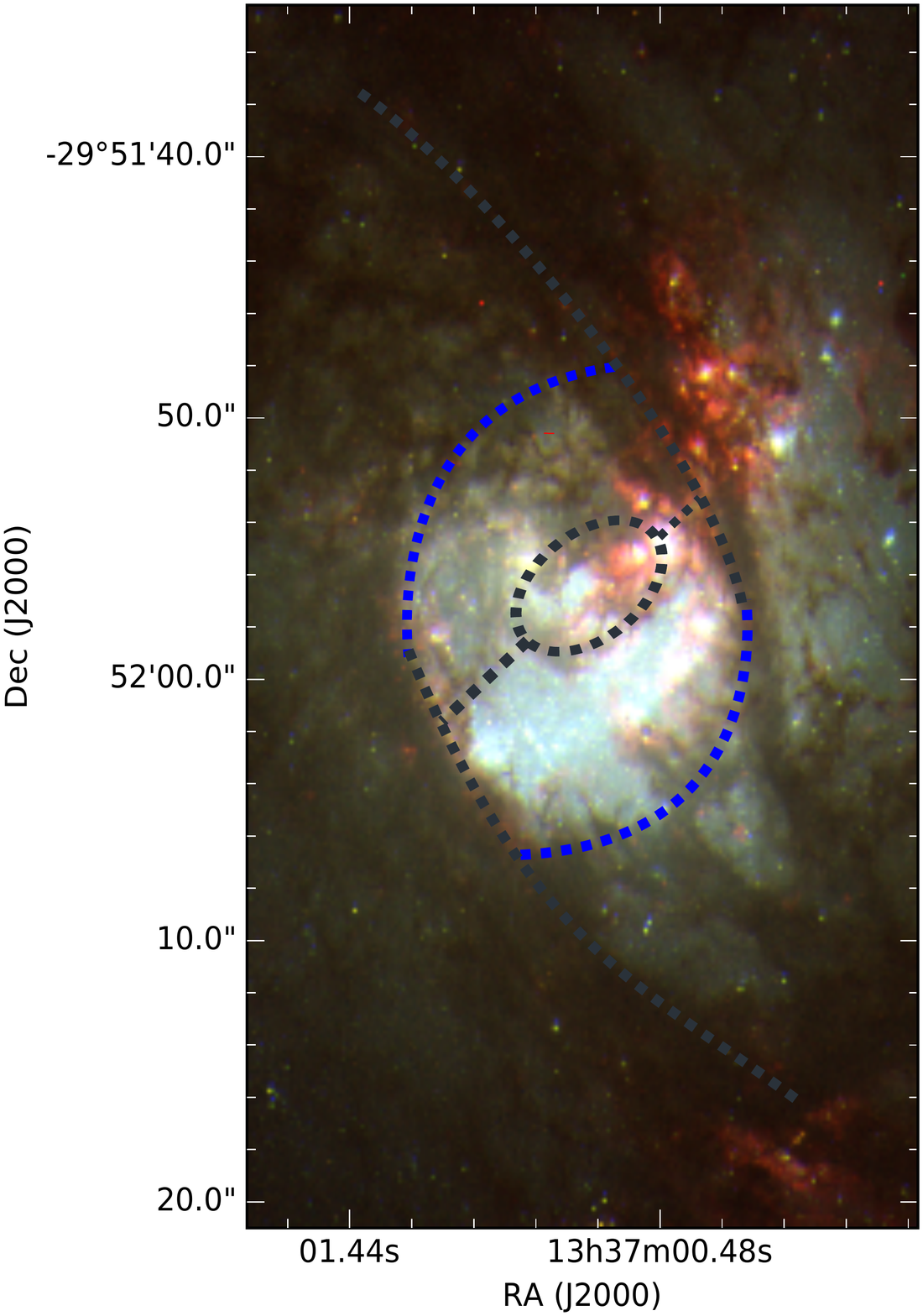}  \hspace{1cm} & \hspace{-1.5cm}  \includegraphics[scale=0.5]{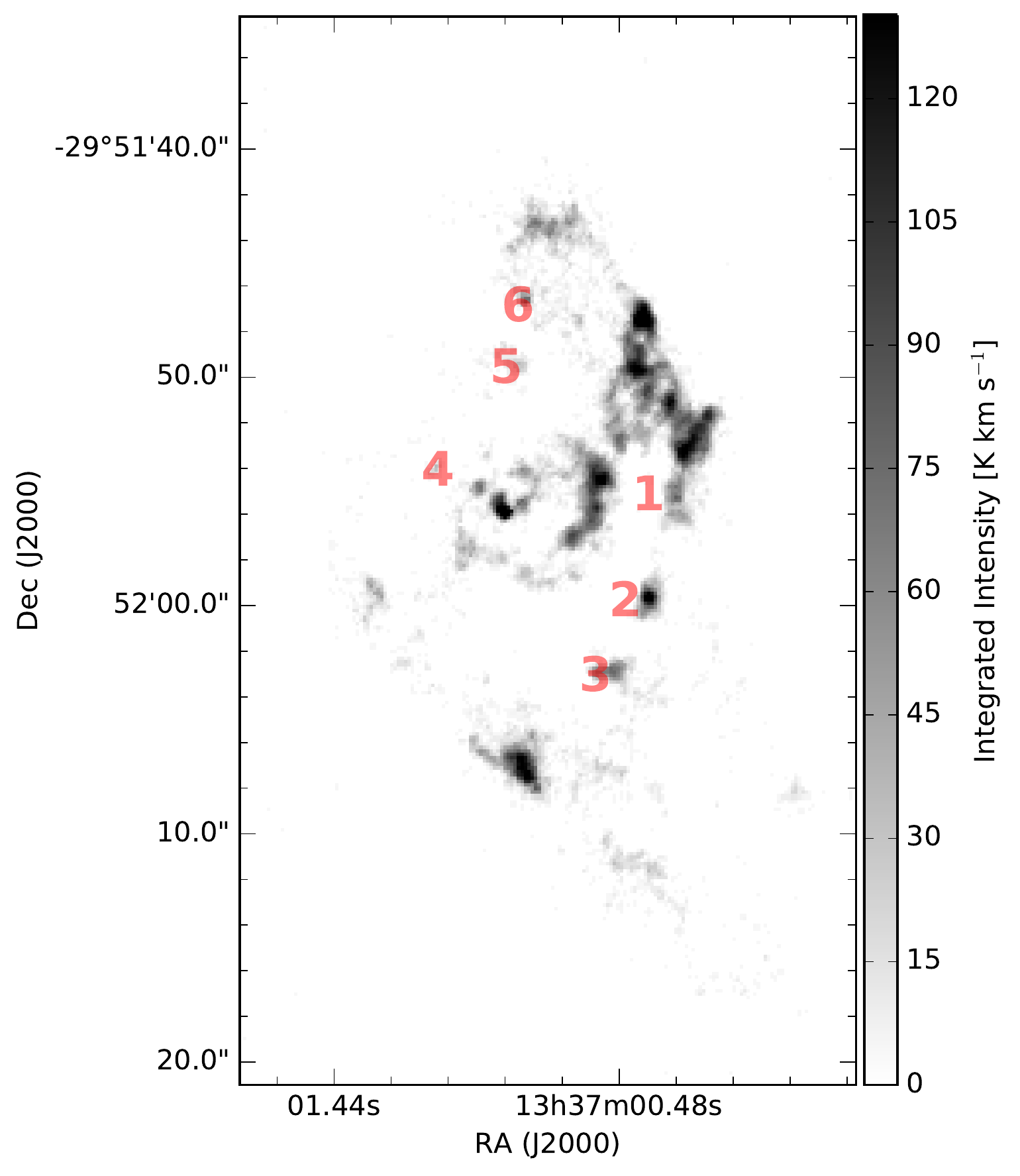} \\
	\end{tabular}
\end{figure*}

\begin{figure*}
	\centering
	\includegraphics[trim={0 7cm 0 0 }, clip, width=0.9\textwidth]{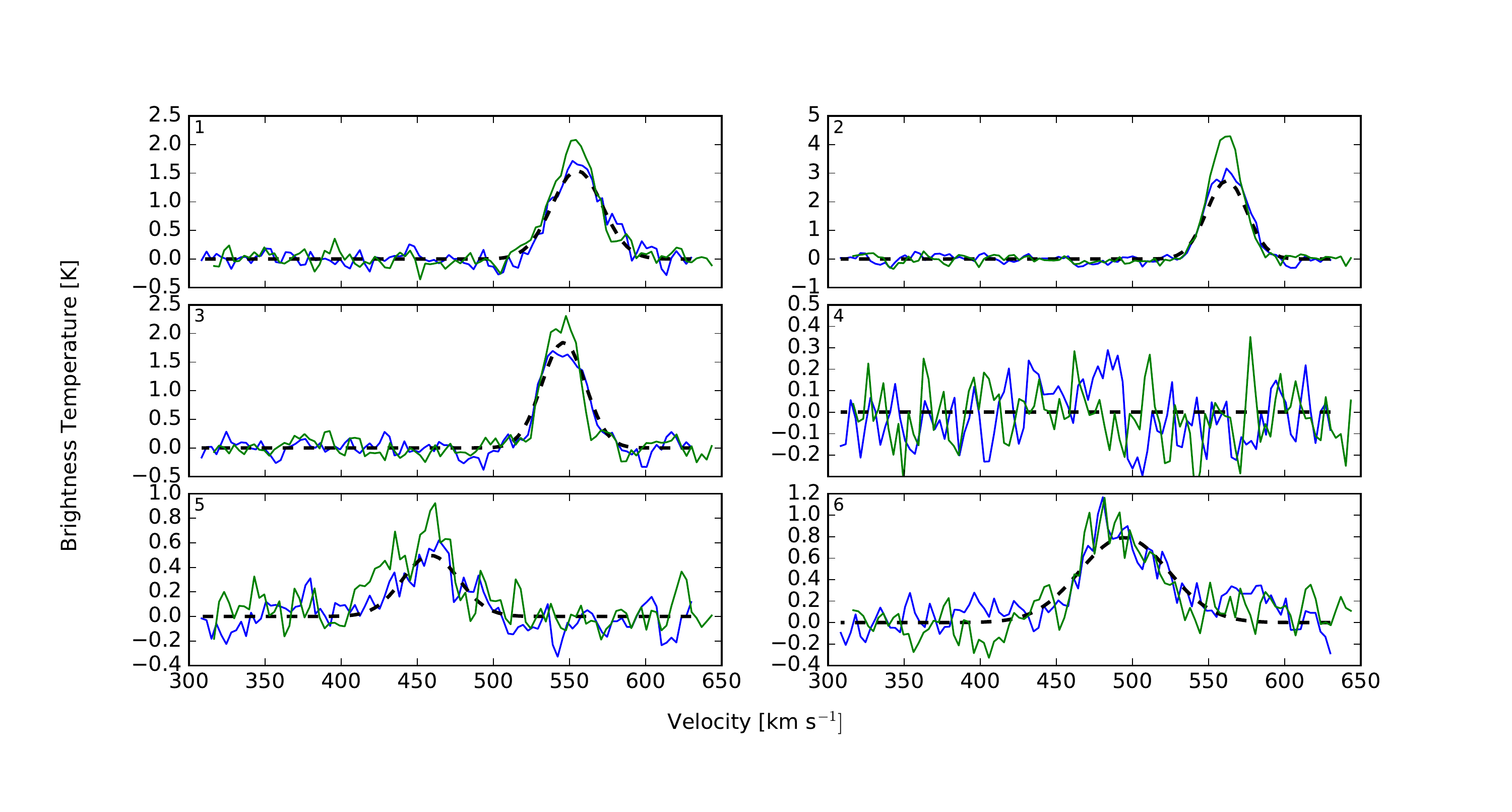}
	\caption{\normalsize HCN ($1-0$) and HCO$^{+}$ ($1-0$) spectrum as in Figure~\ref{fig:in} but extracted from key locations within the outer circumnuclear ring, as shown in blue in the upper left panel.}
\label{fig:outer}
\end{figure*}

\begin{figure*}
	\begin{tabular}{cc}
     \includegraphics[scale=0.3]{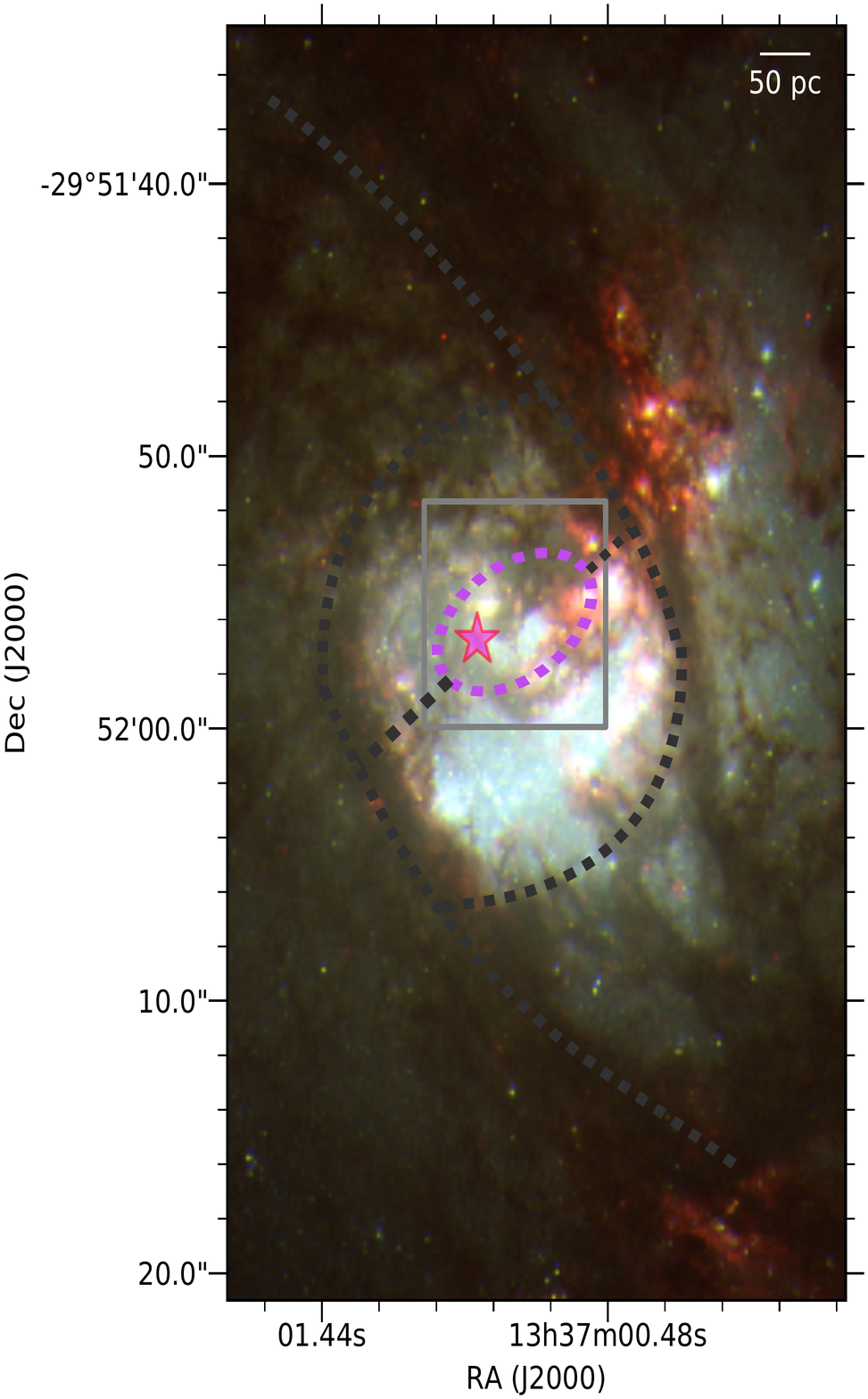} \hspace{1cm} & \hspace{-1cm} \includegraphics[width=0.5\textwidth]{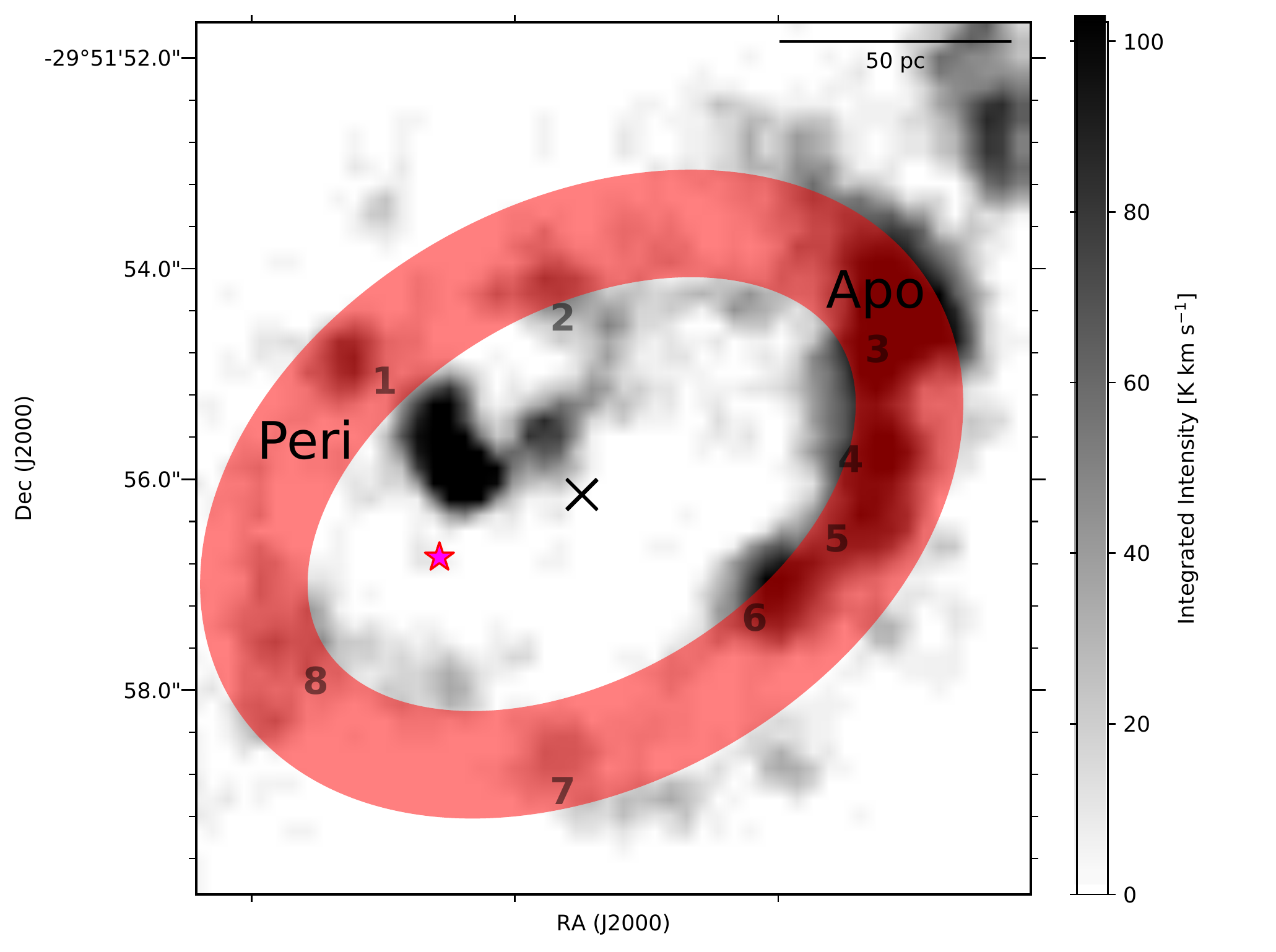} \\
	\end{tabular}
		\includegraphics[trim={0 4cm 0 0 }, clip, width=\textwidth]{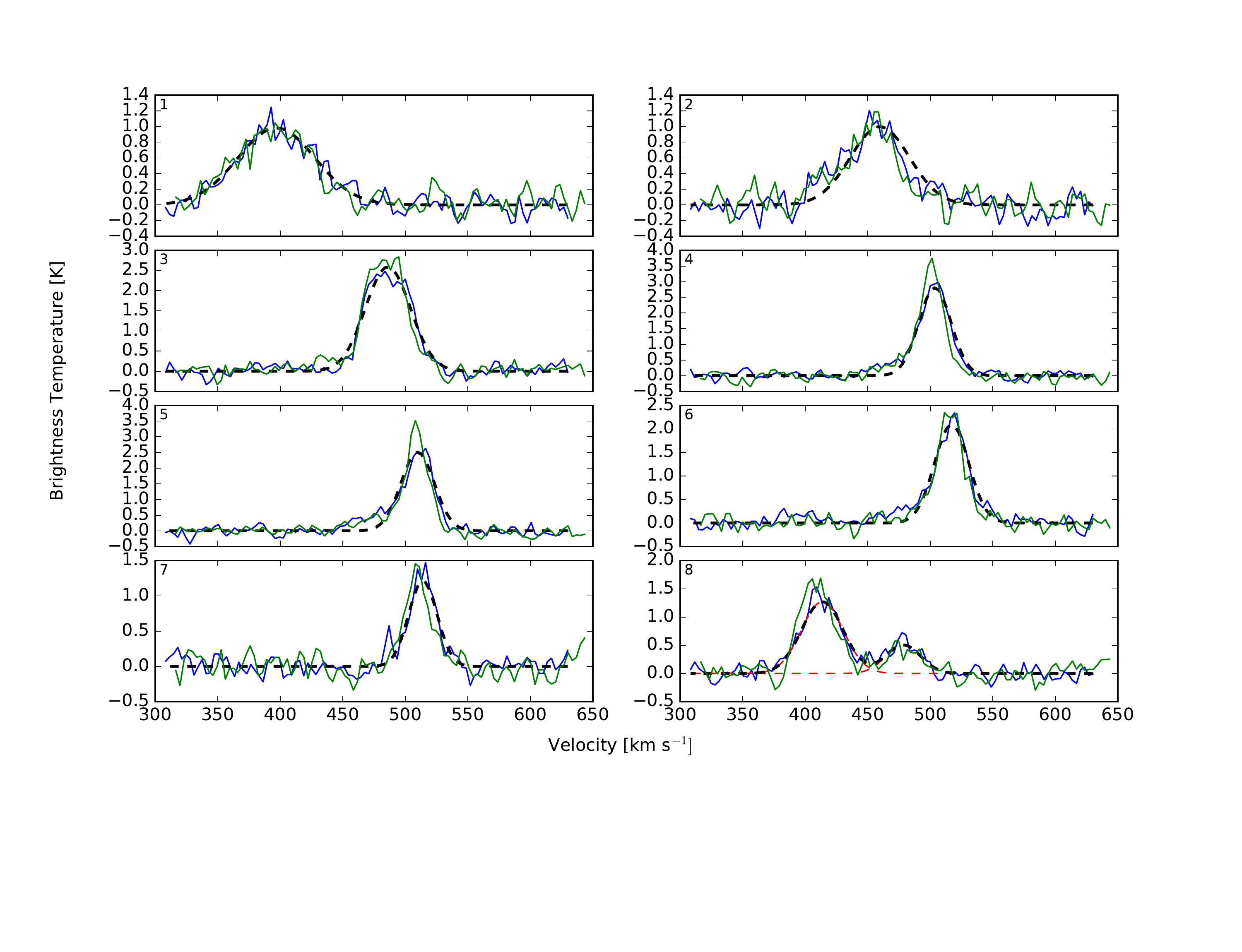}
	\caption{\normalsize HCN ($1-0$) and HCO$^{+}$ ($1-0$) spectrum as in Figure~\ref{fig:in} but extracted from key locations within the inner circumnuclear ring as shown in purple in the upper left panel. The box on the left-hand panel shows the field of view of the right-hand panel. The cross shows the location of the centre of the manually fitted ellipse (red shaded region) to the circumnuclear ring, and the plus is the location of the visible nucleus of M83, which was used as the zero-point for radius and azimuthal angle calculations. The ellipse has semi-major and semi-minor axes of $a=45 $pc and $b = 27 $pc with a position angle of $60^{\circ}$. The locations of peri- and apocentre from the visible nucleus are also labeled.}
\label{fig:inner}
\end{figure*}

\section{Line-fitting of HCN ($1-0$) and HCO$^+$ ($1-0$) with SCOUSE}
\label{sec:scouse}
Due to the possibility of multiple spectral components per sightline, which leads to unreliable results when using moment analysis, the HCN ($1-0$) and HCO$^{+}$ ($1-0$) data were run through the \textbf{S}emi-automated multi-\textbf{CO}mponent \textbf{U}niversal \textbf{S}pectral-line fitting \textbf{E}ngine (SCOUSE) as presented in \citet{2016ascl.soft01003H}. The method and results of the SCOUSE fitting process are described in Appendix \ref{sec:scouse}. We use the results of the line fitting analysis to investigate how the kinematics of the gas varies with position in the central few hundred pc of the galaxy.

\begin{figure*}
	\centering
	\begin{tabular}{cc}
		\includegraphics[trim={2cm 0 0.5cm 0}, clip,width=0.6\textwidth]{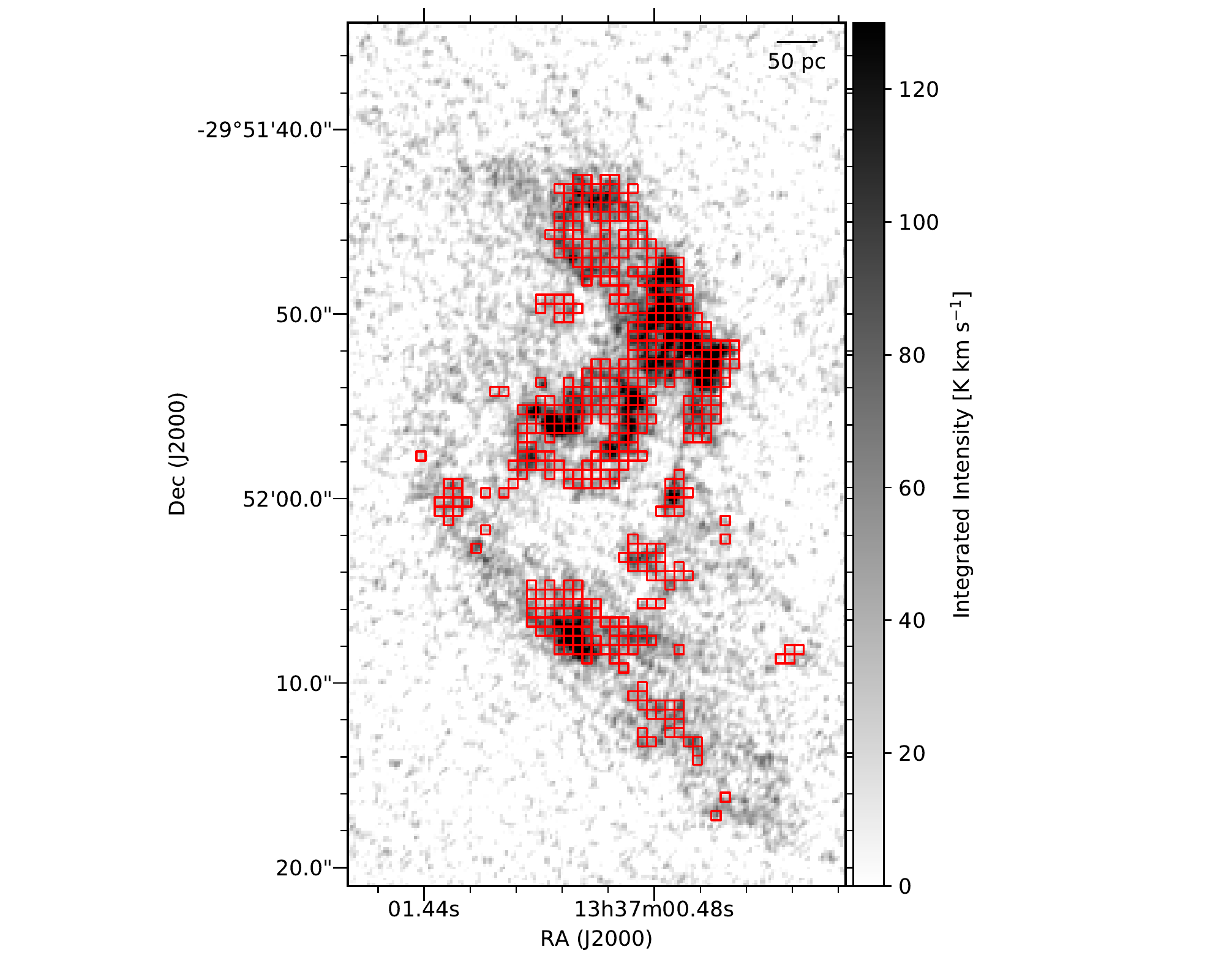} & \hspace{-2.5cm} 
		\includegraphics[trim={2cm 0 0.5cm 0}, clip,width=0.6\textwidth]{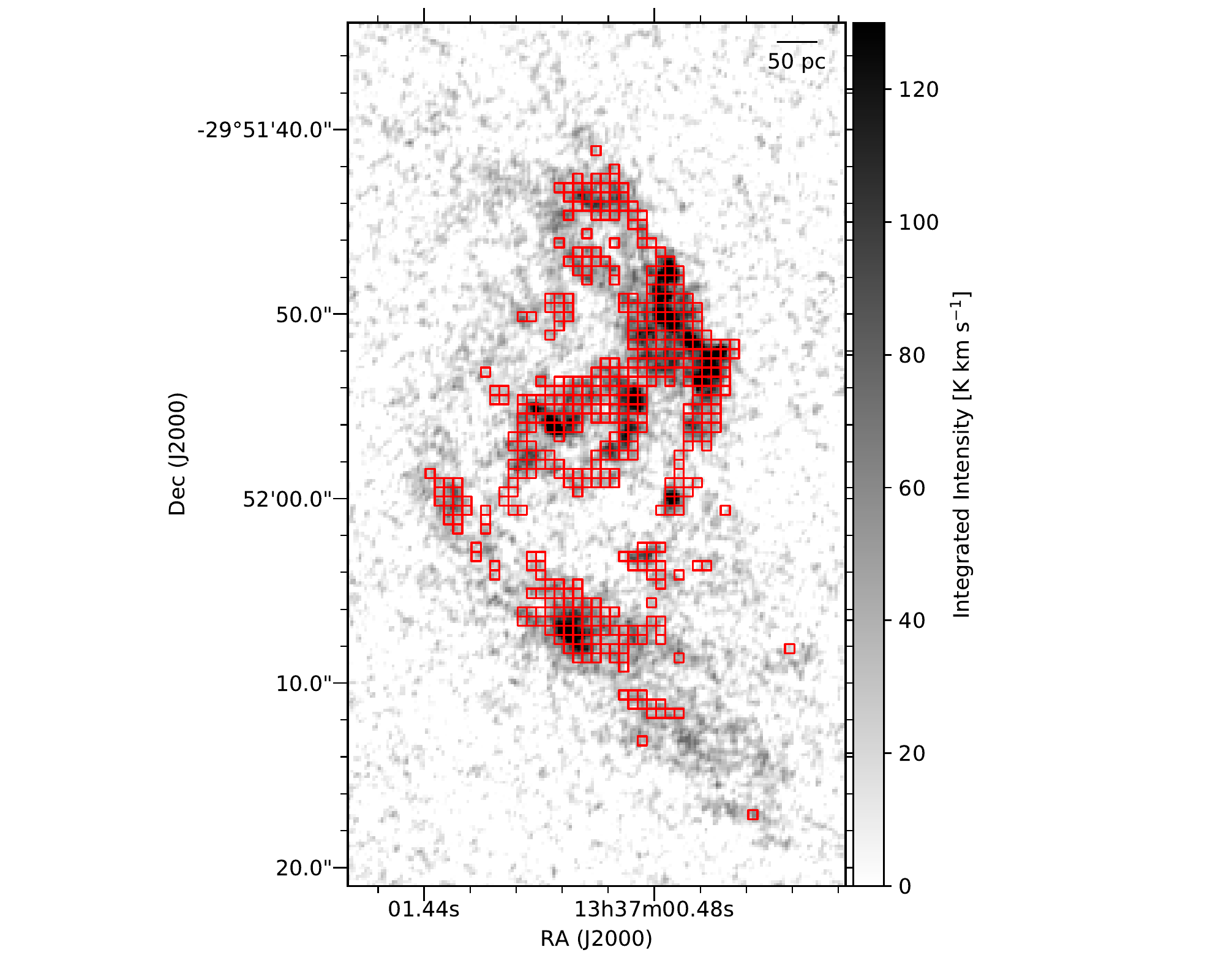}
	\end{tabular}
	\caption{\normalsize Coverage of SCOUSE. Red boxes denote an individual spectral averaging area, overlaid on top of a HCN ($1-0$) [Left] and HCO$^{+}$ ($1-0$) [Right] integrated intensity map.}
	\label{fig:cov}
\end{figure*}

SCOUSE is a line-fitting algorithm that is capable of fitting Gaussian profiles to large spectral-line datasets efficiently. It does this by breaking the input dataset into smaller equally sized regions, rejecting those regions in which less than 50$\%$ of the cells exceed a user-defined noise threshold. For the remaining regions, the signal is averaged over the entire region on per-channel basis to produce a spatially averaged spectrum. Each of these spectra are then manually inspected, and all lines are fit by the user, with the number of guassian components, and their given parameters estimated manually. This fitting process is then used as a template for the spectra of each cell that comprises each spatially averaged area (SAA). These SAAs were selected to be 0.5$^{\prime \prime}$ ($\sim$11 pc) in radius as this is twice the expected cloud size within this environment. We enforced a $5 \sigma$ cut with an RMS of 0.25 K per 3.2 km s$^{-1}$ channel. This allowed us to get maximum coverage over the important emission whilst minimising the time needed to fit all spectra. Figure~\ref{fig:cov} demonstrates how this coverage is defined in SCOUSE. Each red box is a spectral averaging area with a user defined radius of, in this case, 0.5$^{\prime \prime}$.

Moment maps were created using CASA, as well as their corresponding maps using the SCOUSE output. Figure~\ref{fig:moms} shows the continuum and zeroth, first and second order moment maps for HCN ($1-0$) output from CASA. The velocity maps delineate the structure of the gas within this region; particularly demonstrating the contiguous gas lanes from the north and south that appear to be feeding the circumnuclear ring sitting at a distance of $\sim$150 pc from the centre. A second circumnuclear ring is also observed sitting at $\sim50-100$\,pc from the nucleus.

\begin{figure*}
    \Centering
    \begin{tabular}{cc}
         \includegraphics[trim={2cm 0 0.5cm 0},clip,width=0.61\textwidth]{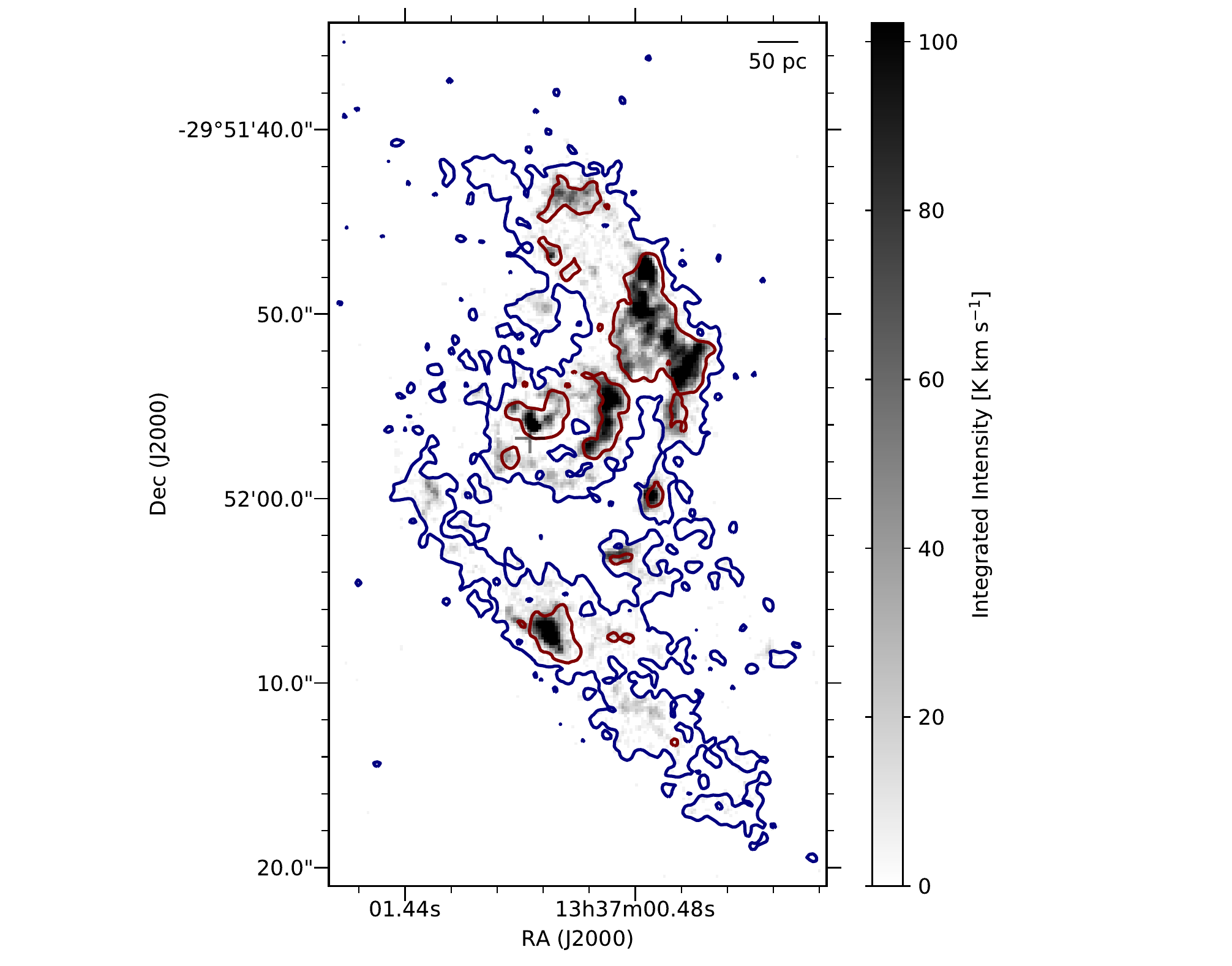} & \hspace{-2.5cm} \includegraphics[trim={2cm 0 0.5cm 0},clip,width=0.61\textwidth]{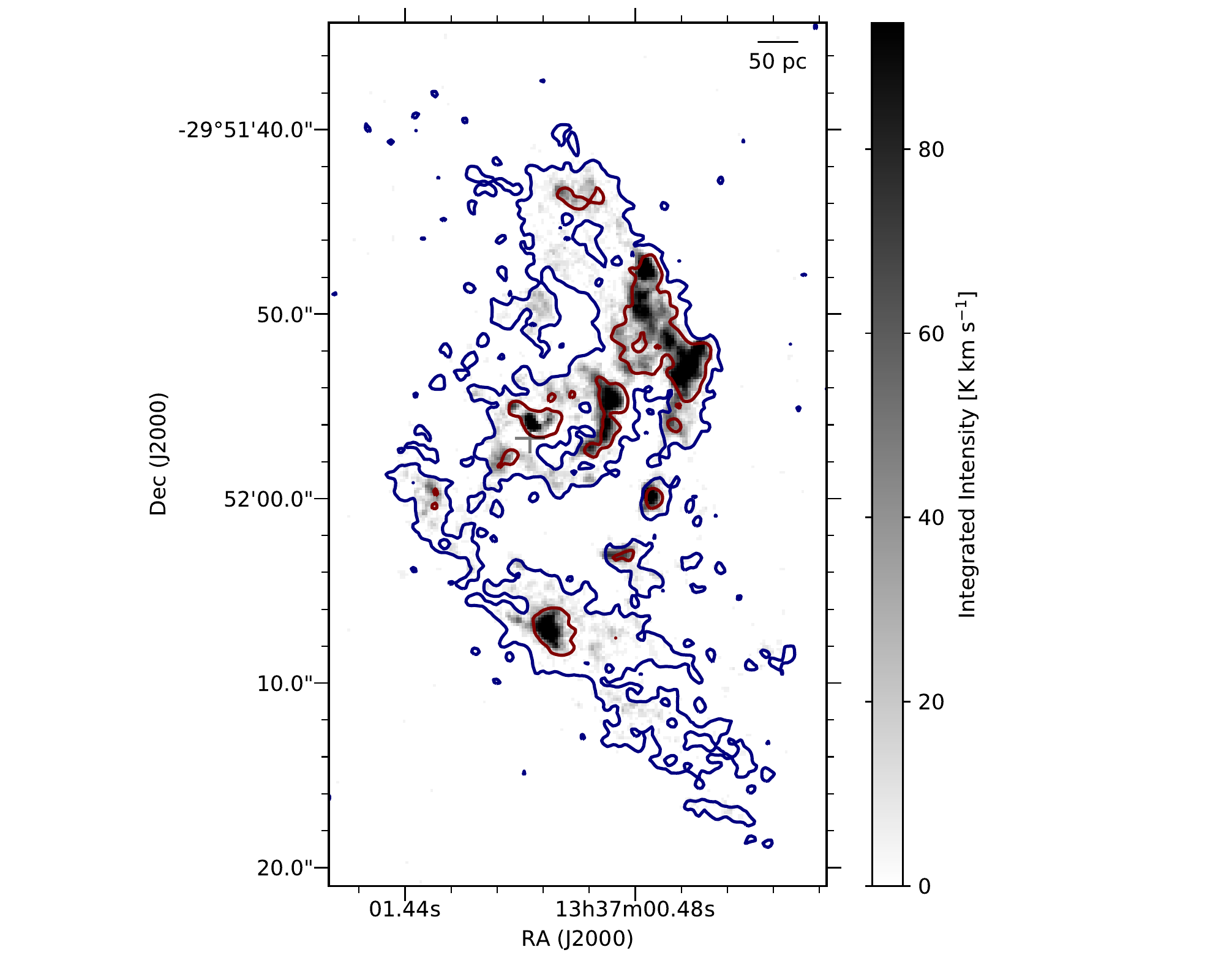}
    \end{tabular}
\caption{\normalsize Integrated intensity map of HCN ($1-0$) and HCO$^{+}$ ($1-0$) produced by SCOUSE. Blue contours show the integrated intensity as produced by CASA at the [25, 75] K km s$^{-1}$ level.}
\label{fig:SCOUSE_HCN_0}
\end{figure*}

\begin{figure*}
	\centering
	\begin{tabular}{cc}
		\includegraphics[trim={2cm 0 0 0},clip,width=0.65\textwidth]{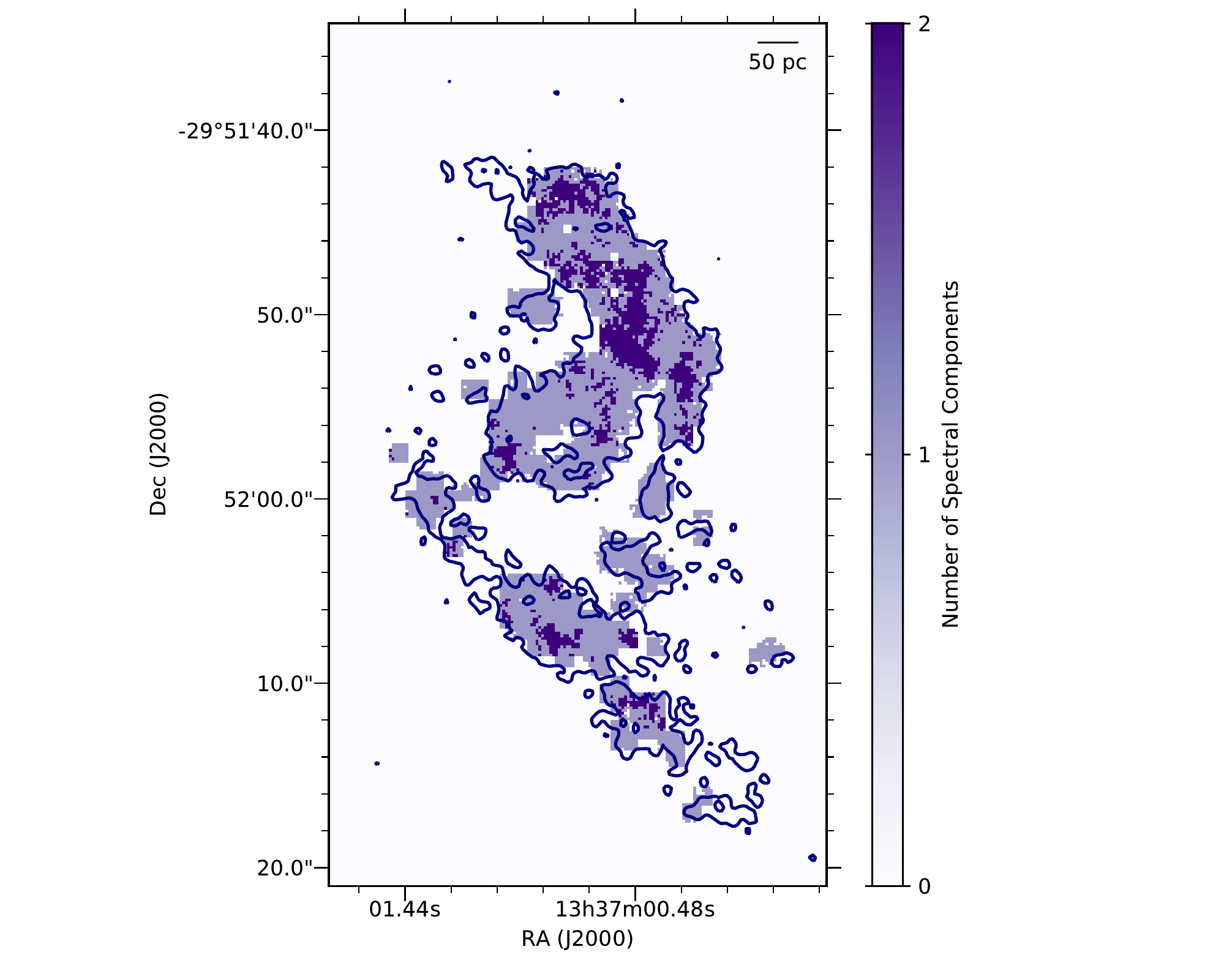} & \hspace{-3cm}
		\includegraphics[trim={2cm 0 0 0},clip,width=0.65\textwidth]{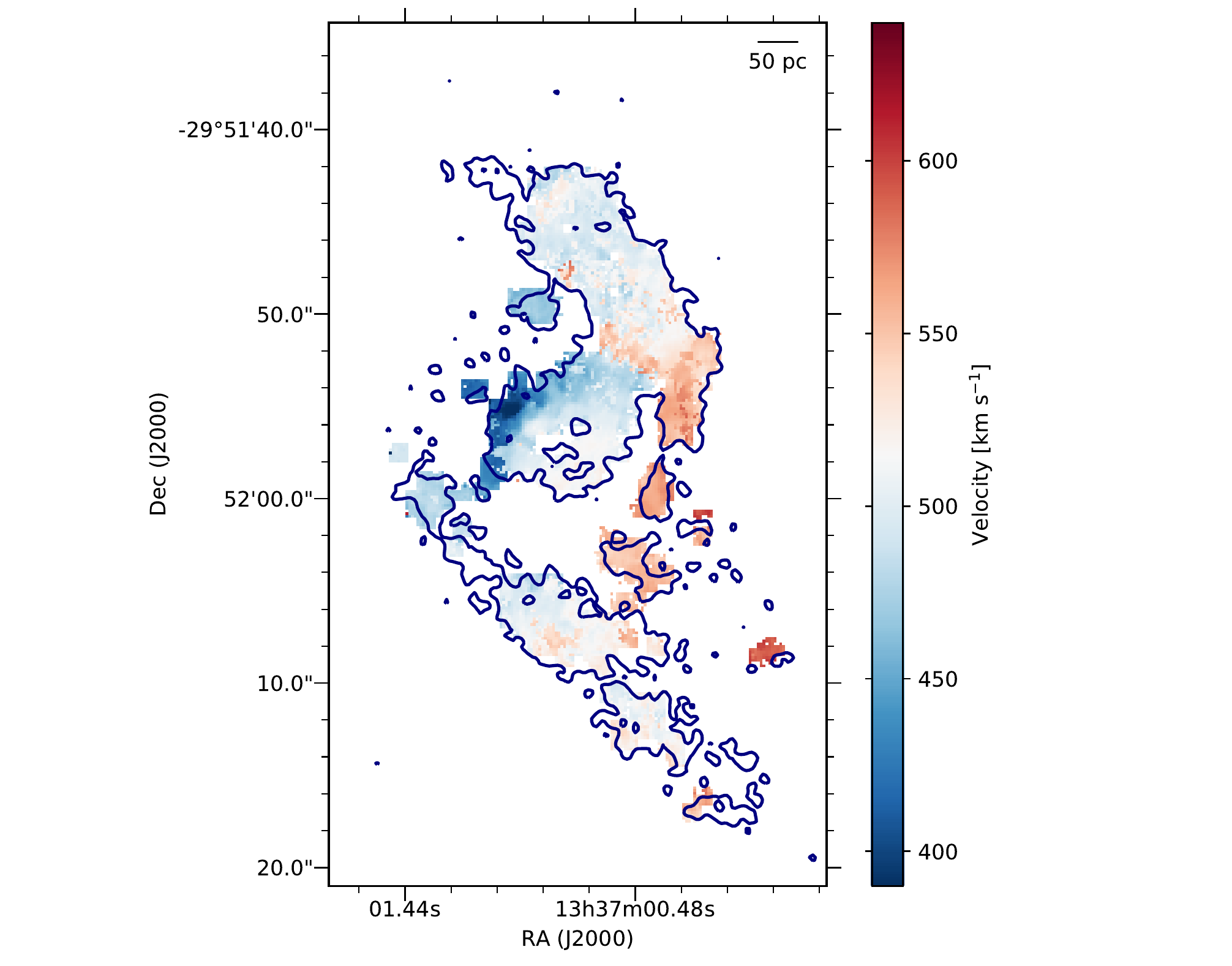} \\
		\includegraphics[trim={2cm 0 0 0},clip,width=0.65\textwidth]{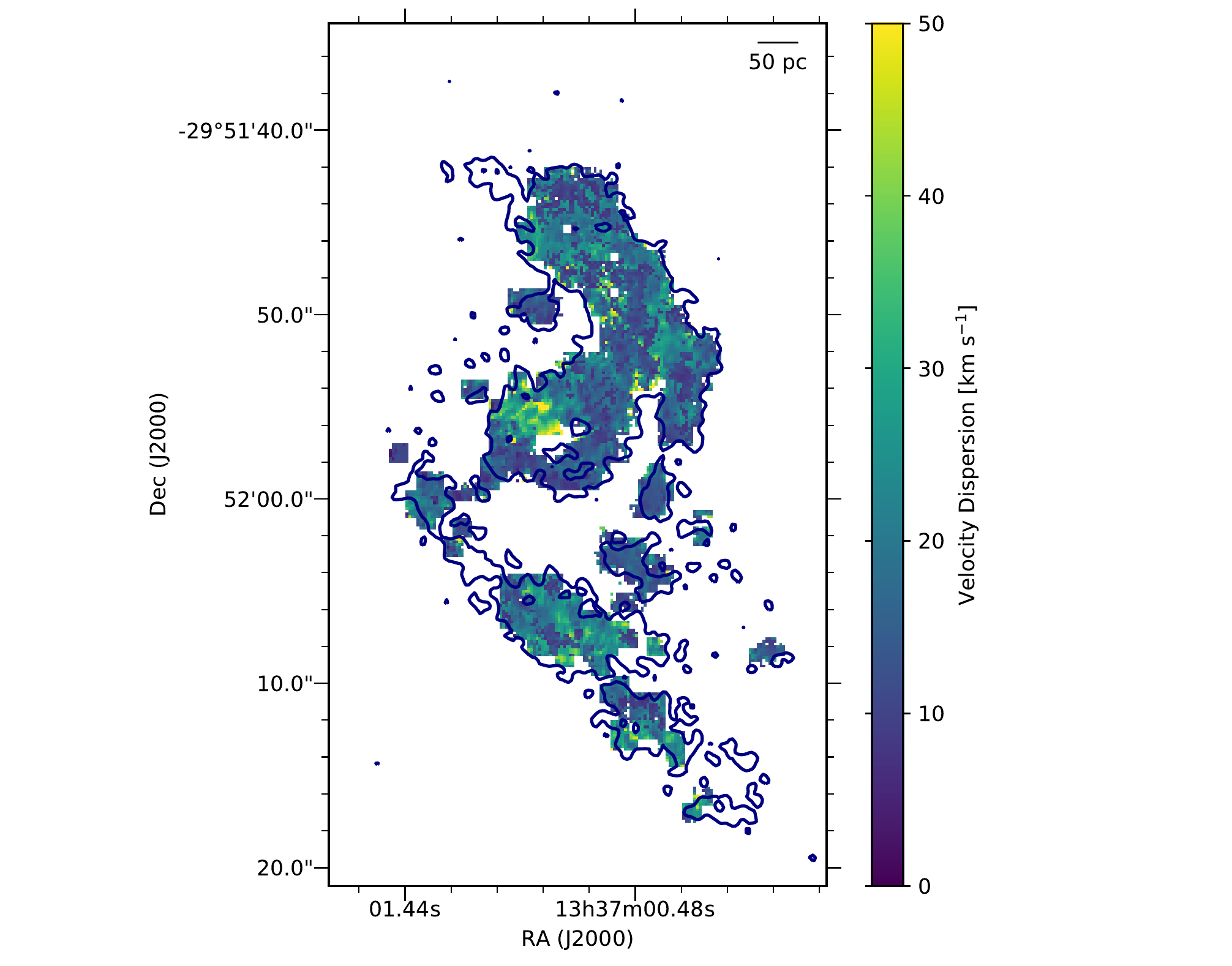} & \hspace{-3cm} \includegraphics[trim={2cm 0 0 0},clip,width=0.65\textwidth]{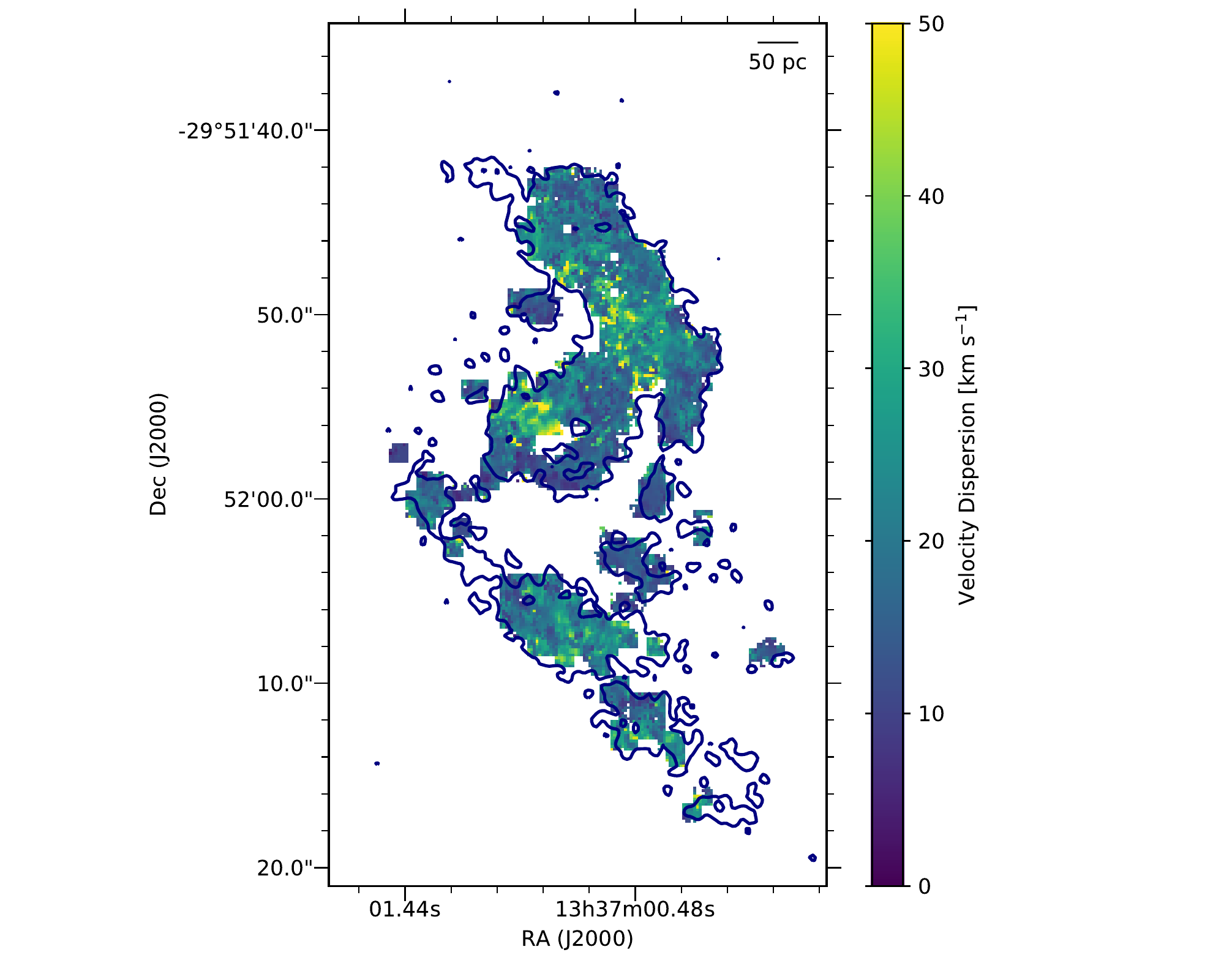}
	\end{tabular}
\caption{\normalsize SCOUSE outputs for HCN ($1-0$). [Top Left]: Number of spectral components per pixel; [Top Right]: centroid velocity; [Bottom Left]: minimum velocity dispersion; [Bottom Right]: maximum velocity dispersion}
\label{fig:HCNG}
\end{figure*}

\begin{figure*}
	\centering
	\begin{tabular}{cc}
		\includegraphics[trim={2cm 0 0 0},clip,width=0.65\textwidth]{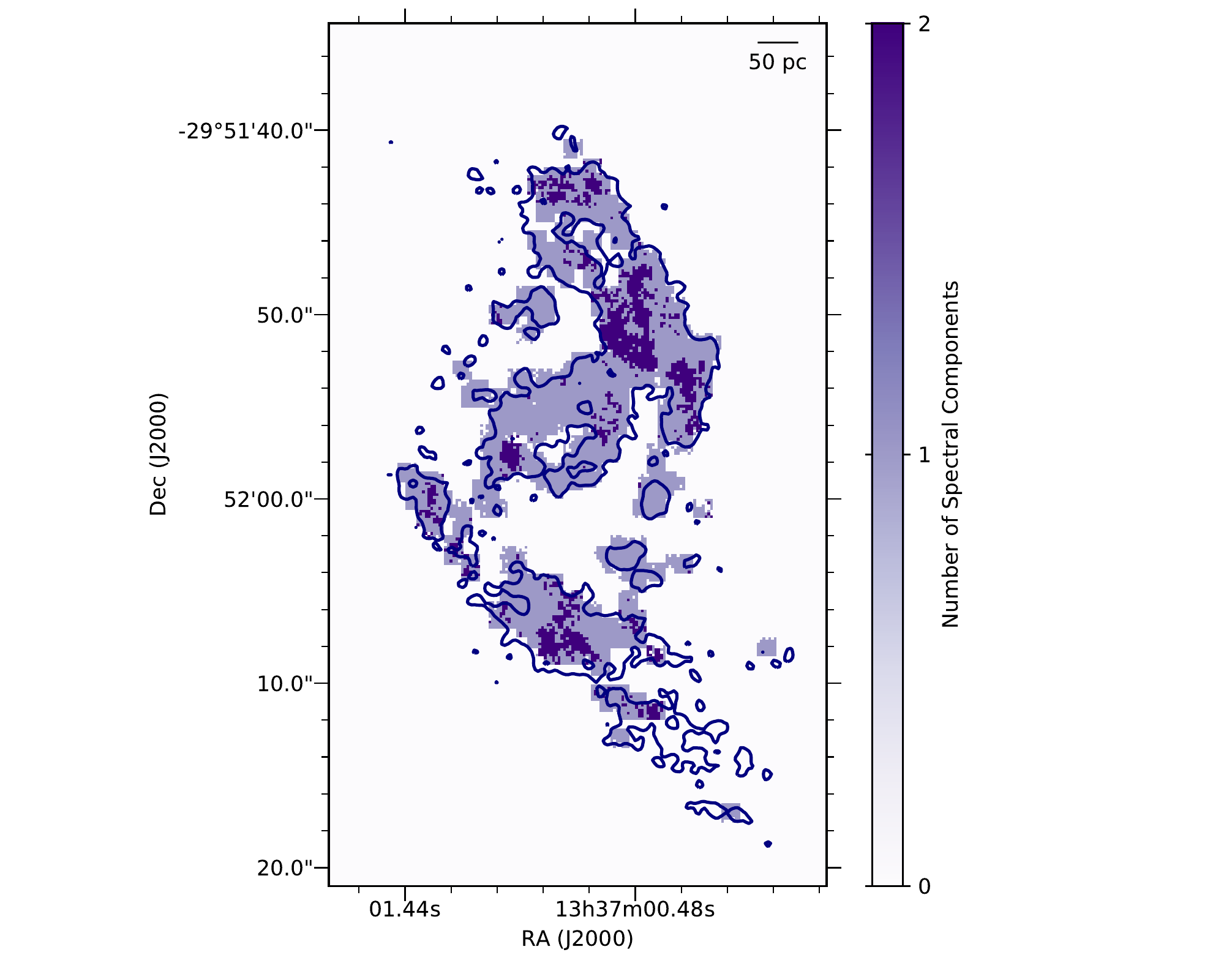} & \hspace{-3cm}
		\includegraphics[trim={2cm 0 0 0},clip,width=0.65\textwidth]{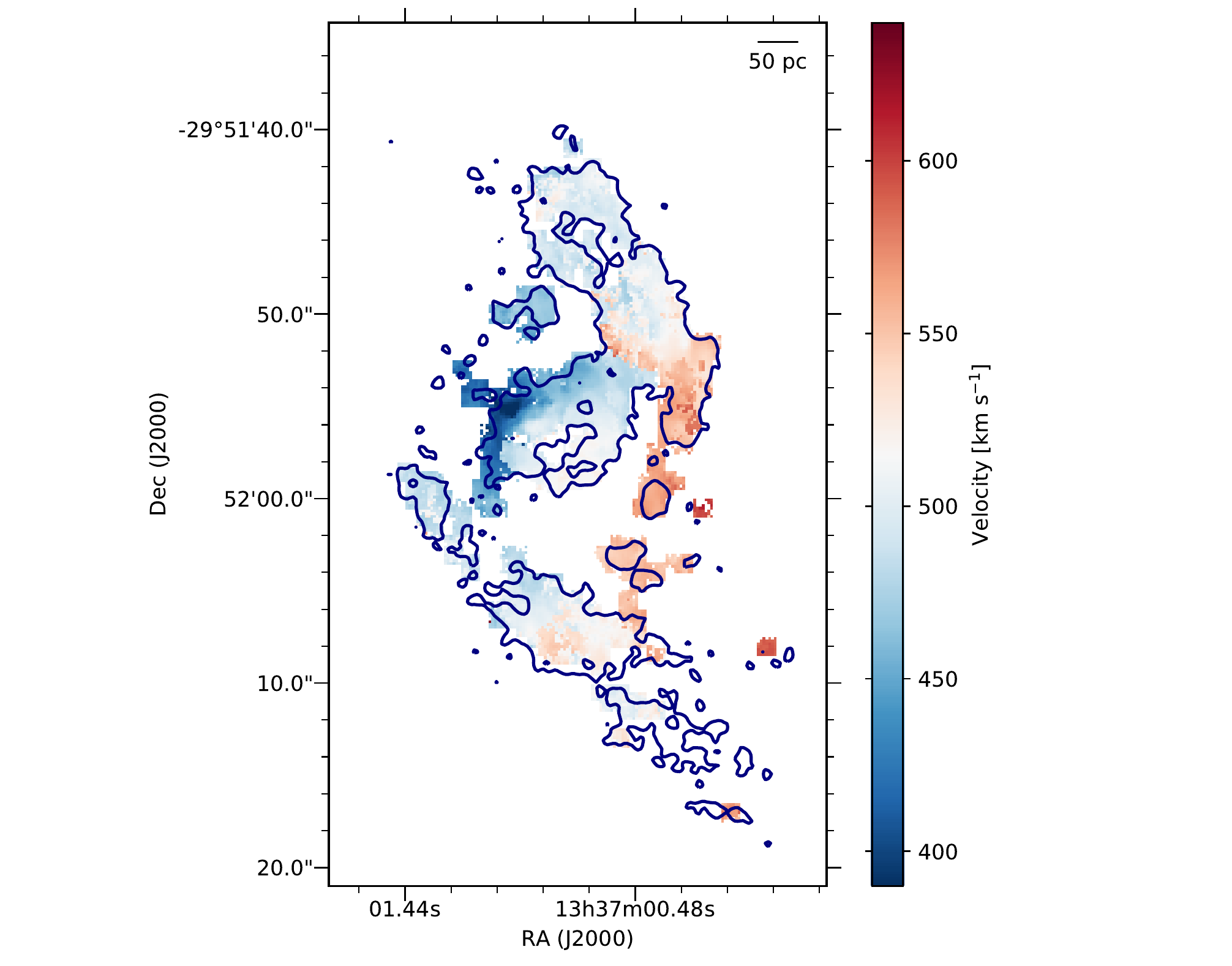} \\
		\includegraphics[trim={2cm 0 0 0},clip,width=0.65\textwidth]{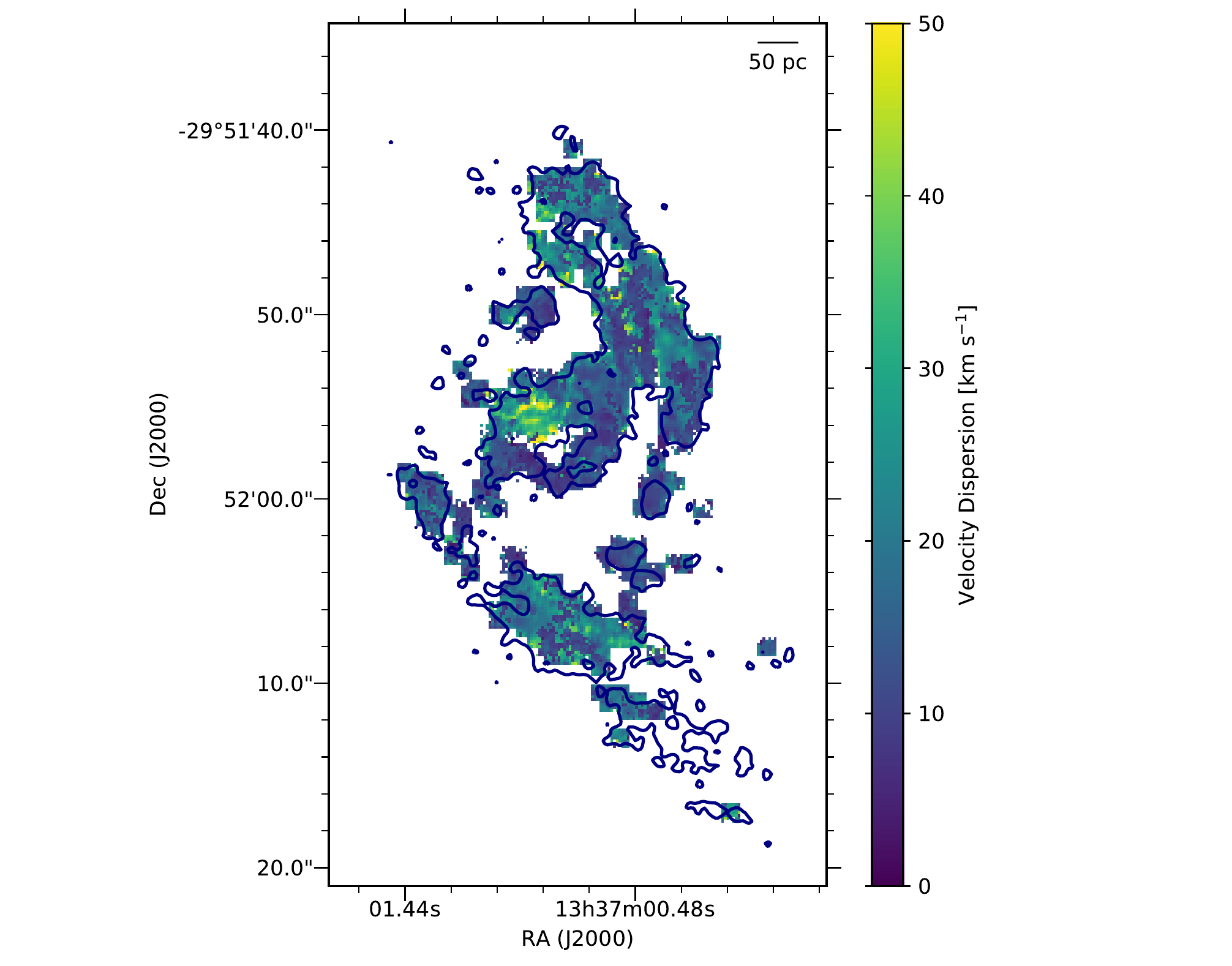} & \hspace{-3cm}
		\includegraphics[trim={2cm 0 0 0},clip,width=0.65\textwidth]{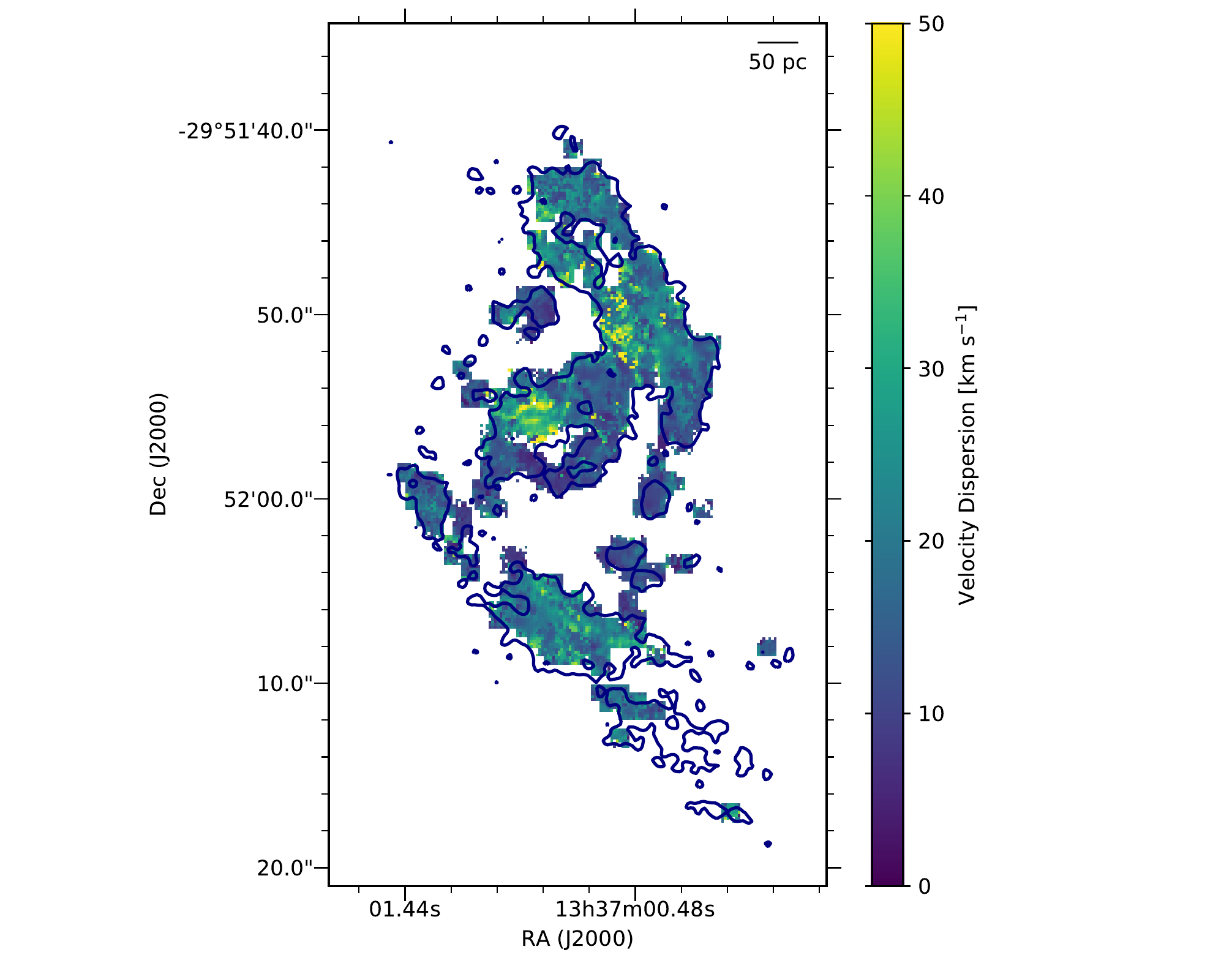}
	\end{tabular}
	\caption{\normalsize SCOUSE outputs for HCO$^{+}$ ($1-0$). [Top Left]: Number of spectral components per pixel; [Top Right]: centroid velocity; [Bottom Left]: minimum velocity dispersion; [Bottom Right]: maximum velocity dispersion}
	\label{fig:HCO+G}
\end{figure*}

Figure~\ref{fig:SCOUSE_HCN_0} shows the integrated intensity maps as produced using the output of SCOUSE for HCN ($1-0$) and HCO$^{+}$ ($1-0$) data, each of which shows almost identical structure to the gas as seen in the integrated intensity maps output from CASA. Centroid velocity and velocity dispersion maps of HCN ($1-0$) and HCO$^{+}$ ($1-0$) were also created using the ouput from SCOUSE. Figure~\ref{fig:HCO+G} shows maps of the number of fitted Gaussian components per pixel, the centroid velocity, and the minimum and maximum velocity dispersions at each pixel (from left to right, and top to bottom) for HCN ($1-0$); Figure~\ref{fig:HCNG} show the same maps for HCO$^{+}$ ($1-0$). To ensure the quality of these fits were sufficient, two rounds of visual inspection were performed. The first being a vital step in the SCOUSE process in which the user is shown each spectra output from the fitting process. The second being an inspection of spectra randomly selected from various locations in the map once the entire SCOUSE process was completed. Both rounds of inspection showed the fitting process had done a good job of recovering the gas kinematics. The output of the fits and fit results are available as an online resource here: XXX.

\bsp	
\label{lastpage}
\end{document}